\documentclass[12pt, a4paper]{article}
 \usepackage[font=small,format=plain,labelfont=bf,up,textfont=normal,up,justification=justified,singlelinecheck=false]{caption}
\usepackage{subcaption}
\usepackage{lmodern}

\usepackage{booktabs}
\usepackage{tabularx}

\usepackage{mathrsfs}
\usepackage[a4paper, left=2cm,right=2cm]{geometry}
\usepackage[colorlinks=true,linkcolor=black,citecolor=teal,urlcolor=MidnightBlue,filecolor=black]{hyperref}
\usepackage{amsfonts}
\usepackage{amsmath,amssymb}
\usepackage{pdflscape}
\usepackage{longtable, array}
\usepackage{setspace}
\usepackage{slashed}
\usepackage{braket}

\usepackage{upgreek} 
\usepackage{hyperref}
\usepackage[dvipsnames]{xcolor}
\definecolor{SchoolColor}{rgb}{0.6471, 0.1098, 0.1882} % Crimson\definecolor{SchoolColor}{rgb}{0.6471, 0.1098, 0.1882} % 猩红
\usepackage{subfloat}
\usepackage{ytableau}
\usepackage{tensor}
\usepackage{cite}
\usepackage{tikz}
\usetikzlibrary{calc}
\usetikzlibrary{patterns}
\usetikzlibrary{arrows.meta}
\usetikzlibrary{decorations.text}
\usepackage[compat=1.0.0]{tikz-feynman}
\usepackage[makeroom]{cancel}
\usepackage{graphicx}% Include figure files\usepackage{graphicx}% 包含图形文件
\usepackage{bm} % For bold math symbols\usepackage{bm} % 用于加粗数学符号
\allowdisplaybreaks[4]
\usepackage{tensor}
\usepackage{cite}
\usepackage{tikz}
\usepackage{graphicx}
\usepackage{graphics}
\graphicspath{{figure/}}
\usepackage{array}
\usepackage{booktabs}

\usepackage{multirow}

\usepackage{dcolumn}% Align table columns on decimal point\usepackage{dcolumn}% 将表格列对齐到小数点
\usepackage{bm}% bold math\usepackage{bm}% 粗体数学

\usepackage{verbatim}

\usepackage{textcomp} % \textless, \textgreater, \textbrokenbar macros\usepackage{textcomp} %  extless,  extgreater,  extbrokenbar 宏
\usepackage{graphicx} % \resizebox macro\usepackage{graphicx} %  esizebox 宏

\numberwithin{equation}{section}
\newcommand{\bea}{\begin{eqnarray}}
\newcommand{\eea}{\end{eqnarray}}
\newcommand{\be}{\begin{equation}}
\newcommand{\ee}{\end{equation}}
\newcommand{\bs}{\begin{subequations}}
\newcommand{\es}{\end{subequations}}
\def\nn{\nonumber}

\newcommand{\beqs}{\begin{eqnarray}}
\newcommand{\eeqs}{\end{eqnarray}}

\numberwithin{equation}{section}

\newcommand{\Rmnum}[1]{\uppercase\expandafter{\romannumeral #1\relax}}
\def\c.c.{\mathrm{c.c.}}

\tikzset{
    dot/.style={circle, draw=black, fill=black, inner sep=1pt},
    n_node/.style={circle, draw=black, fill=gray!10, inner sep=1.2pt, font=\tiny},
    blue_line/.style={blue, thick},
     gray line/.style={lightgray!40, thin},
    red_line/.style={red, thick, dashed, dash pattern=on 2pt off 1.5pt},
    blue arc/.style={blue, thick},
    red curve/.style={red, thick},
    label_style/.style={font=\tiny, gray}
}

\definecolor{preblue}{HTML}{2F80C9}
\definecolor{shellorange}{HTML}{E8843A}
\definecolor{postgreen}{HTML}{3AA76D}
\definecolor{slicemagenta}{HTML}{C13B8A}
\definecolor{mutedgray}{HTML}{66717D}

\tikzset{
  structure/.style={draw=black!82, line width=0.8pt},
  leader/.style={draw=mutedgray!75, line width=0.45pt},
  surface grid/.style={draw=mutedgray!65, dashed, line width=0.45pt},
  shell arrow/.style={draw=shellorange, line width=1.1pt,
    -{Latex[length=2.1mm, width=1.5mm]}},
  null future/.style={draw=postgreen, line width=1.15pt},
  null initial/.style={draw=preblue, line width=1.15pt},
  time slice/.style={draw=slicemagenta, line width=1.5pt},
  crossing/.style={circle, draw=slicemagenta, fill=white,
    line width=0.9pt, inner sep=1.8pt},
  region title/.style={font=\sffamily\small\bfseries},
  region note/.style={font=\sffamily\small, text=mutedgray},
  math halo/.style={fill=white, fill opacity=0.86,
    text opacity=1, inner sep=1.2pt},
}

\begin{document}
\begin{titlepage}

\begin{flushright}\vspace{-3cm}
{\small
%{\tt arXiv:yymm.nnnn} \\
\today }\end{flushright}
\vspace{0.5cm}
\begin{center}
	{{ \LARGE{\bf{Spacelike reduction of the gravitational \\ \vspace{8pt} topological terms and  associated helicity densities}}}}%\vspace{8pt}\\in higher dimensional CFT}}}} 
	\vspace{8mm}
	\centerline{
Hongxi Huang\footnotemark[1],
Jiang Long\footnotemark[2],
Gan Zhao\footnotemark[3]\ \&
Xin-Hao Zhou\footnotemark[4]
}
\footnotetext[1]{\texttt{u202410192@hust.edu.cn}}
\footnotetext[2]{\texttt{longjiang@hust.edu.cn}}
\footnotetext[3]{\texttt{zhaogan@hust.edu.cn}}
\footnotetext[4]{\texttt{zhouxinhao01@hust.edu.cn}}
	\vspace{1mm}
	\normalsize
	\bigskip\medskip 
    
\textit{School of Physics, Huazhong University of Science and Technology, \\ Luoyu Road 1037, Wuhan, Hubei 430074, China
	}
	%\vfil
	%\pacs{04.70.Dy}
	
	\vspace{25mm}
	\begin{abstract}
		\noindent
		Helicity in gravity is a multifaceted concept, and several inequivalent definitions exist in the literature. In this work, we reduce the topological Pontryagin and Nieh-Yan terms to a constant time spacelike hypersurface. Remarkably, upon applying the SVT decomposition at the linearized level, we obtain four distinct helicity densities. The helicity density arising from the Nieh-Yan term is termed the spin-1 and spin-2 gravitomagnetic helicity density, depending on whether it originates from vector or tensor modes. Meanwhile, the helicity density arising from the Pontryagin term is termed the spin-1 and spin-2 gravito-current helicity density, according to the corresponding mode contributions. We study the mode and multipole expansions of these helicity functionals and apply them to leading order Newtonian two-body systems, weak field boosted Kerr, a locally defined slowly varying gyratonic pp-wave model, and a linearized gravitational Hopfion.
		\end{abstract}

\end{center}
\end{titlepage}
\tableofcontents
\section{Introduction}
Helicity plays related but inequivalent roles in different areas of
physics. In the representation theory of massless particles, it labels
irreducible representations of the little group
\cite{1939AnMat..40..149W}. In free Maxwell theory, optical helicity can
be identified with the Noether charge associated with electromagnetic
duality rotations \cite{Calkin1965,DeserTeitelboim1976}. For a barotropic
ideal fluid, the velocity field $\mathbf v$ and vorticity $\boldsymbol\omega$ define the fluid helicity
\(\int d^3x\,\mathbf v\cdot\boldsymbol\omega\)
\cite{Moreau1961,Moffatt1969}. Its magnetohydrodynamic counterpart is the
magnetic helicity
\(\int d^3x\,\mathbf a\cdot\mathbf b\)
\cite{Elsasser1956Hydromagnetic,Woltjer1958Theorem,Moffatt1969,Blackman:2014kxa}  where $\mathbf a$ is the magnetic potential and $\mathbf b$  the magnetic field. On a domain with
a boundary, the gauge invariance of magnetic helicity requires suitable
boundary conditions or the use of relative helicity
\cite{BergerField1984}. 
Gravitoelectromagnetism, i.e., the analogy between linearized gravity and electromagnetism in the weak field and slow motion regime, has been reviewed and developed in \cite{2003gr.qc....11030M,Costa:2012cw}. Within this framework, a gravitomagnetic helicity construction was proposed in \cite{Bini:2021gdb}. Related parity odd spin-1 and spin-2 quantities arise in duality based descriptions of linearized gravity \cite{HenneauxTeitelboim2005,Barnett2014,Aghapour:2018pup,Toth:2021dut}. Electromagnetic analogues include Lipkin's zilch and optical chirality \cite{Lipkin1964Conservation,CameronBarnettYao2012,Philbin2013}, while parity odd tensor structures associated with the gravitational Pontryagin density also appear in parity violating gravity and cosmology \cite{Alexander:2004wk,Kamada:2019ewe}.

%Helicity assumes similar but inequivalent roles across different domains. In the representation theory of massless particles, it serves as a label for little-group representations \cite{1939AnMat..40..149W}.  In free Maxwell theory, it can be identified as the Noether charge associated with electromagnetic duality rotations \cite{Calkin1965,DeserTeitelboim1976}. In fluid dynamics, one can use the fluid velocity and its vorticity to construct fluid helicity \cite{Elsasser1956Hydromagnetic} and this has been extended to the famous magnetic helicity \cite{Woltjer1958Theorem,Blackman:2014kxa} via  a formal analogy in which the velocity field is replaced by the vector potential and the vorticity by the magnetic field. Furthermore, Gravitoelectromagnetism (GEM) \cite{2003gr.qc....11030M,Costa:2012cw},  a profound formal analogy between linearized general relativity in the weak field and slow motion limit and classical electrodynamics, has been used to propose gravitomagnetic helicity \cite{Bini:2021gdb}. More interesting definitions can be found in \cite{Alves:2018wku,Facundo:2025dgz,Barnett2014,Aghapour:2018pup,Toth:2021dut,Alexander:2004wk,Kamada:2019ewe,Lipkin1964Conservation}. 

The present work is motivated by two recent developments. The first concerns the radiative electromagnetic helicity flux at future null infinity \(\mathscr I^+\) \cite{Heng:2025kmr,Liu:2023qtr,Liu:2024rvz}. This quantity should be distinguished from the magnetic, electric, and optical helicities defined as charges on spacelike hypersurfaces. The second development involves the comparison of the asymptotic descendants of the gravitational Pontryagin and Nieh-Yan densities \cite{Long:2025fbb}. Under the standard Bondi-Sachs falloff conditions \cite{Bondi1962,Sachs1962,MadlerWinicour2016}, that analysis finds that the complete Pontryagin Chern-Simons three-form alone does not reproduce the proposed radiative spin-2 helicity flux \cite{Liu:2023gwa}, whereas the teleparallel Nieh-Yan descendant can yield a parity odd \(\epsilon^{AB}C_{AC}\partial_u {C_{B}}^C\)-type flux. Here \(C_{AB}\) denotes the Bondi shear and \(u\) is retarded time.
%The direct motivation for the present project arises from two lines of recent work. The first \cite{Heng:2025kmr} concerns the definition of radiative electromagnetic helicity flux density at future null infinity ($\mathscr I^+$)\cite{Liu:2023qtr,Liu:2024rvz}, which emphasises its distinction from the magnetic/electric/optical helicities defined on ordinary spatial slices. The second \cite{Long:2025fbb} involves a comparison between the descent of Pontryagin and Nieh-Yan boundary terms in gravity, showing that under the standard Bondi falloff conditions, the full Pontryagin Chern-Simons boundary form cannot directly yield the target helicity flux \cite{Liu:2023gwa}, whereas the teleparallel Nieh-Yan boundary term can descend to a $C\dot C$-type helicity flux. 

A useful observation is that several of the helicity densities relevant to the present work arise from the descent of exact characteristic classes. In electromagnetism, the Abelian Pontryagin density, or equivalently the second Chern character, satisfies
\begin{equation}
da\wedge da  =d(a\wedge da),
\end{equation}
and the pullback of \(a\wedge da\) to a spacelike hypersurface gives the magnetic helicity. In gravity,
\begin{equation}
R^a{}_b\wedge R^b{}_a=dQ_{3},
\end{equation}
where \(Q_{3}\) is the gravitational Chern-Simons three-form. Likewise, the Nieh-Yan density is exact \cite{Nieh:1981ww},
\begin{equation}
N_{\rm NY}
=d\!\left(e^a\wedge T_a\right)
=
T^a\wedge T_a
-e^a\wedge e^b\wedge R_{ab}.
\end{equation}
Here \(e^a\) denotes the vielbein (or coframe) one-form, while \(T^a\) and \(R^a{}_b\) denote the torsion and curvature two-forms associated with the spin connection \(\omega^a{}_b\), respectively. In teleparallel equivalent of general relativity (TEGR), \(R^a{}_b(\omega)=0\), since \(\omega^a{}_b\) is a flat teleparallel spin connection. This should not be confused with the generally non-vanishing curvature of the Levi-Civita connection. The Pontryagin and Nieh-Yan densities can enter chiral anomaly relations \cite{Kimura1969,DelbourgoSalam1972,ChandiaZanelli1997,Obukhov1997}. The Nieh-Yan coefficient, however, generally depends on the regularization prescription or ultraviolet scale and should not be regarded as universal in the same sense as the Pontryagin contribution.

In this paper, we investigate the pullback of the Nieh-Yan boundary form to a spacelike hypersurface in the linearized TEGR. Using the scalar-vector-tensor (SVT) decomposition \cite{Flanagan:2005yc}, we show that, modulo codimension-two surface terms, the resulting bulk term separates into spin-1 and spin-2 contributions. The spin-1 contribution has the same  form as gravitomagnetic helicity, up to convention dependent normalization, whereas the spin-2 contribution is closely related to terms appearing in duality based descriptions of linearized gravitational waves \cite{Barnett2014,HenneauxTeitelboim2005}. To our knowledge, the latter has not previously been isolated as the spin-2 sector of the spatial Nieh-Yan descendant.

We also perform the corresponding reduction of the gravitational
Pontryagin density. Its spatial descendant again separates, modulo
codimension-two terms, into spin-1 and spin-2 bulk sectors. Their derivative-weighted parity odd structure is analogous to current helicity and to Lipkin's zilch or optical chirality \cite{Lipkin1964Conservation,CameronBarnettYao2012,Philbin2013}. Since these bulk expressions are written entirely in terms of the SVT gauge invariant variables, they are invariant under linearized diffeomorphisms.Their separation from the surface terms, as well as
their conservation properties, nevertheless depends on the equations
of motion and the imposed falloff conditions. We therefore use the
term "helicity" for these parity odd quadratic functionals without
assuming that every one of them is an independently conserved Noether
charge.

%The interesting part is that different proposals on helicities (spin-1 or spin-2) are actually boundary terms of topological quantities on different type of hypersurfaces while these topological terms are ultimately related to the chiral anomalies\cite{}. For example, the boundary term of the second Chern class in electromagnetic theory is a Chern-Simons term. The Chern-Simons term leads to the magnetic helicity on a spacelike hypersurface and spin-1 helicity flux at future null infinity. Similarly, the Nieh-Yan boundary term in TEGR leads to the spin-2 helicity flux.
%The remaining question is to explore the physical meaning of this boundary Nieh-Yan term on a spacelike hypersurface. In this paper, we study this problem and show that it indeed leads to a concept of helicity. However, according to the scalar-vector-tensor(SVT) decomposition of the gravitational field, the helicity is decomposed into a summation of the spin-1 and spin-2 part. The spin-1 part 
%matches exactly the gravitomagnetic helicity \cite{Bini:2021gdb} while the spin-2 part has not been mentioned in the literature, though this term appears as a part of the "gravitational-wave analogue of the electromagnetic helicity" from duality rotation \cite{Barnett2014}. Based on this novel result, we also reduce the Pontryagin term to a spacelike hypersurface. Interestingly, the boundary term still decomposes into a spin-1 and a spin-2 part. Formally, they are analogies of Lipkin's zilch charge of electromagnetism \cite{Lipkin1964Conservation}. Both of these terms are gauge invariant at the linearized level. 

The structure of this paper is organized as follows. Section \ref{setup} presents the SVT  decomposition of the gravitational field. Using the six gauge invariant modes, Section \ref{topredop} defines four helicity densities by reducing the Pontryagin and Nieh-Yan terms to a constant time hypersurface. Section \ref{mode} elaborates on the mode expansions of these densities and Section \ref{multipoleexpansion} focus on their multipole expansions. These formulations are then applied to two-body systems, weak field boosted Kerr, a slowly varying gyratonic pp-wave model and a gravitational Hopfion  before concluding.

%Review on magnetic helicity \cite{Blackman:2014kxa}. Fluid helicity \cite{Elsasser1956Hydromagnetic}. Magnetic helicity \cite{Woltjer1958Theorem}. spin-1 NY \cite{Bini:2021gdb}. Gravitomagnetic helicity (spin 1) \cite{Alves:2018wku,Facundo:2025dgz}. spin-2 NY \cite{Barnett2014Maxwellian，Aghapour:2018pup,Toth:2021dut}. Spin-2 Pontragin \cite{Alexander:2004wk,Kamada:2019ewe}. Spin-1 Pontragin is part of the Lipkin zilch density \cite{Lipkin1964Conservation}
\section{Setup}\label{setup}
In this section, we will follow \cite{Flanagan:2005yc} to review the SVT decomposition of the gravitational field in flat spacetime. The SVT decomposition in a broader context can be found in \cite{Bardeen:1980kt,Kodama:1984ziu}. We will focus on linearized gravity  where the metric $g_{\mu\nu}$ is considered as a small deformation of the Minkowski metric $\eta_{\mu\nu}$
\be 
g_{\mu\nu}=\eta_{\mu\nu}+ h_{\mu\nu}
\ee where $\eta_{\mu\nu}=\text{diag}(-,+,+,+)$ and the magnitude of the gravitational field $h_{\mu\nu}$ is rather small $|h_{\mu\nu}|\ll 1$. The linearized coordinate transformation of $h_{\mu\nu}$ is 
\be 
\delta_\xi h_{\mu\nu}=-\partial_\mu\xi_\nu-\partial_\nu\xi_\mu=-2\partial_{(\mu}\xi_{\nu)}
\ee where $\xi^\mu$ denotes the infinitesimal coordinate transformation $x^\mu\to x'^{\mu}=x^\mu+\xi^\mu$ and its indices are lowered by $\eta_{\mu\nu}$. In an arbitrary gauge, we decompose $h_{\mu\nu}$ according to the irreducible representation of the spatial $\text{SO}(3)$ group 
\bs\begin{align}
    h_{00}&=2\phi,\\ 
    h_{0i}&=\beta_i+\partial_i\gamma,\\
    h_{ij}&=h_{ij}^{\text{TT}}+\frac{1}{3}H \delta_{ij}+\partial_{(i}\epsilon_{j)}+(\partial_i\partial_j-\frac{1}{3}\delta_{ij}\nabla^2)\lambda.
\end{align}\es We use normalized symmetrization and Laplacian as follows\footnote{A similar
anti-symmetrization will be defined as
\[
    X_{[i}Y_{j]}
\equiv
\frac12\left(X_iY_j-X_jY_i\right).
\]
}
\begin{equation}
X_{(i}Y_{j)}
\equiv
\frac12\left(X_iY_j+X_jY_i\right), \qquad \nabla^2=\delta^{ij}\partial_i\partial_j.
\end{equation} Here $H=\delta^{ij}h_{ij}$ is the spatial trace of $h_{ij}$ while  $h_{ij}^{\text{TT}}$ is a symmetric, transverse and traceless spin-2 mode in the sense that 
\be 
h_{ij}^{\text{TT}}=h_{ji}^{\text{TT}},\quad \partial^i h_{ij}^{\text{TT}}=0,\quad \delta^{ij}h_{ij}^{\text{TT}}=0.
\ee Therefore, only two of the components of $h_{ij}^{\text{TT}}$ are independent. 
Similarly, $\beta_i$ and $\epsilon_i$ are transverse spin-1 modes due to the constraints 
\be 
\partial^i\beta_i=0,\quad \partial^i\epsilon_i=0.
\ee Thus each spin-1 mode contributes two independent components. 
Combining with the four $\text{SO}(3)$ scalar $\phi,\gamma, H,\lambda$ modes, the number of the independent components is exactly ten
\be 
2+2\times 2+4=10.
\ee A parallel decomposition of the vector field $\xi^\mu$ is 
\be 
\xi_\mu=(A,B_i+\partial_i C)
\ee where $B_i$ is transverse 
\be 
\partial^i B_i=0.
\ee It follows that the following combinations are gauge invariant at the linearized level 
\be 
\Phi=-\phi+\dot\gamma-\frac{1}{2}  \ddot{\lambda},\quad \Theta=\frac{1}{3}(H-\nabla^2\lambda),\quad \Xi_i=\beta_i-\frac{1}{2}\dot\epsilon_i,\quad h_{ij}^{\text{TT}}.
\ee Note that $\Xi_i$ is transverse $\partial^i\Xi_i=0$. In the following, we will refer $h_{ij}^{\text{TT}}$ as the spin-2 modes, $\Xi_i$ the spin-1 modes. The remaining two quantities $\Phi,\Theta$ are gauge invariant scalar modes. 
We assume that all perturbations fall off sufficiently rapidly at
spatial infinity so that the SVT decomposition is
unique and harmonic zero modes can be excluded. Given the linearized metric $h_{\mu\nu}$, the gauge invariant modes are obtained via the formula 
\bs\label{inverse}\begin{align}
\Phi
={}&
-\frac12h_{00}
+\nabla^{-2}\partial^i\dot h_{0i}
-\frac34\nabla^{-4}
\partial^i\partial^j  \ddot h_{ij}
+\frac14\nabla^{-2}  \ddot h_{ii},
\\[1mm]
\Theta
={}&
\frac12
\left(
h_{ii}
-\nabla^{-2}\partial^i\partial^j h_{ij}
\right),
\\[1mm]
\Xi_i
={}&
\hat P_i{}^j
\left(
h_{0j}
-\nabla^{-2}\partial^k\dot h_{jk}
\right),
\\[1mm]
h_{ij}^{\mathrm{TT}}
={}&
\hat\Pi_{ij,kl}h_{kl}.
\end{align}\es where the projector is defined with the inverse Laplacian
\begin{equation}
\hat P_{ij}=\delta_{ij}-\frac{\partial_i\partial_j}{\nabla^2},
\qquad
\hat\Pi_{ij,kl}
=\frac12\left(\hat P_{ik}\hat P_{jl}+\hat P_{il}\hat P_{jk}-\hat P_{ij}\hat P_{kl}\right).
\label{eq:TT-projector-position}
\end{equation}
Unless stated otherwise, the spatial SVT decomposition is defined on a
complete constant time slice
\(V\simeq\mathbb R^3\) of the inertial Minkowski foliation.
All perturbations and admissible infinitesimal diffeomorphisms are assumed
to satisfy the same regularity and falloff conditions at spatial infinity.
The inverse Laplacian is defined by the Green function that vanishes at
spatial infinity,
\be 
(\nabla^{-2}f)(\mathbf x)
=
-\frac{1}{4\pi}
\int_{\mathbb R^3}d^3\mathbf y\,
\frac{f(\mathbf y)}{|\mathbf x-\mathbf y|},\label{inverselap}
\ee 
or, equivalently, for nonzero Fourier modes,
\[
\widehat{\nabla^{-2}f}(\mathbf k)
=
-\frac{\widehat f(\mathbf k)}{|\mathbf k|^2},
\qquad
\mathbf k\neq0,
\qquad
\nabla^{-4}=(\nabla^{-2})^2.
\]
No independent zero or harmonic modes are retained. They are fixed by
regularity and the boundary conditions at spatial infinity. Consequently,
the term "gauge invariant" below means invariant under infinitesimal
diffeomorphisms that preserve these conditions. Since the SVT projectors
contain inverse Laplacians, the resulting modes are nonlocal fields defined
on the entire slice \(\mathbb R^3\). Whenever an idealized example does not
satisfy the assumed global falloff conditions, we interpret the displayed
SVT modes only as local or asymptotic expressions, supplemented implicitly
by an appropriate infrared completion.

The linearized Christoffel symbol is  
\bs\begin{align}
    \Gamma^0_{00}&=-\dot\phi,\\
    \Gamma^0_{0i}&=-\partial_i\phi,\\
\Gamma^0_{ij}&
=\frac12\dot h_{ij}-\partial_{(i}\beta_{j)}-\partial_i\partial_j\gamma,\\
\Gamma^i_{00}&=\dot{\beta}_i+\partial_i\dot\gamma-\partial^i\phi,\\
\Gamma^i_{0j}&
=\frac12\dot h_{ij}
+\frac12(\partial_j\beta_i-\partial_i\beta_j),\\ 
\Gamma^i_{jk}&
=\frac12(\partial_jh_{ki}+\partial_kh_{ji}-\partial_ih_{jk})
\end{align}\es while the Riemann tensor is 
\bs\begin{align}
    R_{0i0j}
&=\partial_i\partial_j\Phi
+\partial_{(i}\dot\Xi_{j)}
-\frac12\ddot h^{\rm TT}_{ij}
-\frac12\delta_{ij}\ddot\Theta ,\\
R_{0ijk}
&=\frac12(\partial_k\dot h^{\rm TT}_{ij}
-\partial_j\dot h^{\rm TT}_{ik})
+\frac12\partial_i(\partial_j\Xi_k-\partial_k\Xi_j)
+\frac12(\delta_{ij}\partial_k\dot\Theta
-\delta_{ik}\partial_j\dot\Theta),\\ 
R_{ijkl}
&=\frac12(\partial_k\partial_jh^{\rm TT}_{il}
+\partial_l\partial_ih^{\rm TT}_{jk}
-\partial_k\partial_ih^{\rm TT}_{jl}
-\partial_l\partial_jh^{\rm TT}_{ik})
+\frac12(\delta_{il}\partial_j\partial_k\Theta
+\delta_{jk}\partial_i\partial_l\Theta
-\delta_{jl}\partial_i\partial_k\Theta
-\delta_{ik}\partial_j\partial_l\Theta).
\end{align}\es  As expected, the Riemann tensor is expressed as the six gauge invariant quantities. The six gauge invariant metric components should not be confused with six
propagating degrees of freedom. In linearized Einstein gravity, the
scalar and vector sectors are constrained, and only the two tensor
polarizations propagate in vacuum.

The SVT decomposition and the above curvature identities are
kinematical and do not rely on a particular gravitational action.
They therefore apply to metric perturbations about Minkowski spacetime
in any diffeomorphism invariant metric theory. The equations of motion,
constraint structure, and number of propagating degrees of freedom are,
however, theory dependent. In TEGR, the metric decomposition must also
be supplemented by the tetrad and inertial spin connection sectors
associated with local Lorentz invariance \cite{Krssak:2018ywd}.

\section{Reduction of topological terms}\label{topredop}
%In this section, we will reduce the Pontryagin term and Nieh-Yan term to a spacelike hypersurface. 

In this section, we reduce the gravitational Pontryagin and Nieh-Yan
four-forms to a constant time spacelike hypersurface \(V\). Working
to quadratic order in perturbations about Minkowski spacetime, we use
the SVT decomposition to express their hypersurface
descendants in terms of the gauge invariant variables
\(\Phi\), \(\Theta\), \(\Xi_i\), and \(h_{ij}^{\rm TT}\). In both
cases, the bulk contribution separates into a transverse vector
sector and a transverse traceless tensor sector, while the scalar and
mixed sectors contribute only through codimension-two surface terms.
The vector and tensor contributions obtained from the Nieh-Yan
descendant have the one-derivative structures of spin-1 and spin-2
gravitomagnetic helicities, whereas the corresponding Pontryagin
contributions contain two additional derivatives and have the
structures of gravito-current helicities. We also identify all
codimension-two terms and state the boundary conditions under which
they may consistently be neglected.

For the Nieh-Yan reduction, we employ the covariant formulation of
TEGR, in which the tetrad
and the flat inertial spin connection are treated as a Lorentz covariant
pair. This makes it possible to separate the physical SVT perturbations
from infinitesimal diffeomorphisms and local Lorentz transformations.
For the Pontryagin tensor sector, we retain the complete off-shell
hypersurface descendant and carefully distinguish it from the
kinetic, derivative-weighted helicity  that can be introduced
on-shell. Their relation involves an additional quantity \(Q^T\) and
therefore depends on both the initial data and the flux through the
codimension-two boundary.
\subsection{Pontryagin term}
We choose a constant time slice $V$ with future directed
unit normal $n_\mu=(-1,0,0,0)$ . The four-dimensional orientation is chosen as \(\epsilon^{0123}=-1\). Accordingly, the induced Levi-Civita symbol on \(V\) is defined by
\[
\epsilon^{ijk}\equiv n_\mu\epsilon^{\mu ijk},
\qquad
\epsilon^{123}=+1.
\] Let
\(
\Gamma^\alpha_{\mu\beta}\,dx^\mu
\)
denote the Levi-Civita connection, the gravitational
Chern-Simons term on $V$ is
\be 
H_{CS}=\int_V d^3x n_\mu \epsilon^{\mu\nu\rho\sigma}\left(\Gamma^\alpha_{\nu\beta}\partial_\rho\Gamma^\beta_{\sigma\alpha}+\frac23\Gamma^\alpha_{\nu\beta}\Gamma^\beta_{\rho\gamma}\Gamma^\gamma_{\sigma\alpha} \right).\label{cs}
\ee  
It is the hypersurface descendant of the Pontryagin four-form,
\[
P_4
\equiv
\operatorname{Tr}(\boldsymbol R\wedge\boldsymbol R)
=
dQ_3(\boldsymbol\Gamma),\qquad \boldsymbol{\Gamma}^{\alpha}{}_{\beta}
=
\Gamma^\alpha_{\mu\beta}\,dx^\mu
\]
where
\[
Q_3(\boldsymbol\Gamma)
=
\operatorname{Tr}
\left(
\boldsymbol\Gamma\wedge d\boldsymbol\Gamma
+
\frac23\boldsymbol\Gamma\wedge
\boldsymbol\Gamma\wedge\boldsymbol\Gamma
\right).
\] and the curvature two-form is 
\[\mathbf R=d\mathbf\Gamma+\mathbf\Gamma\wedge\mathbf\Gamma.\] Although the connection is retained only to linear order in
\(h_{\mu\nu}\), the leading non-vanishing Chern-Simons term is
quadratic in the perturbation. Therefore,
\be
H_{\rm CS}^{(2)}
=
\int_{V}d^3x\,
\epsilon^{ijk}
\Gamma^{(1)\alpha}_{i\beta}
\partial_j\Gamma^{(1)\beta}_{k\alpha},
\label{csquadratic}
\ee
while the cubic connection term contributes only at
\(\mathcal O(h^3)\).
%At the linearized level, only the quadratic terms are important. Since the Chern-Simons term \eqref{cs} is the boundary term of $\tau=\int R^{ab}\wedge R_{ab}$ while the Riemann tensor is composed by the 6 gauge invariant modes, the boundary term itself should also be fixed by the 6 modes. 

Assuming boundary conditions that remove the zero modes of the spatial
SVT decomposition, a convenient gauge invariant representative of the
metric perturbation is
\be [h_{00}]=-2\Phi,\quad [h_{0i}]=\Xi_i,\quad [h_{ij}]=h_{ij}^{\text{TT}}+\Theta \delta_{ij}
\ee where each term is gauge invariant. A general metric perturbation can then be written as 
\be 
h_{\mu\nu}=[h_{\mu\nu}]+2\partial_{(\mu}\zeta_{\nu)}
\ee where 
\be 
\zeta_0=\gamma-\frac{1}{2}\dot\lambda,\quad \zeta_i=\frac12(\epsilon_i+\partial_i\lambda).
\ee It is important that the Pontryagin four-form is gauge invariant at
this order, whereas the Chern-Simons three-form is invariant only up
to an exact form. Indeed,
\be
\Gamma^{(1)\alpha}_{\mu\beta}[h]
=
\Gamma^{(1)\alpha}_{\mu\beta}[[h]]
+
\partial_\mu\partial_\beta\zeta^\alpha.
\label{gammadecomposition}
\ee
Substitution into \eqref{csquadratic} gives
\be
H_{\rm CS}^{(2)}[h;V]
=
H_{\rm CS}^{(2)}[[h];V]
+
\int_{\partial V}d^2y\,\sqrt q\,
r_i\mathcal J_\zeta^i,
\label{csgaugevariation}
\ee
where
\be
\mathcal J_\zeta^i
=
\epsilon^{ijk}
(\partial_\beta\zeta^\alpha)
\partial_j
\Gamma^{(1)\beta}_{k\alpha}[[h]].
\label{Jzeta}
\ee
Consequently, the bulk part of the hypersurface descendant is fixed
by the gauge invariant SVT modes, but the Chern-Simons term on
a hypersurface with boundary retains a codimension-two surface
ambiguity. This term vanishes when \(\partial V=\varnothing\), when
the relevant boundary conditions are imposed, or when the
gauge invariant representative \(\zeta_\mu=0\) is chosen. Here the coordinates \(y^A\), \(A=1,2\), are intrinsic coordinates on
\(\partial V\), \(r_k\) is the outward-pointing unit normal to
\(\partial V\) within \(V\), and
\[
q=\det(q_{AB})
\]
is the determinant of the induced two-dimensional metric. Since the
integrand is already quadratic in the perturbations, \(r_k\) and
\(q_{AB}\) may be evaluated using the background geometry at this
order. 

Substituting the gauge invariant SVT representative
\([h_{\mu\nu}]\) into \eqref{csquadratic} and organizing the result
according to its spatial \(SO(3)\) content, we find that only the
vector-vector and tensor-tensor sectors contribute to the bulk
integral. The pure
scalar sector gives no bulk contribution, whereas the nonvanishing
terms involving \(\Theta\), as well as the mixed vector-tensor terms,
can be written as total spatial divergences. After performing the spatial integrations by parts according to the
convention adopted below, the quadratic Chern-Simons term
evaluated on the gauge invariant representative takes the form
\be
H_{\rm CS}^{(2)}[[h]]
=
H_{\rm CS}^{V}
+
H_{\rm CS,0}^{T}
+
B_{\rm CS}^{\rm SVT}.
\label{CSdecomposition}
\ee The vector and tensor bulk contributions are
%According to the SVT decomposition,  the Chern-Simons term is decomposed into the several different types. For example, the contribution from $\Phi\Phi$ is obtained by setting $\Xi_i=0,\ \Theta=0,\ h_{ij}^{\text{TT}}=0$. The result is 
%\be 
%H_{CS}\Big|_{\Phi\Phi}=0.
%\ee 
%The complete  result is 
%\be 
%H_{CS}=H_{CS}^{V}+H_{CS}^{T}+B_{CS}
%\ee  where $H_{CS}^V$ is the contribution from the spin-1 modes and $H_{CS}^T$ is the contribution from the spin-2 modes 
\bs\begin{align}
    H_{CS}^V&=\frac{1}{2}\int d^3x \epsilon^{ijk}\partial_j\Xi_i \nabla^2\Xi_k,\label{hv}\\ 
    H_{CS,0}^T&
=\frac12\int d^3x \epsilon^{ijk}\left[
\dot h^{\mathrm{TT}}_{i\ell}
\partial_j\dot h^{\mathrm{TT}}_{k\ell}
+h^{\mathrm{TT}}_{i\ell}
\partial_j\nabla^2h^{\mathrm{TT}}_{k\ell}\right].\label{ht}
\end{align}\es We add a subscript "0" to the tensor sector for now, We add a subscript "0" to the tensor sector for now. The notation \(H_{\rm CS}^{T}\) introduced below is reserved for an improved on-shell kinetic helicity.
The term \(B_{\rm CS}^{\rm SVT}\) is supported on the
codimension-two boundary \(\partial V\)
\be
B_{\rm CS}^{\rm SVT}
=
\int_{\partial V}
d^2y\,\sqrt q\,r_k\mathcal B_{\rm SVT}^k,
\label{Bk}
\ee
where
\be
\mathcal B_{\rm SVT}^k
=
\mathcal B_{\Xi\Xi}^k
+
\mathcal B_{hh}^k
+
\mathcal B_{\dot\Theta\Xi}^k
+
\mathcal B_{\dot h\Xi}^k
+
\mathcal B_{\Theta h}^k.
\label{Bk2}
\ee
The explicit expressions for the five contributions in
\eqref{Bk2} will be given below. For a general representative with \(\zeta_\mu\neq0\), the
Chern-Simons term contains an additional gauge-dependent
codimension-two term:
\be
H_{\rm CS}^{(2)}[h]
=
H_{\rm CS}^{V}
+
H_{\rm CS,0}^{T}
+
B_{\rm CS}^{\rm SVT}
+
B_\zeta,
\label{CSgeneraldecomposition}
\ee 
where
\be
B_\zeta
=
\int_{\partial V}
d^2y\,\sqrt q\,r_i\mathcal J_\zeta^i.\ee  Thus, the bulk terms are completely determined by the gauge invariant
SVT modes, while the dependence on the choice of representative is
confined to \(\partial V\). We now discuss the vector bulk contribution, the tensor bulk
contribution, and the codimension-two boundary terms separately.

\paragraph{Spin-1 sector.} 
Since the gauge invariant vector mode is transverse,
\[
\nabla\cdot\boldsymbol\Xi=0,
\]
we define its associated gravitomagnetic field and gravito-current by
\be
\mathbf B_\Xi
\equiv
\nabla\times\boldsymbol\Xi,
\qquad
\mathbf J_\Xi
\equiv
\nabla\times\mathbf B_\Xi.
\label{BXJXdefinition}
\ee Transversality implies
\be
\mathbf J_\Xi
=
\nabla\times(\nabla\times\boldsymbol\Xi)
=
-\nabla^2\boldsymbol\Xi.
\label{JXlaplacian}
\ee
Equation~\eqref{hv} can therefore be written as
\bea
H_{\rm CS}^{V}
=
\frac12
\int_{V}d^3x\,
\mathbf B_\Xi\cdot\mathbf J_\Xi
=
-\frac12
\int_{V}d^3x\,
(\nabla\times\boldsymbol\Xi)
\cdot\nabla^2\boldsymbol\Xi.
\label{hcsV}
\eea Thus, \(H_{\rm CS}^{V}\) has the algebraic structure of a
current helicity \cite{Brandenburg:2004jv} rather than that of a magnetic helicity. In
magnetohydrodynamics, the magnetic and current helicity densities are,
respectively,
\bs
\begin{align}
h_{\rm m}
&=
\mathbf a\cdot\mathbf b,
\qquad
\mathbf b=\nabla\times\mathbf a,
\\
h_{\rm c}
&=
\mathbf b\cdot(\nabla\times\mathbf b).
\end{align}
\es
Consequently, the correspondence relevant here is
\[
\mathbf b\longleftrightarrow\mathbf B_\Xi,
\qquad
\nabla\times\mathbf b\longleftrightarrow\mathbf J_\Xi.
\]
This motivates referring to \(H_{\rm CS}^{V}\) as the
spin-1 gravito-current helicity. By contrast, the gravitomagnetic helicity discussed in the literature
has the magnetic helicity structure
\be
\text{gravitomagnetic helicity}\propto
\int_{V}d^3x\,
\boldsymbol\Xi\cdot
(\nabla\times\boldsymbol\Xi)
\label{gramh}
\ee
up to normalization and sign conventions
\cite{Bini:2021gdb}. In the conventions adopted in the present work,
this structure appears in the spin-1 Nieh-Yan
descendant,
\be
H_{\rm NY}^{V}
=
-\frac14
\int_{V}d^3x\,
\boldsymbol\Xi\cdot\mathbf B_\Xi.
\ee

The Pontryagin expression \eqref{hcsV} contains two additional spatial
derivatives relative to this spin-1 gravitomagnetic  helicity.

The gauge invariant bulk density selected by \eqref{hv} is
\be
\mathcal H_{\rm CS}^{V}
=
\frac12
\mathbf B_\Xi\cdot\mathbf J_\Xi
=
-\frac12
(\nabla\times\boldsymbol\Xi)
\cdot\nabla^2\boldsymbol\Xi .
\label{hcsvdensity}
\ee
Because \(\Xi_i\) is invariant under linearized diffeomorphisms, this
bulk representative is also gauge invariant at the linearized level.
Nevertheless, \(\mathcal H_{\rm CS}^{V}\) is a density defined on the
chosen spatial foliation rather than a four-dimensional scalar.
Moreover, as usual for a hypersurface descendant, it is defined only
up to a total spatial divergence, whose integral contributes on
\(\partial V\). The expression \eqref{hcsvdensity} is also related algebraically to the
magnetic part of Lipkin's electromagnetic zilch. In one commonly used
normalization, the vacuum electromagnetic zilch density is\cite{Lipkin1964Conservation,CameronBarnettYao2012,Philbin2013}
\be
\rho_{\rm zilch}
=
\frac12
\left[
\mathbf e\cdot(\nabla\times\mathbf e)
+
\mathbf b\cdot(\nabla\times\mathbf b)
\right].
\label{zilchdensity}
\ee Here $\mathbf e$ denotes the electric field. 
Our spin-1 density corresponds in form only to the second, magnetic
contribution under the replacement
\[
\mathbf b\longleftrightarrow\mathbf B_\Xi.
\]
It should therefore not be identified with one half of the complete
zilch density in general. Such an identification is possible only
under additional conditions, for example after an averaging procedure
for which the electric and magnetic contributions are equal. The
analogy is algebraic, since the spin-1 gravitational mode obeys a
constraint equation rather than the source-free Maxwell evolution
equations. For linearized Einstein gravity, the vector constraint in the
normalization used here is
\be
\nabla^2\Xi_i
=
-16\pi G\,T_{0i}^{\rm T}.
\label{Xivectorconstraint}
\ee
It follows that
\be
\mathcal H_{\rm CS}^{V}
=
8\pi G\,
\mathbf B_\Xi\cdot\mathbf T_0^{\rm T},
\qquad
(\mathbf T_0^{\rm T})_i=T_{0i}^{\rm T}.
\label{HCSVsource}
\ee
Hence the bulk density vanishes pointwise in every smooth source-free
region:
\be
T_{0i}^{\rm T}=0
\quad\Longrightarrow\quad
\mathcal H_{\rm CS}^{V}=0.
\ee Here "source-free" means \(T_{0i}^{\rm T}=0\), rather than merely
\(T_{0i}=0\) on an open set. Because the transverse projector is spatially
nonlocal, \(T_{0i}=0\) outside the support of the physical stress tensor does
not in general imply \(T_{0i}^{\rm T}=0\) there. The pointwise implication above for
\(\mathcal H_{\rm CS}^{V}\) therefore applies only after the projected source
has been shown to vanish.
%This does not imply that \(\Xi_i\) or \(\mathbf B_\Xi\) must vanish in
%an exterior vacuum region. They may be nonzero harmonic fields
%generated by sources in the excluded interior. Nor does it imply that
%the complete spin-1 contribution to the Chern-Simons functional
%vanishes, since the surface term
%\(B_{\rm CS}^{\rm SVT}\), including
%\(B_{\Xi\Xi}\), may remain nonzero. If \(\Xi_i\) is smooth and regular
%on all of \(\mathbb R^3\), decays at spatial infinity, and obeys
%\(\nabla^2\Xi_i=0\) everywhere, then these conditions instead imply
%\(\Xi_i=0\).
In a modified theory of gravity, the vector constraint must be derived
from the corresponding field equations. If it permits
\(\nabla^2\Xi_i\neq0\) in a region without ordinary matter, then
\(\mathcal H_{\rm CS}^{V}\) can be nonvanishing there.

\paragraph{Spin-2 sector.} For any symmetric spatial tensor \(t_{ij}\), we define the symmetrized
curl operator
\be
(Ct)_{ij}
\equiv
\epsilon_{kl(i}\partial^k t_{j)}{}^l
=
\frac12\left(
\epsilon_{kli}\partial^k t_j{}^l
+
\epsilon_{klj}\partial^k t_i{}^l
\right).
\label{Coperator}
\ee
The operator \(C\) maps a transverse traceless tensor into another
transverse traceless tensor and commutes with
\(\partial_t\) and \(\nabla^2\). Equation~\eqref{ht} becomes
\be
H_{\rm CS,0}^{T}
=
\frac12\int_{V}d^3x
\left[
\dot h^{\rm TT}_{ij}(C\dot h^{\rm TT})_{ij}
+
h^{\rm TT}_{ij}\nabla^2(Ch^{\rm TT})_{ij}
\right].
\label{hcsT}
\ee This is the complete quadratic spin-2 contribution to the
Pontryagin Chern-Simons hypersurface descendant. No field equation
has been used in deriving \eqref{hcsT}. The result is therefore
kinematical within linear perturbation theory around Minkowski
spacetime. Closely related parity odd tensor structures occur in
cosmological calculations of the gravitational Pontryagin density
\cite{Alexander:2004wk,Satoh:2007gn,Kamada:2019ewe}, although their
normalizations, backgrounds, and treatments of total derivatives
need not coincide with those used here.

In a source-free region of linearized Einstein gravity,  the
tensor mode satisfies
\be
\ddot h^{\rm TT}_{ij}-\nabla^2h^{\rm TT}_{ij}=0.
\label{TTwaveequation}
\ee
This equation is not universal in modified theories of gravity. On
solutions of \eqref{TTwaveequation}, define
\be
Q^T
=
\frac12\int_{V}d^3x
\left[
\dot h^{\rm TT}_{ij}(C\dot h^{\rm TT})_{ij}
-
h^{\rm TT}_{ij}\nabla^2(Ch^{\rm TT})_{ij}
\right].
\label{QTdefinition}
\ee
\iffalse Using \eqref{TTwaveequation} and the commutativity of \(C\) with
\(\partial_t\) and \(\nabla^2\), one obtains
\be
\frac{dQ^T}{dt}
=
\frac12\int_{\partial V}d^2y\,\sqrt q\,r_k
\left[
(\partial^kh^{\rm TT}_{ij})(C\dot h^{\rm TT})_{ij}
-
h^{\rm TT}_{ij}\partial^k(C\dot h^{\rm TT})_{ij}
\right].
\label{dQT}
\ee Thus, \(Q^T\) is conserved only when the surface integral in \eqref{dQT}
vanishes. This holds, for example, when \(V\) has no boundary, when
the fields vanish in a neighbourhood of \(\partial V\), or when
boundary conditions are imposed directly so that the surface integral vanishes.\fi
We promote $Q^T$ to a 3-form $\mathcal{J}_Q^T$:
\begin{align}
\mathcal{J}_Q^T &= \frac{1}{2} \Big[ \dot{h}_{ij}^{\text{TT}}(C\dot{h}^{\text{TT}})^{ij} - h_{ij}^{\text{TT}} \nabla^2 (Ch^{\text{TT}})^{ij} \Big] \mathrm{d}^3x \notag\\
&\quad + \frac{1}{4} \mathrm{d}t \wedge \epsilon_{k\ell m} \Big[ (\partial^k h_{ij}^{\text{TT}}) (C\dot{h}^{\text{TT}})^{ij} - h_{ij}^{\text{TT}} \partial^k (C\dot{h}^{\text{TT}})^{ij} \Big] \mathrm{d}x^\ell \wedge \mathrm{d}x^m. 
\end{align}
The first term corresponds to the quantity $Q^T$, while the second term vanishes upon pullback to a spacelike hypersurface. Furthermore, $\mathcal{J}_Q^T$ is closed under exterior differentiation,
\be 
\mathrm{d} \mathcal{J}_Q^T \approx 0,
\ee 
where $\approx$ denotes equality on-shell (i.e., after imposing the equations of motion). Hence, we can define a new 3-form $Q_3^{\mathrm{new}}$ as
\be 
    Q_3^{\mathrm{new}}=Q_3+\mathcal{J}_Q^T,
\ee 
such that 
\be 
    \mathrm{d}Q_3^{\mathrm{new}}\approx P_4.
\ee 
Therefore, pulling back $Q^{\text{new}}_3$ to a spacelike hypersurface yields the density $\mathcal H_{\rm CS,\text{kin}}^{T}$,
\be 
\mathcal H_{\rm CS,\text{kin}}^{T}
=
\dot h^{\rm TT}_{ij}(C\dot h^{\rm TT})_{ij}.\label{kinhelicity}
\ee  
For a fixed coordinate region \(V\), Eq.~\eqref{TTwaveequation} gives
\be 
\frac{dQ^T}{dt}
=\frac12\int_{\partial V}d^2y\,\sqrt q\,r_k
\left[
(\partial^kh^{\rm TT}_{ij})(C\dot h^{\rm TT})_{ij}
-h^{\rm TT}_{ij}\partial^k(C\dot h^{\rm TT})_{ij}
\right].
\ee 
Thus $Q^T$ is conserved only if this flux vanishes—for instance, on a boundaryless slice, for fields supported away from the boundary, or under suitable boundary/falloff conditions that make the displayed integral zero. %The algebraic identity \eqref{HCSrelation} remains valid
%without this assumption, but at a finite open boundary or in a radiative
%limit the flux must be retained. A time-dependent spatial boundary would in
%addition generate the usual transport term.

In the preceding derivation of $\mathcal H_{\rm CS,\text{kin}}^T$, the pullback is taken to the constant-time spacelike hypersurface $V$. Under this pullback, the term containing $\mathrm{d}t$ vanishes.   Note that the presence of $Q^T$ modifies the bulk density.

\iffalse At regular spatial infinity, a sufficient set of falloff conditions is
\be
h^{\rm TT}_{ij}=O(r^{-1}),\qquad
\partial_kh^{\rm TT}_{ij}=O(r^{-2}),\qquad
\dot h^{\rm TT}_{ij}=O(r^{-2}),\qquad
\partial_k\dot h^{\rm TT}_{ij}=O(r^{-3}).
\label{spatialfalloff}
\ee
These imply
\[
(C\dot h^{\rm TT})_{ij}=O(r^{-3}),
\qquad
\partial_k(C\dot h^{\rm TT})_{ij}=O(r^{-4}),
\]
so that the surface integral in \eqref{dQT} vanishes as \(O(r^{-3})\).
The condition \(h^{\rm TT}_{ij}=O(r^{-1})\) alone is not sufficient.
For an eternal radiative field
\[
h^{\rm TT}_{ij}
=
\frac{F_{ij}(t-r,\mathbf n)}{r}+O(r^{-2}),
\]
radial and time derivatives can remain \(O(r^{-1})\), and the flux
through a large sphere can approach a finite nonzero value.\fi

Combining \eqref{hcsT} and \eqref{QTdefinition}, we obtain the exact
identity
\be
H_{\rm CS,0}^{T}
=
H_{\rm CS,\text{kin}}^{T}-Q^T,
\qquad
H_{\rm CS,\text{kin}}^{T}
\equiv
\int_{V_t}d^3x\,
\dot h^{\rm TT}_{ij}(C\dot h^{\rm TT})_{ij}.
\label{HCSrelation}
\ee
The corresponding bulk densities \eqref{hcsT} and \eqref{kinhelicity} should be
distinguished. This subtraction is an on-shell improvement of the complete descendant, not an identification of the two densities. The complete spin-2 Pontryagin hypersurface descendant remains the density~\eqref{hcsT}, while \(\dot h^{\rm TT}_{ij}(C\dot h^{\rm TT})_{ij}\) is a physically distinct on-shell kinetic helicity. The latter is not the direct spin-2 Pontryagin descendant. As we shall justify in the next section, the second density provides a more appropriate description of helicity than the first. Here "more appropriate" refers to the helicity observable, not to the topological descent. The descent itself selects \eqref{hcsT}. The kinetic density is obtained only after using the wave equation and subtracting a conserved improvement term.
We therefore omit the subscript "kin" and introduce the spin-2 gravito-current helicity
\be
H_{\rm CS}^T=\int d^3x  \dot h^{\rm TT}_{ij}(C\dot h^{\rm TT})_{ij},
\ee
whose density reads
\be
\mathcal H_{\rm CS}^T=\dot h^{\rm TT}_{ij}(C\dot h^{\rm TT})_{ij}.\label{hcsTnew}
\ee

\paragraph{
Boundary term.
} For the gauge invariant representative \(\zeta_\mu=0\), the
codimension-two contribution in \eqref{CSdecomposition} is
\be
B_{\rm CS}^{\rm SVT}
=
\int_{\partial V}d^2y\,\sqrt q\,r_m
\mathcal B_{\rm SVT}^m,
\qquad
\mathcal B_{\rm SVT}^m
=
\mathcal B_{\Xi\Xi}^m+\mathcal B_{hh}^m
+\mathcal B_{\dot\Theta\Xi}^m
+\mathcal B_{\dot h\Xi}^m
+\mathcal B_{\Theta h}^m,
\ee
where
\bs
\begin{align}
\mathcal B^m_{\Xi\Xi}
&=
\frac12\epsilon^{ijk}\Xi_i
\left(
\partial^m\partial_j\Xi_k
-\delta^m{}_j\nabla^2\Xi_k
\right),
\\
\mathcal B^m_{hh}
&=
\frac12\epsilon^{ijk}h^{\rm TT}_{i\ell}
\left(
\partial_j\partial_\ell h^{\rm TT}_{mk}
-\partial_j\partial^m h^{\rm TT}_{k\ell}
\right),
\\
\mathcal B^m_{\dot\Theta\Xi}
&=
-\frac12\epsilon^{mij}\Xi_i\partial_j\dot\Theta,
\\
\mathcal B^m_{\dot h\Xi}
&=
-\frac12\epsilon^{ijk}
\left(
\dot h^{\rm TT}{}_i{}^m\partial_j\Xi_k
+\Xi_i\partial_j\dot h^{\rm TT}{}_k{}^m
\right),
\\
\mathcal B^m_{\Theta h}
&=
\frac12\epsilon^{ijk}
(\partial_j\Theta)
(\partial_kh^{\rm TT}{}_i{}^m).
\end{align}
\es
Here \(m\) is a free spatial index, whereas \(r_m\) denotes the
outward-pointing unit normal to \(\partial V\).
The first two terms arise from the vector-vector and tensor-tensor
sectors, respectively. The remaining terms originate from the
scalar-vector, vector-tensor, and scalar-tensor sectors.
For a general representative with \(\zeta_\mu\neq0\), the complete
surface contribution is instead
\be
B_{\rm CS}^{\rm total}
=
B_{\rm CS}^{\rm SVT}+B_\zeta,
\qquad
B_\zeta
=
\int_{\partial V_t}d^2y\,\sqrt q\,r_m
\mathcal J_\zeta^m,
\ee
where \(\mathcal J_\zeta^m\) is given in \eqref{Jzeta}.
Whether these terms vanish depends on the hypersurface, the asymptotic
region, and the boundary conditions. We have discussed the conditions for a weak field, slowly moving
isolated binary in its center-of-mass frame in Appendix~\ref{falloff0}. \iffalse give
\bs
\begin{align}
\mathcal B_{\Xi\Xi}^m&=O(r^{-6}),
&
\mathcal B_{hh}^m&=O(r^{-4}),
&
\mathcal B_{\dot\Theta\Xi}^m&=O(r^{-6}),
\\
\mathcal B_{\dot h\Xi}^m&=O(r^{-5}),
&
\mathcal B_{\Theta h}^m&=O(r^{-4}).
\end{align}
\es
Since the area element grows as \(r^2d\Omega\), all the corresponding
surface integrals vanish at regular spatial infinity.

This conclusion does not automatically extend to future null
infinity. For a radiative field
\[
h^{\rm TT}_{ij}
=
\frac{F_{ij}(u,\mathbf n)}{r}+O(r^{-2}),
\qquad u=t-r,
\]
radial derivatives of \(F_{ij}(u,\mathbf n)\) remain \(O(r^{-1})\).
Consequently,
\[
\mathcal B_{hh}^m=O(r^{-2}),
\]
and its integral over a large sphere can have a finite nonzero limit.
The same caution applies to an idealized eternal periodic binary,
which does not obey the stronger regular-spatial-infinity conditions.\fi

In the following, we focus on the local gauge invariant bulk
representatives \(\mathcal H_{\rm CS}^{V}\) and
\(\mathcal H_{\rm CS}^{T}\). This is a restriction of scope,
rather than a consequence of gauge invariance. Whenever the complete
integrated Chern-Simons descendant is considered for a finite region
or at null infinity, the surface terms must be restored explicitly.

\iffalse 
The boundary term $B_{CS}$ is given in \eqref{Bk}-\eqref{Bk2} with \bs 
\begin{align}
\mathcal B^r_{\Xi\Xi}
&=\frac12\epsilon^{ijk}\Xi_i
\left(
\partial^r\partial_j\Xi_k
-\delta^r{}_j\nabla^2\Xi_k
\right),\\
\mathcal B^r_{hh}
&=\frac12\epsilon^{ijk}h^{\rm TT}_{i\ell}
 \left(
 \partial_j\partial_\ell h^{\rm TT}_{rk}
-\partial_j\partial_r h^{\rm TT}_{k\ell}
 \right),\\
\mathcal B^r_{\dot\Theta\Xi}
&=-\frac12\epsilon^{rij}\Xi_i\partial_j\dot\Theta,\\
\mathcal B^r_{\dot h\Xi}
&=-\frac12\epsilon^{ijk}
 \left(
 \dot h^{\rm TT}{}_i{}^r\partial_j\Xi_k
 +\Xi_i\partial_j\dot h^{\rm TT}{}_k{}^r
 \right),\\
\mathcal B^r_{\Theta h}
&=\frac12\epsilon^{ijk}
 (\partial_j\Theta)(\partial_kh^{\rm TT}{}_i{}^r).
\end{align}
\es 
The first term originates solely from the spin-1 modes, while the second term arises from the spin-2 modes. The remaining three terms come from mixing contributions. Whether these terms are significant should be assessed on a case-by-case basis. In Appendix \ref{falloff0}, we show that these boundary terms vanish for a weakly coupled, slowly moving binary system in its center-of-mass frame. In the following discussion, we will consistently ignore these boundary terms, since the helicity densities are local, gauge invariant quantities and are already independent of them.\fi
\subsection{Nieh-Yan term}
The Nieh-Yan form is naturally defined in Riemann-Cartan geometry,
where the coframe and the metric are related by
\be
g_{\mu\nu}=\eta_{ab}e^a{}_{\mu}e^b{}_{\nu},
\qquad
e^a=e^a{}_{\mu}dx^\mu.
\ee
For a metric-compatible spin connection,
\(\omega_{ab}=-\omega_{ba}\), the torsion and curvature two-forms are
\be
T^a=de^a+\omega^a{}_b\wedge e^b,
\qquad
R^a{}_b=d\omega^a{}_b+\omega^a{}_c\wedge\omega^c{}_b.
\label{torsion-curvature-forms}
\ee
The Nieh-Yan four-form is the exact form \cite{Nieh:1981ww}
\be
N_{\rm NY}
\equiv
T^a\wedge T_a-e^a\wedge e^b\wedge R_{ab}
=d\!\left(e^a\wedge T_a\right).
\label{NYidentity}
\ee
 Exactness alone does not imply that its integral is
a nontrivial or quantized topological invariant. If
\(e^a\wedge T_a\) is globally defined and smooth on a compact manifold
without boundary, the integral of \(N_{\rm NY}\) vanishes. Nonzero
values can instead depend on boundaries, singularities, torsional
defects, or global properties of the coframe. This qualification is
also important in applications to the chiral anomaly, where the
coefficient of the Nieh-Yan contribution can depend on the regulator
scale \cite{ChandiaZanelli1997,Obukhov1997}.

For the torsion-free Levi-Civita connection used in metric general
relativity, \(T^a=0\), and hence \(N_{\rm NY}=0\). Teleparallel gravity
provides a different geometric representation of the same Einstein
dynamics where the teleparallel spin connection is flat, while its torsion
is generally nonzero. The historical development of teleparallel
gravity can be traced to Einstein's work on distant parallelism
\cite{Einstein1928} and to subsequent tetrad and translational gauge
formulations \cite{Cho:1975dh,Hayashi:1967se,Hayashi:1979qx}. Modern
reviews include
\cite{Maluf:2013gaa,Krssak:2018ywd,Bahamonde:2021gfp}. Recent work has
also used the teleparallel Nieh-Yan boundary form to construct a
spin-2 helicity flux at \(\mathscr I^+\) \cite{Long:2025fbb}. The
quantity considered below is instead the descendant on a spacelike
constant time hypersurface. 

In TEGR,
\be
R^a{}_b=0,
\qquad
N_{\rm NY}=T^a\wedge T_a=d\!\left(e^a\wedge T_a\right).
\ee
For a spacetime \(M\) bounded by two constant time hypersurfaces
\(V_{t_1}\) and \(V_{t_2}\), together with an outer boundary
\(\mathcal B\), Stokes' theorem gives
\be
\int_M N_{\rm NY}
=H_{\rm NY}[V_{t_2}]-H_{\rm NY}[V_{t_1}]
+\int_{\mathcal B}e^a\wedge T_a,
\qquad
H_{\rm NY}[V]
\equiv
\int_{V}e^a\wedge T_a,
\label{NYdescent}
\ee
where every three-form is understood to be pulled back to the
corresponding boundary hypersurface. 

At linear order, a coframe reproducing the SVT decomposition of the
metric can be written as
\bs
\begin{align}
e^{\hat0}{}_0
&=1-\phi,
\\
e^{\hat0}{}_i
&=-\frac12\left(\beta_i+\partial_i\gamma\right)-\mathfrak b_i,
\\
e^{\hat i}{}_0
&=\delta^{\hat i i}
\left[
\frac12\left(\beta_i+\partial_i\gamma\right)-\mathfrak b_i
\right],
\\
e^{\hat i}{}_j
&=\delta^{\hat i}{}_j
+\delta^{\hat i i}\left[
\frac12h^{\rm TT}_{ij}
+\frac16H\delta_{ij}
+\frac12\partial_{(i}\epsilon_{j)}
+\frac12\left(
\partial_i\partial_j-\frac13\delta_{ij}\nabla^2
\right)\lambda
+\epsilon_{ij}{}^k\mathfrak r_k
\right].
\end{align}
\es
Here \(\mathfrak b_i\) and \(\mathfrak r_i\) parameterize the
independent infinitesimal Lorentz transformation of the coframe
\be
\ell_{\hat0\hat i}=\mathfrak b_i,
\qquad
\ell_{\hat i\hat j}=\delta_{i \hat{i}}\delta_{j \hat{j}}\epsilon^{ijk}\mathfrak r_k,
\qquad
\ell_{ab}=-\ell_{ba}.
\label{ell-definition}
\ee Let
\be
\bar e^{\hat0}=dt,
\qquad
\bar e^{\hat i}=\delta^{\hat i}_{i}dx^i
\ee
be the background coframe. A general linearized coframe can be
organized as
\be
e^a=[e^a]+d\zeta^a+\Lambda^a{}_b\bar e^b+O(2),
\qquad
\Lambda_{ab}=\ell_{ab}+\partial_{[a}\zeta_{b]}.
\label{coframe-decomposition}
\ee
Here
\be
\zeta_0=\gamma-\frac12\dot\lambda,
\qquad
\zeta_i=\frac12\left(\epsilon_i+\partial_i\lambda\right),
\qquad
\zeta^a=\bar e^a{}_{\mu}\zeta^\mu.
\label{zeta-coframe}
\ee
The quantity \(\Lambda_{ab}\) is the total antisymmetric
part that remains after the diffeomorphism contribution has been
written as \(d\zeta^a\). The representative constructed from the gauge invariant SVT modes is
\bs
\begin{align}
[e^{\hat0}]
&=(1+\Phi)dt-\frac12\Xi_i dx^i,
\\
[e^{\hat i}]
&=\frac12\Xi^{\hat i}dt
+\left[
\left(1+\frac12\Theta\right)\delta^{\hat i}{}_j
+\frac12h^{\rm TT\,\hat i}{}_j
\right]dx^j.
\end{align}
\label{GI-coframe}
\es
Its coefficients are invariant under linearized diffeomorphisms,
subject to the boundary conditions used to define the SVT
decomposition. The coframe itself is still a Lorentz vector and must
be accompanied by the corresponding inertial spin connection. To determine that connection, one switches off the gravitational
modes in \eqref{coframe-decomposition} and introduces the reference
coframe
\be
E^a=\bar e^a+d\zeta^a+\Lambda^a{}_b\bar e^b+\text{higher orders}.
\ee
The inertial spin connection is required to satisfy
\be
dE^a+\omega^a{}_b\wedge E^b=0,
\qquad
R^a{}_b(\omega)=0.
\ee
Locally, a flat spin connection can be written as
\be
\omega^a{}_b=(L\,dL^{-1})^a{}_b,
\qquad
L^a{}_b=\delta^a{}_b+\Lambda^a{}_b+\text{higher orders},
\ee
and hence
\be
\omega^a{}_b=-d\Lambda^a{}_b+\text{higher orders}.
\label{inertial-spin-connection}
\ee
This is the covariant reference-coframe prescription of TEGR
\cite{KrssakPereira2015,Krssak:2018ywd}. Flatness implies that the
connection is pure gauge locally. A global Weitzenbock gauge can be
obstructed by nontrivial holonomy. It is important that the symmetric coframe choice \(\ell_{ab}=0\)
does not generally imply \(\omega^a{}_b=0\), because then
\(\Lambda_{ab}=\partial_{[a}\zeta_{b]}\). The Weitzenbock gauge instead
requires
\be
\Lambda_{ab}=0
\qquad\Longleftrightarrow\qquad
\ell_{ab}=-\partial_{[a}\zeta_{b]}.
\label{Weitzenbock-condition}
\ee
The simultaneous choice \(\zeta^a=0\) and \(\ell_{ab}=0\) is a
particularly simple representative for which both
\(\Lambda_{ab}=0\) and \(\omega^a{}_b=0\). Define the gauge invariant coframe perturbation by
\be
\delta e^a_{\rm GI}\equiv[e^a]-\bar e^a.
\ee
Using \eqref{coframe-decomposition} and
\eqref{inertial-spin-connection}, the linearized torsion is
\be
T^{a(1)}=d\delta e^a_{\rm GI}.
\label{linear-GI-torsion}
\ee
The terms containing \(d\Lambda^a{}_b\) cancel between \(de^a\) and
\(\omega^a{}_b\wedge e^b\), while \(d^2\zeta^a=0\). More generally,
under a simultaneous Lorentz transformation of the coframe and spin
connection,
\be
e^a\longrightarrow L^a{}_b e^b,
\qquad
T^a\longrightarrow L^a{}_b T^b,
\ee
so that \(e^a\wedge T_a\) is Lorentz invariant. Because of this Lorentz invariance, the quadratic descendant may be
evaluated locally in the Weitzenbock gauge \(\Lambda_{ab}=0\). In this
gauge,
\be
e^a=[e^a]+d\zeta^a+\text{higher orders},
\qquad
\omega^a{}_b=0,
\qquad
T^{a(1)}=d\delta e^a_{\rm GI}.
\ee
Evaluating the same expression in an arbitrary Lorentz frame gives
the same result when the coframe and inertial spin connection are
retained consistently to the required perturbative order.

Keeping the quadratic functional constructed from the first-order
perturbations, the pullback of the Nieh-Yan boundary form to the hypersurface \(V\)
becomes
\bs
\begin{align}
H_{\rm NY}^{(2)}[V]
&=
\int_{V}\delta e^a_{\rm GI}\wedge d\delta e_{{\rm GI},a}
+\int_{V}d\zeta^a\wedge d\delta e_{{\rm GI},a}
\nn\\
&=
\int_{V}\delta e^a_{\rm GI}\wedge d\delta e_{{\rm GI},a}
+\int_{\partial V}\zeta^a d\delta e_{{\rm GI},a}
\nn\\
&=H_{\rm NY}^{V}+H_{\rm NY}^{T}+B_{\rm NY}^{\zeta}.
\end{align}
\label{NY-SVT-decomposition}
\es
All Lorentz-frame dependence has canceled through the inertial spin
connection. The remaining surface term records the use of a general
diffeomorphism representative on a hypersurface with boundary. Neither scalar mode contributes to the bulk expression. The
codimension-two term is
\be
B_{\rm NY}^{\zeta}
=\int_{\partial V}d^2y\,\sqrt q\,r_m\mathcal J^m_{NY},
\label{BNY}
\ee
where
\be
\mathcal J^m_{NY}
=\frac12\epsilon^{mjk}
\left[
\zeta^{\hat0}\partial_j\Xi_k
+\zeta^{\hat\ell}\partial_j
\left(\Theta\delta_{\ell k}+h^{\rm TT}_{\ell k}\right)
\right].
\label{JNY}
\ee
The condition \(\zeta^a|_{\partial V}=0\) is sufficient, but not
necessary, for \(B_{\rm NY}^{\zeta}\) to vanish. Transformations that
do not vanish at the boundary can change the fixed-hypersurface
representative by this surface term, which must then be retained. The corresponding bulk densities are
\bs
\begin{align}
\mathcal H_{\rm NY}^{V}
&=-\frac14\boldsymbol\Xi\cdot
\left(\boldsymbol\nabla\times\boldsymbol\Xi\right),
\label{HNY-vector-density}
\\
\mathcal H_{\rm NY}^{T}
&=\frac14h^{\rm TT}_{ij}(Ch^{\rm TT})_{ij}.
\label{HNY-tensor-density}
\end{align}
\es
The first expression has the structure of gravitomagnetic helicity
\cite{Bini:2021gdb}. Unlike the electromagnetic vector potential,
\(\Xi_i\) is invariant under linearized diffeomorphisms. The second
expression is a spin-2 curl pairing closely related to
duality-based helicity constructions for linearized gravity
\cite{HenneauxTeitelboim2005,Barnett2014,Aghapour:2018pup}. Both are
gauge invariant bulk representatives on the chosen spatial
foliation. They are not four-dimensional scalar densities and remain
defined modulo total spatial divergences.

\subsection{Summary}
Table~\ref{sum} summarizes the four parity odd,
gauge invariant bulk sectors that arise in the quadratic reduction of
the Nieh-Yan and Pontryagin forms to a constant time hypersurface
\(V\). They should be distinguished from the complete hypersurface
descendants, which also contain codimension-two contributions. The Nieh-Yan bulk functionals contain one derivative, whereas the
Pontryagin bulk functionals contain three derivatives. Motivated by
the magnetic  and current helicity terminology, we refer to
the former as spin-\(s\) gravitomagnetic helicities and to the latter
as spin-\(s\) gravito-current helicities. For the spin-1 sector, this
terminology is consistent with the gravitomagnetic helicity
construction of \cite{Bini:2021gdb}. The spin-2
terminology is a proposal of the present work, motivated by analogous
curl pairings in duality-based descriptions of linearized gravity
\cite{HenneauxTeitelboim2005,Barnett2014,Aghapour:2018pup}.
 
\begin{table}
\hspace{-1cm}\begin{center}
\renewcommand\arraystretch{1.3}
 \begin{tabular}{|c||c||c|}\hline
Topological origin &Expression& Terminology\\\hline\hline
Nieh-Yan& $H_{NY}^V=-\frac{1}{4}\int d^3x \epsilon_{ijk}\Xi_i\partial_j\Xi_k$ &spin-1 gravitomagnetic helicity\\\hline
Nieh-Yan& $H_{NY}^T=\frac{1}{4}\int d^3x h_{ij}^{\text{TT}}Ch_{ij}^{\text{TT}}$ &spin-2 gravitomagnetic helicity\\\hline
Pontryagin& $H_{CS}^V=-\frac{1}{2}\int d^3x \epsilon_{ijk}\partial_j\Xi_k\nabla^2\Xi_i$ &spin-1 gravito-current helicity\\\hline
Pontryagin& $H_{CS}^T=\int d^3x \dot h_{ij}^{\text{TT}}C\dot h_{ij}^{\text{TT}}$ &spin-2 gravito-current helicity\\\hline
\end{tabular}
\caption{The first three rows are sectors of quadratic
spacelike hypersurface descendants, modulo codimension-two surface
terms. The final row is a separately defined on-shell kinetic
functional related to the complete spin-2 Pontryagin descendant by
Eq.~\eqref{HCSrelation}.}\label{sum}
\end{center}
\end{table}

The word "spin" in Table~\ref{sum} labels the irreducible spatial
\(SO(3)\) sector of the SVT decomposition. It does not imply that
every listed field is an independent propagating degree of freedom. Indeed, 
in linearized Einstein gravity, \(\Xi_i\) is determined by a vector
constraint, whereas \(h_{ij}^{\rm TT}\) carries the two radiative
degrees of freedom. 
An important qualification concerns the spin-2 Pontryagin sector.
With the bulk reduction adopted above, the density selected by
the Pontryagin descent is \eqref{hcsT}.
This modified spin-2 bulk descendant must be distinguished from the
separately defined density \eqref{hcsTnew}.
The latter is algebraically related to the spin-2 gravitomagnetic helicity density by
\be
\mathcal H_{\rm CS}^{T}
=
4\left.
\mathcal H_{\rm NY}^{T}
\right|_{h^{\rm TT}_{ij}\rightarrow\dot h^{\rm TT}_{ij}} .
\label{NY-CS-kin-comparison}
\ee
The factor of four follows from the normalization.

The terminology adopted above admits a representation theoretic
interpretation.  With Pauli-Lubanski operator
\[
W^\mu
=
-\frac12\epsilon^{\mu\nu\rho\sigma}
P_\nu M_{\rho\sigma},
\]
the temporal component is \(W^0=\bm P\cdot\bm J\).  On the transverse
spin-1 and TT spin-2 sectors, respectively,
\be 
(W^0_{(1)}\Xi)_i=(\bm\nabla\times\bm\Xi)_i,
\qquad
(W^0_{(2)}h^{\rm TT})_{ij}=2(Ch^{\rm TT})_{ij}.
\ee 
The factor of two in the second relation arises because the rotation
generator acts on both indices of a symmetric tensor.  Consequently,
\[
\mathcal H_{\rm NY}^{V}
=
-\frac14\Xi_i(W^0_{(1)}\Xi)_i,
\qquad
\mathcal H_{\rm NY}^{T}
=
\frac18h_{ij}^{\rm TT}
(W^0_{(2)}h^{\rm TT})_{ij}.
\]
The corresponding Pontryagin inspired expressions contain two
additional derivatives:
\[
\mathcal H_{\rm CS}^{V}
=
-\frac12(W^0_{(1)}\Xi)_i\nabla^2\Xi_i,
\qquad
\mathcal H_{\rm CS}^{T}
=
\frac12\dot h_{ij}^{\rm TT}
(W^0_{(2)}\dot h^{\rm TT})_{ij}.
\]
Thus the Nieh-Yan quantities measure first  moments of the
helicity spectrum, whereas the Pontryagin inspired quantities measure
higher wave number or frequency moments of the same spectrum.
This interpretation should not be taken to mean that the local
densities are unique Pauli-Lubanski charge densities.  The SVT
decomposition is tied to a spacelike foliation, and local densities
remain sensitive to total-divergence improvements.  Moreover, in
linearized Einstein gravity the vector mode \(\Xi_i\) is constrained
rather than an independent massless spin-1 particle.  The genuine
one-particle Pauli-Lubanski relation
\(W^\mu=\lambda P^\mu\) applies directly to the TT graviton modes,
with \(\lambda=\pm2\). The same approach can be extended to other spinning fields. For more details, please read Appendix \ref{pauli1}.
A representation-theoretic extension to
free massless fields of arbitrary integer spin is also discussed in the
concluding section.

\section{Mode expansion}\label{mode}
In Einstein gravity, the spin-1 and spin-2 modes satisfy the following equations 
\bs\begin{align}
    \nabla^2\Xi_i&=-16\pi G T_{0i}^{\text{T}},\label{Xieqn}\\ 
    \Box h_{ij}^{\text{TT}}&=-16\pi G T_{ij}^{\text{TT}}\label{hijeqn}
\end{align}\es 
where $T_{0i}^{\text{T}}$ is the transverse part of the momentum density and $T_{ij}^{\text{TT}}$ is the transverse and traceless part of the stress tensor $T_{ij}$. The first equation \eqref{Xieqn} is a Poisson equation and thus $\Xi_i$ is not the propagating modes. The second equation \eqref{hijeqn} shows that $h_{ij}^{\text{TT}}$ are truly propagating degrees of freedom. 
The solutions are 
\bs\begin{align}
    \Xi_i(t,\mathbf x)&=4G\int d^3\mathbf x' \frac{T_{0i}^{\text{T}}(t,\mathbf x')}{|\mathbf x-\mathbf x'|}+\Xi_i^{\text{hom}},\\ 
    h_{ij}^{\text{TT}}(t,\mathbf x)&=4G\int d^3\mathbf x' \frac{T_{ij}^{\text{TT}}(t-|\mathbf x-\mathbf x'|,\mathbf x')}{|\mathbf x-\mathbf x'|}+h_{ij}^{\text{TT},\text{hom}}.\label{hijsol}
\end{align}\es The first solution is instantaneous because the vector equation is a
constraint, while the second is retarded. Since the TT projection is
spatially nonlocal, \eqref{hijsol} should be understood as the
retarded Green function solution of the already projected equation. When there is no source in the whole space, the solutions are completely determined by the homogeneous solutions. 

The homogeneous solution $\Xi_i^{\text{hom}}$ obeys the equation 
\be 
\nabla^2\Xi_i^{\text{hom}}=0,\quad \partial_i\Xi_i^{\text{hom}}=0.
\ee 
In the whole space, assuming $\Xi_i^{\text{hom}}\to 0$ as $r\to \infty$, the unique solution is $\Xi_i^{\text{hom}}=0$. 
We conclude that the spin-1 gravitomagnetic/gravito-current helicities are always zero in the absence of matter
\be 
\mathcal H_{NY}^V=\mathcal H_{CS}^V=0
\ee 
For the spin-2 modes, 
the homogeneous solution $h_{ij}^{\text{TT},\text{hom}}$ satisfies the equation 
\be 
\Box h_{ij}^{\text{TT},\text{hom}}=0,\quad \partial^i h_{ij}^{\text{TT},\text{hom}}=0,\quad \delta^{ij}h_{ij}^{\text{TT},\text{hom}}=0. 
\ee A real solution can be
expanded as \cite{Maggiore:2007ulw}
\be 
h_{ij}^{\text{TT},\text{hom}}=\sum_{\sigma=\pm  }\int \frac{d^3\mathbf k}{(2\pi)^3}\left[h_\sigma(\mathbf k)e^{-i\omega t}+h^*_{\sigma}(-\mathbf k)e^{i\omega t}\right]e^\sigma_{ij}(\mathbf k) e^{i\mathbf k\cdot\mathbf x},\quad \omega=k=|\mathbf k|
\ee where the polarization tensor obeys the condition
\begin{align}
e_{ij}^{(\sigma)}=e_{ji}^{(\sigma)},\qquad
k_i e_{ij}^{(\sigma)}=0,\qquad
\delta_{ij}e_{ij}^{(\sigma)}=0,
\qquad
e_{ij}^{(\sigma)*}e_{ij}^{(\sigma')}
=2\delta_{\sigma\sigma'},\qquad
e_{ij}^{(\sigma)*}(\mathbf k)=e_{ij}^{(-\sigma)}(\mathbf k)=e_{ij}^{(\sigma)}(-\mathbf k).
\end{align} The completeness relation is 
\be \frac12\sum_{\sigma=\pm}
e_{ij}^{(\sigma)}
e_{kl}^{(\sigma)*}
=
\Pi_{ij,kl}(\hat k),\ee where \be \Pi_{ij,kl}
(\hat k)=
\frac12
\left(
P_{ik}(\hat k)P_{jl}(\hat k)+P_{il}(\hat k)P_{jk}(\hat k)-P_{ij}(\hat k)P_{kl}(\hat k)
\right),
\qquad
P_{ij}(\hat k)=\delta_{ij}-\hat k_i\hat k_j,\qquad \hat k=\frac{\mathbf k}{k}.\ee
Note that in the mode expansion, \(\sigma\in\{+1,-1\}\) is a chirality label, while the physical spin-2 helicity is given by \(s=2\sigma\in\{+2,-2\}\).
Then the spin-2 gravitomagnetic helicity is 
\be 
H_{\mathrm{NY}}^T
=
\frac12\sum_{\sigma=\pm}\sigma
\int\frac{d^3\mathbf k}{(2\pi)^3}
k\,\left|\mathcal H_\sigma(t,\mathbf k)\right|^2 \label{hnytmode}
\ee   and the gravito-current helicity is
\be 
H_{\mathrm{CS}}^T
=
2\sum_{\sigma=\pm }\sigma
\int\frac{d^3\mathbf k}{(2\pi)^3}
k\,\left|\dot{\mathcal H}_\sigma(t,\mathbf k)\right|^2 .\label{hcstmode}
\ee We have defined 
\be 
\mathcal H_\sigma(t,\mathbf k)=h_{\sigma}(\mathbf k)e^{-i\omega t}+h_{\sigma}^*(-\mathbf k)e^{i\omega t}.
\ee Therefore, 
\begin{align} \left|\mathcal H_\sigma\right|^2
&=
|h_\sigma(\mathbf k)|^2+|h_\sigma(-\mathbf k)|^2
+2\operatorname{Re}\left[
h_\sigma(\mathbf k)h_\sigma(-\mathbf k)e^{-2i\omega t}
\right],\\
\left|\dot{\mathcal H}_\sigma\right|^2
&=
k^2\left\{
|h_\sigma(\mathbf k)|^2+|h_\sigma(-\mathbf k)|^2
-2\operatorname{Re}\left[
h_\sigma(\mathbf k)h_\sigma(-\mathbf k)e^{-2i\omega t}
\right]
\right\}
\end{align} The terms depending on time $t$ reflect the interference effect. The oscillatory terms may  disappear after an explicit time average, or through
dephasing under suitable  assumptions for a continuous
spectrum. We define
\be
\overline{X}
\equiv
\lim_{\mathcal T\rightarrow\infty}
\frac1{\mathcal T}
\int_{t_0}^{t_0+\mathcal T}dt\,X(t).
\label{time-average-definition}
\ee
For a monochromatic mode, averaging over an integer number of periods
gives the same result. Applying \eqref{time-average-definition} to
\eqref{hnytmode} and \eqref{hcstmode}, and changing
\(\mathbf k\rightarrow-\mathbf k\) where needed, yields 
\bs\label{averageHh}\begin{align}
\overline{H_{\mathrm{NY}}^T}
&=
\int\frac{d^3\mathbf k}{(2\pi)^3}\,
k\left(
|h_{+}|^2-|h_{-}|^2
\right),
\\
\overline{H_{\mathrm{CS}}^T}
&=
4\int\frac{d^3\mathbf k}{(2\pi)^3}\,
k^3\left(
|h_{+}|^2-|h_{-}|^2
\right).
\end{align}\es 
These formulas are signed moments of the classical chiral spectrum.
They should not be identified directly with graviton occupation
numbers unless the field is canonically normalized and quantized.
After canonical normalization of the quadratic Einstein action,
\(k|h_\sigma|^2\) is proportional, with convention-dependent factors
of \(G\), \(\hbar\), and the normalization volume, to the occupation
number in the helicity-\(2\sigma\) sector. In that sense
\(\overline{H_{\rm NY}^{T}}\) is proportional to the net chiral
occupation, whereas \(\overline{H_{\rm CS}^{T}}\) weights the same
imbalance by two additional powers of frequency. For comparison, mode expansion of the original spin-2 Pontryagin bulk descendant \eqref{hcsT} is 
\bea 
-4\sum_{\sigma=\pm}\sigma
\int\frac{d^3\mathbf k}{(2\pi)^3}
k^3
\operatorname{Re}\!\left[
h_\sigma(\mathbf k)
h_\sigma(-\mathbf k)e^{-2i\omega t}
\right]
\eea whose time average is zero. Therefore, \eqref{hcsT} cannot be the candidate quantity of helicity. This conclusion relies on the criterion adopted here for a time-averaged free-field spectral helicity; nevertheless, the vanishing average does not render the complete Pontryagin descendant mathematically trivial.

%The average spin-2 gravitomagnetic helicity is exactly the number difference between the positive and negative gravitons. On the other hand, the average spin-2 gravito-current helicity is the frequency weighted number difference between the positive and negative gravitons. 

\paragraph{Electromagnetic analogue.} The spectral structure has a useful electromagnetic analogue.  The magnetic and  current helicity are defined as 
\bs\begin{align}
    H_{\text{m}}&=\int d^3x\ \mathbf a\cdot\mathbf b,\\ 
    H_{\text{c}}&=\int d^3x\ \mathbf b\cdot(\nabla\times\mathbf b).
\end{align}\es In
Coulomb gauge, we use the mode expansion of the magnetic potential
\be 
a_i(t,\mathbf x)
=
\sum_{s=\pm}
\int\frac{d^3\mathbf k}{(2\pi)^3}
\left[
c_s(\mathbf k)e^{-ikt}
+
c_s^*(-\mathbf k)e^{ikt}
\right]
e_i^{(s)}(\mathbf k)
e^{i\mathbf k\cdot\mathbf x},
\qquad k=|\mathbf k|
\ee where the polarization vector $e_i^{(s)}$ satisfies the condition 
\be 
k_i e_i^{(s)}=0,
\qquad
e_i^{(s)*}e_i^{(s')}
=
\delta_{ss'},
\qquad 
e_i^{(s)*}(\mathbf k)
=
e_i^{(-s)}(\mathbf k),
\qquad
e_i^{(s)}(-\mathbf k)
=
e_i^{(s)*}(\mathbf k),
\ee  and the completeness relation
\be 
\sum_{s=\pm}
e_i^{(s)}(\mathbf k)
e_j^{(s)*}(\mathbf k)
=
P_{ij}(\mathbf k)
=
\delta_{ij}-\hat k_i\hat k_j.
\ee 
Then the magnetic helicity and current helicity become 
\bs\begin{align}
H_{\mathrm m}
&=
2\sum_{s=\pm}s
\int\frac{d^3\mathbf k}{(2\pi)^3}k
\Big[
|c_s(\mathbf k)|^2
+
\operatorname{Re}\!\left(
c_s(\mathbf k)c_s(-\mathbf k)e^{-2ikt}
\right)
\Big],\\
H_{\mathrm c}
&=
2\sum_{s=\pm}s
\int\frac{d^3\mathbf k}{(2\pi)^3}k^3
\Big[
|c_s(\mathbf k)|^2
+
\operatorname{Re}\!\left(
c_s(\mathbf k)c_s(-\mathbf k)e^{-2ikt}
\right)
\Big].
\end{align}
\es Average over a period, we find 
\bs\begin{align}
\overline{ H_{\mathrm m}}
&=
2
\int\frac{d^3\mathbf k}{(2\pi)^3}
k\,\left(|c_+(\mathbf k)|^2-|c_-(\mathbf k)|^2\right),\\
\overline{ H_{\mathrm c}}
&=
2
\int\frac{d^3\mathbf k}{(2\pi)^3}
k^3\,\left(|c_+(\mathbf k)|^2-|c_-(\mathbf k)|^2\right).
\end{align}\es  They have the same spectral form as
\eqref{averageHh}, up to normalization, which motivates the terminology of 
spin-2 gravitomagnetic and gravito-current helicity.

\paragraph{Energy-helicities inequalities.}  
The quadratic Hamiltonian of the free TT field is
\begin{equation}
E
=
\frac{1}{64\pi G}
\int d^3\mathbf x\,
\left[
\dot h_{ij}^{\rm TT}\dot h_{ij}^{\rm TT}
+
(\partial_k h_{ij}^{\rm TT})
(\partial_k h_{ij}^{\rm TT})
\right].
\label{GW-energy-position}
\end{equation}
Substituting the mode expansion gives
\begin{align}
E
&=
\frac{1}{32\pi G}
\sum_{\sigma=\pm}
\int\frac{d^3\mathbf k}{(2\pi)^3}
\left[
\left|\dot{\mathcal H}_\sigma(t,\mathbf k)\right|^2
+
k^2\left|\mathcal H_\sigma(t,\mathbf k)\right|^2
\right]
\nonumber\\
&=
\frac{1}{8\pi G}
\sum_{\sigma=\pm}
\int\frac{d^3\mathbf k}{(2\pi)^3}
k^2|h_\sigma(\mathbf k)|^2.
\label{GW-energy-momentum}
\end{align}
In the second equality, the oscillatory terms cancel between the kinetic
and gradient contributions. Consequently, the total energy is conserved
without performing a time average.

Using \(d^3\mathbf k=k^2dk\,d\Omega_{\mathbf k}\), we define the
angle-integrated energy spectrum by
\begin{equation}
E=\int_0^\infty dk\,\mathscr E(k),
\end{equation}
where
\begin{equation}
\mathscr E(k)
=
\frac{k^4}{8\pi G(2\pi)^3}
\int d\Omega_{\mathbf k}\,
\left[
|h_+(k,\Omega_{\mathbf k})|^2
+
|h_-(k,\Omega_{\mathbf k})|^2
\right].
\label{energy-spectrum}
\end{equation}
Similarly, the time averaged helicities can be written as
\begin{equation}
\overline{H_{\rm NY}^{T}}
=
\int_0^\infty dk\,\mathscr H_{\rm NY}(k),
\qquad
\overline{H_{\rm CS}^{T}}
=
\int_0^\infty dk\,\mathscr H_{\rm CS}(k),
\end{equation}
with
\begin{align}
\mathscr H_{\rm NY}(k)
&=
\frac{k^3}{(2\pi)^3}
\int d\Omega_{\mathbf k}\,
\left[
|h_+(k,\Omega_{\mathbf k})|^2
-
|h_-(k,\Omega_{\mathbf k})|^2
\right],
\label{NY-spectrum}
\\
\mathscr H_{\rm CS}(k)
&=
\frac{4k^5}{(2\pi)^3}
\int d\Omega_{\mathbf k}\,
\left[
|h_+(k,\Omega_{\mathbf k})|^2
-
|h_-(k,\Omega_{\mathbf k})|^2
\right].
\label{CS-spectrum}
\end{align}
In particular,
\begin{equation}
\mathscr H_{\rm CS}(k)
=
4k^2\mathscr H_{\rm NY}(k).
\label{spectral-NY-CS-relation}
\end{equation}

Since
\begin{equation}
\left|
|h_+|^2-|h_-|^2
\right|
\leq
|h_+|^2+|h_-|^2,
\end{equation}
the spectral densities obey the pointwise inequalities
\bs 
\begin{align}
|\mathscr H_{\rm NY}(k)|
&\leq
\frac{8\pi G}{k}\,\mathscr E(k),
\label{NY-spectral-inequality}
\\
|\mathscr H_{\rm CS}(k)|
&\leq
32\pi G\,k\,\mathscr E(k).
\label{CS-spectral-inequality}
\end{align}\es 
Applying the triangle inequality to the integrals over \(k\), we obtain \bs 
\begin{align}
\left|
\overline{H_{\rm NY}^{T}}
\right|
&\leq
8\pi G
\int_0^\infty\frac{dk}{k}\,\mathscr E(k),
\label{NY-energy-inequality}
\\
\left|
\overline{H_{\rm CS}^{T}}
\right|
&\leq
32\pi G
\int_0^\infty dk\,k\,\mathscr E(k).
\label{CS-energy-inequality}
\end{align}\es 
Equivalently, writing \(dE(k)=\mathscr E(k)\,dk\), these inequalities
take the compact form
\bs 
\begin{align}
\left|
\overline{H_{\rm NY}^{T}}
\right|
&\leq
8\pi G
\int_0^\infty\frac{dE(k)}{k},
\\
\left|
\overline{H_{\rm CS}^{T}}
\right|
&\leq
32\pi G
\int_0^\infty k\,dE(k).
\end{align}\es 

If the spectrum has support in a finite interval
\(k_{\min}\leq k\leq k_{\max}\), the preceding relations imply
\bs \begin{align}
\left|
\overline{H_{\rm NY}^{T}}
\right|
&\leq
\frac{8\pi G}{k_{\min}}E,
\label{NY-hard-bound}
\\
\left|
\overline{H_{\rm CS}^{T}}
\right|
&\leq
32\pi G\,k_{\max}E.
\label{CS-hard-bound}
\end{align}\es 
The first inequality requires a nonzero infrared cutoff, whereas the
second requires control of the ultraviolet part of the spectrum.
The spectral inequalities are saturated when only one circular
polarization is present. Saturation of the integrated inequalities also
requires the helicity asymmetry to have the same sign throughout the
spectrum. For a monochromatic, purely circularly polarized wave of
wave number \(k_0\), one therefore has
\begin{equation}
\left|
\overline{H_{\rm NY}^{T}}
\right|
=
\frac{8\pi G}{k_0}E,
\qquad
\left|
\overline{H_{\rm CS}^{T}}
\right|
=
32\pi G\,k_0E,
\end{equation}
together with
\begin{equation}
\overline{H_{\rm CS}^{T}}
=
4k_0^2\overline{H_{\rm NY}^{T}}.
\end{equation}

The inequality for \(H_{\rm CS}^{T}\) applies to the spin-2
gravito-current helicity defined in Eq.~\eqref{hcstmode}. It does not
apply without modification to the complete Pontryagin
Chern-Simons descendant, which contains the additional
\(h^{\rm TT}_{ij}\nabla^2Ch^{\rm TT}_{ij}\) contribution.

\section{Multipole expansion}\label{multipoleexpansion}
Suppose that the physical stress energy tensor is spatially localized,
\be
T_{\mu\nu}(t,\mathbf x')=0,
\qquad
|\mathbf x'|>r_c.
\label{compact-physical-source}
\ee
We impose regularity in the interior and asymptotic decay on the
constraint sector, together with the retarded, no-incoming-radiation
boundary condition on the tensor sector. These conditions set the
independent homogeneous solutions to zero. The resulting formal
solutions are
\bs\begin{align}
    \Xi_i(t,\mathbf x)&=4G\int d^3\mathbf x' \frac{T_{0i}^{\text{T}}(t,\mathbf x')}{|\mathbf x-\mathbf x'|},\label{xiT}\\ 
    h_{ij}^{\text{TT}}(t,\mathbf x)&=4G\int d^3\mathbf x' \frac{T_{ij}^{\text{TT}}(t-|\mathbf x-\mathbf x'|,\mathbf x')}{|\mathbf x-\mathbf x'|}.\label{hijsol2}
\end{align}\es We have ignored the plane wave solution for the spin-2 modes.
\paragraph{Spin-1 modes.} The SVT decomposition separates the momentum density 
\(T_{0i}\) into transverse and longitudinal modes
\be 
T_{0i}=T_{0i}^{\text{T}}+\partial_iS,\quad \partial^i T_{0i}^{\text{T}}=0
\ee 
where the transverse mode is 
\be 
T_{0i}^{\text{T}}=\left(\delta_{ij}-\frac{\partial_i\partial_j}{\nabla^2}\right)T_{0j}. \label{proT0i}
\ee 
Here the inverse Laplacian and its zero mode are fixed by the same
boundary conditions used in the SVT decomposition.  For fields that
vanish sufficiently rapidly at spatial infinity, according to \eqref{inverselap}, 
the projector in \eqref{proT0i} is consequently
nonlocal in position space \cite{Flanagan:2005yc}. Suppose that the stress tensor has support inside a ball of
radius \(r_c\).  The vector constraint
can then be solved either in terms of the projected source or directly
in terms of \(T_{0i}\)\bs
\begin{align}
\Xi_i(t,\mathbf x)
&=4G\int d^3\mathbf x'\,
\frac{T_{0i}^{\rm T}(t,\mathbf x')}
{|\mathbf x-\mathbf x'|}
=2G\int d^3\mathbf x'\,
K_{ij}(\mathbf x-\mathbf x')T_{0j}(t,\mathbf x'),
\label{sourcetoXi}
\end{align}
\es
where
\be
K_{ij}(\mathbf R)
=\frac{\delta_{ij}+N_iN_j}{R},
\qquad
\mathbf R=\mathbf x-\mathbf x',
\qquad
R=|\mathbf R|,
\qquad
N_i=\frac{R_i}{R}.
\label{transverse-green-kernel}
\ee
%Due to the $\nabla^{-2}$ operator, the transverse mode is non-vanishing even in the sourceless region $r>r_c$. Using Fourier transform, we solve the spin-1 mode as follows 
%\be 
%\Xi_i(t,\mathbf x)=2G\int d^3\mathbf x' \frac{\delta_{ij}+N_i N_j}{R}T_{0j}(t,\mathbf x')\label{sourcetoXi}
%\ee where 
%\be 
%\mathbf R=\mathbf x-\mathbf x',\quad R=|\mathbf R|,\quad \mathbf N=\frac{\mathbf R}{R}
%\ee and the integral kernel is 
%\be 
%K_{ij}(\mathbf x-\mathbf x')=\frac{\delta_{ij}+N_i N_j}{R}.
%\ee 

For $r>r_c$, introduce the multipole moments 
\be \mathcal J_{j|L}(t)
=
\int d^3\mathbf x'\,
x'_{i_1}\cdots x'_{i_\ell}
T_{0j}(t,\mathbf x').
\qquad L=i_1\cdots i_\ell,\ee  then we find the multipole expansion in the region $r>r_c$
\be \Xi_i(t,\mathbf x)
=
2G
\sum_{\ell=0}^{\infty}
\frac{(-1)^\ell}{\ell!}
\mathcal J_{j|L}(t)
\partial_L K_{ij}(\mathbf x).\ee
The moments \(\mathcal J_{j|L}\) are symmetric in the multi-index
\(L\), but they have not been projected to their STF parts.  In
contrast to the scalar \(1/r\) expansion, their traces cannot in
general be discarded because
\be
\nabla^2K_{ij}(\mathbf x)
=\frac{2}{r^3}\left(\delta_{ij}-3n_in_j\right),\qquad n_i=\frac{x_i}{r},
\qquad r\ne0.
\label{kernel-not-harmonic}
\ee Thus these moments should not be identified directly with the
standard irreducible source multipoles used in gravitational radiation
theory.  Their decomposition into irreducible Cartesian tensors can
be organized by the usual STF methods
\cite{Thorne:1980ru,Damour:1990gj,2014LRR....17....2B}.
 The first few multipole moments are  
\bs\begin{align}
P_i&=\int d^3\mathbf x'\,T_{0i}(t,\mathbf x'),\\
D_{ij}
&=\int d^3\mathbf x'\,x'_jT_{0i}(t,\mathbf x').\end{align}\es
The $P_i$ is the momentum of the system. Using the conservation law of stress tensor, $P_i$ is simplified to 
\be 
P_i=\int d^3\mathbf x' \partial_j'(x_i' T_{0j}(t,\mathbf x'))-\int d^3\mathbf x' x_i' \partial'_j T_{0j}(t,\mathbf x')=-\frac{d}{dt}I_i
\ee 
where $I_i$ is the dipole moment 
\be 
I_i=\int d^3\mathbf x' x'_i T_{00}(t,\mathbf x').
\ee 
We have ignored the boundary term by assuming the source is located inside the ball $r<r_c$. Thus the $\ell=0$ moment contributes to the spin-1 mode as follows 
\be 
\Xi_i^{(\ell=0)}(t,\mathbf x)=\frac{2G}{r}\left(P_i+n_i(\mathbf P\cdot \mathbf n)\right)=-\frac{2G}{r}\left(\dot I_i+n_i n_j\dot I_j \right).
\ee In the center-of-mass frame of an isolated system, the momentum is zero and one should consider the next order. The moment with $\ell=1$ can be decomposed into symmetric traceless, anti-symmetric and trace terms 
\be 
D_{ij}=D_{\langle ij\rangle}+D_{[ij]}+\frac{1}{3}\delta_{ij}D_{kk}.
\ee For the anti-symmetric part, we define 
\be 
D_{[ij]}
=
-\frac12\epsilon_{ijk}S_k\quad \Rightarrow\quad S_i=\epsilon_{ijk}D_{kj}=\epsilon_{ijk}\int d^3\mathbf x' x'_j T_{0k}(t,\mathbf x').
\ee  

Therefore, $S_i$ is the total angular momentum of the system. For the symmetric traceless part, we use the conservation of stress tensor 
\bea 
D_{\langle ij\rangle}&=&\int d^3\mathbf x' x'_{\langle j}T_{0i\rangle}(t,\mathbf x')=-\frac{1}{2}\dot{I}_{ij}
\eea where $I_{ij}$ is the STF  quadrupole moment 
\be 
I_{ij}=\int d^3\mathbf x' (x'_i x'_j-\frac{1}{3}\delta_{ij}x'^2) T_{00}(t,\mathbf x').\label{defIij}
\ee 
Since 
\be 
\partial_k K_{ij}=\frac{1}{r^2}\left(-3n_i n_jn_k-\delta_{ij}n_k+\delta_{ik}n_j+\delta_{jk}n_i\right),
\ee the $\ell=1$ term is 
\bea 
\Xi_i^{(\ell=1)}=\frac{2G}{r^2}\left(3D_{\langle jk\rangle} n_i n_j n_k+\epsilon_{ijk}S_j n_k\right).
\eea The trace term does not contribute to the expansion.
Therefore, the leading two orders of  the multipole expansion of $\Xi_i$ are 
\be 
\Xi_i=\frac{2G}{r}(P_i+n_i n_j P_j)+\frac{2G}{r^2}\left(3D_{\langle jk\rangle} n_i n_j n_k+\epsilon_{ijk}S_j n_k\right)+\mathcal{O}(r^{-3}).\label{Xiexpansion}
\ee 
\begin{enumerate}
    \item 
In a general reference of frame, $P_i\not=0$, the leading terms are 
\bs\label{leadingp}\begin{align}
\mathcal H_{\mathrm{NY}}^V
&=
-\frac{G^2}{r^4}
\Big[
P_iS_i
+
3(P_in_i)(S_jn_j)
+
6P_i\epsilon_{ijk}
D_{\langle j\ell\rangle}
n_\ell n_k
\Big]
+\mathcal O(r^{-5}),\\ 
\mathcal H_{\mathrm{CS}}^V
&=
\frac{4G^2}{r^6}
\Big[
P_iS_i
+
3(P_in_i)(S_jn_j)
+
6P_i\epsilon_{ijk}
D_{\langle j\ell\rangle}
n_\ell n_k
\Big]
+\mathcal O(r^{-7}).
\end{align}\es Interestingly, the leading order of these two quantities are proportional to each other  
\be 
\mathcal H_{CS}^V=-\frac{4}{r^2} \mathcal H_{NY}^V+\mathcal{O}(r^{-7}).
\ee  This proportionality is not an exact identity between the two local
densities. Their angular integrals at fixed \(r\) are
\bea 
\int d\Omega \mathcal H_{NY}^V=-\frac{8\pi G^2}{r^4}\mathbf P\cdot\mathbf S+\mathcal{O}(r^{-5}).
\eea  These are angular integrals of the volume densities, not the
codimension-two surface integrals; the latter contain the area factor
\(r^2\).
The pseudoscalar \(\mathbf P\cdot\mathbf S\) is invariant under a
spatial translation.  Indeed,
\be
x^i\longrightarrow x^i-a^i,
\qquad
\mathbf S\longrightarrow\mathbf S-\mathbf a\times\mathbf P,
\qquad
\mathbf P\cdot\mathbf S\longrightarrow\mathbf P\cdot\mathbf S.
\ee
 It has the kinematic form of
a spin projection along the momentum, but for a massive system it is the temporal component of the Pauli-Lubanski
pseudovector.  Thus this term is a
frame-dependent boosted contribution, rather than an "extra
helicity". It vanishes in the center-of-mass frame. We will discuss this result using boosted Kerr solution.

\item In the center-of-mass frame, $P_i=0$, the leading results are 
\bs\begin{align}
\mathcal H_{\mathrm{NY}}^V
&=
-\frac{6G^2}{r^5}
S_iD_{\langle ij\rangle}n_j
+\mathcal O(r^{-6}),\\
\mathcal H_{\mathrm{CS}}^V
&=
\frac{24G^2}{r^7}
\Big[
S_iD_{\langle ij\rangle}n_j
+
2(S_in_i)
D_{\langle jk\rangle}n_jn_k
\Big]
+\mathcal O(r^{-8}).\label{nonvanishinghcsv}
\end{align}\es 
\end{enumerate}
Both angular integrals vanish because the displayed angular
structures are odd under \(\mathbf n\rightarrow-\mathbf n\).  The leading contribution is a
coupling between the current-dipole moment \(S_i\) and the time
derivative of the mass quadrupole \(I_{\langle ij\rangle}\). %They are zero when integrated over the unit sphere. At the leading order, the non-vanishing result is from the interaction term of the spin and the quadrupole. 

The nonzero exterior value of
\(\mathcal H_{\rm CS}^{V}\) in \eqref{nonvanishinghcsv}  is not a contradiction.  Due to the non-local projector 
\eqref{proT0i}, the projected constraint
\be
\nabla^2\Xi_i=-16\pi G\,T_{0i}^{\rm T}
\ee
does not imply \(\nabla^2\Xi_i=0\) outside the support of the
unprojected stress tensor.  The apparent tension arises only if the
nonlocal projected source is treated as though it had the same
compact support as \(T_{\mu\nu}\).  

%The above expansion indicates that $\mathcal H_{CS}^V\not=0$ even in the region $r>r_c$ where the source vanishes. It seems that this result  contradicts with the definition 
%\be 
%\mathcal H_{CS}^V=-\frac{1}{2}\epsilon_{ijk}\partial_j\Xi_k \nabla^2\Xi_i=-\frac{1}{2}\left(\nabla\times\mathbf \Xi \right)\cdot\nabla^2\mathbf\Xi
%\ee while the spin-1 mode is determined by the source via \eqref{xiT}. The point is that $T_{0i}^{\text{T}}$ does not vanishes outside  the ball $r=r_c$ even when $T_{\mu\nu}$ vanishes for $r>r_c$.

\paragraph{Spin-2 modes.} 
%It can be shown that $V_i$ contributes to the spin-1 mode via a gradient 
%\be 
%\Xi_i=\frac{6G}{5r^3}\left[3n_in_j V_j-V_i\right]+\cdots=-\frac{6G}{5}\partial_i\left(\frac{n_j V_j}{r^3}\right)+\cdots.
%\ee Since the curl of a gradient is zero, $V_i$ contributes nothing to the helicity densities. 

We now turn to the radiative tensor sector.  In the slow motion
approximation and at leading order in the wave zone expansion, the
transverse-traceless field is
\be 
h_{ij}^{\text{TT}}(t,\mathbf x)=\frac{2G}{r}\Pi_{ij,kl}(\mathbf n)\ddot{I}_{kl}(u)+\cdots,\qquad
u=t-r
\label{hijTTquad}
\ee 
in the slow motion limit. Here $I_{ij}$ is the STF quadrupole moment of the source.
Equation~\eqref{hijTTquad} is simultaneously the leading \(1/r\)
term and the leading mass quadrupole term in the slow motion
multipole expansion.
The ellipsis therefore includes both subleading powers of \(1/r\)
and higher radiative multipoles, the latter of which can also occur
at order \(1/r\).

To discuss the causal support on a constant time hypersurface, first
consider an idealized source that is stationary before \(u_i\) and
after \(u_f\).  Stationarity does not require the quadrupole itself
to vanish.  Rather, one may write
\be
I_{ij}(u)
=
\begin{cases}
I_{ij}^{\rm in}, & u<u_i,\\
I_{ij}^{\rm dyn}(u), & u_i<u<u_f,\\
I_{ij}^{\rm out}, & u>u_f,
\end{cases}
\qquad
I_{ij}^{(p)}=0
\quad
\text{outside }(u_i,u_f),\quad p\geq1.
\label{stationary-quadrupole-support}
\ee
This compact support assumption concerns the radiative time
derivatives.  It is stronger than necessary for the helicity
densities below, but makes the causal discussion transparent.  We
also assume sufficient smoothness at \(u_i\) and \(u_f\), so that the
idealized switching does not introduce distributional wave fronts.

Consider the constant time hypersurface
\be
V=\{t=\mathrm{constant}\},
\qquad
t>u_f.
\ee
The outgoing null hypersurfaces \(u=u_i\) and \(u=u_f\) intersect
\(V\) on spheres of radii
\be
r_i=t-u_i,
\qquad
r_f=t-u_f,
\qquad
r_f<r_i.
\ee
The part of \(V\) on which the wave zone approximation is
valid is consequently divided into
\be
\boxed{
\begin{aligned}
r<r_f
&\quad\Longleftrightarrow\quad
u>u_f,
&&\text{post-radiation region},
\\
r_f<r<r_i
&\quad\Longleftrightarrow\quad
u_i<u<u_f,
&&\text{active-radiation-shell},
\\
r>r_i
&\quad\Longleftrightarrow\quad
u<u_i,
&&\text{pre-radiation region}.
\end{aligned}}
\label{causal-radiation-regions}
\ee
Every line in \eqref{causal-radiation-regions} is to be intersected
with the wave zone condition
\be
r\gg r_c,
\qquad
r\gg\lambda_{\rm GW},
\qquad
r\gg GM.
\label{wave zone-conditions}
\ee
In particular, the first region need not contain a wave zone portion
unless \(r_f\) is sufficiently large. 

\begin{figure}[t]
\centering
\includegraphics[width=0.82\textwidth]
{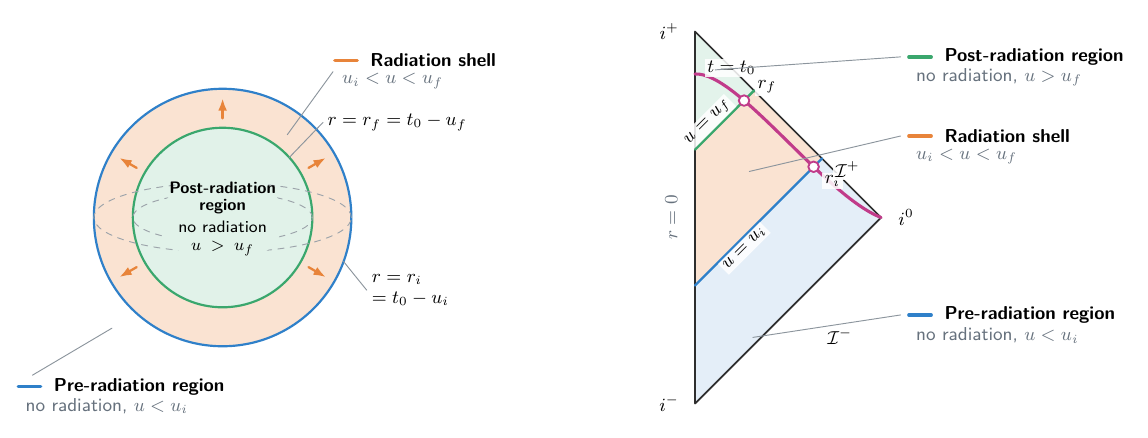}
\caption{Intersection of a constant time hypersurface with the
outgoing null hypersurfaces \(u=u_i\) and \(u=u_f\).  The displayed
three region decomposition applies to the finite duration radiative
approximation.}
\label{region}
\end{figure}

At fixed \(t\), smaller radii correspond to later retarded times,
whereas larger radii probe earlier states of the source.  Thus
\be
r\longrightarrow\infty
\quad\text{at fixed finite }t
\qquad\Longrightarrow\qquad
u=t-r\longrightarrow-\infty.
\label{spatial-infinity-limit}
\ee
Spatial infinity on such a slice therefore lies in the
pre-radiation region. In Figure \ref{region}, we have shown the three regions.  For a compact pulse in linearized gravity on
four-dimensional Minkowski spacetime, this sharp shell follows from
Huygens propagation.  On black hole backgrounds, curvature backscattering produces late time tails, as in Price's analysis of perturbations of Schwarzschild spacetime \cite{Price:1971fb}. The sharp compact shell used here should therefore be regarded as a Minkowski space idealization rather than as a generic description of propagation on curved backgrounds.

Since we focus on linearized gravity, only in the radiation-shell, the helicity densities are non-vanishing. In this case, we have
\bs\begin{align}
    \mathcal H_{NY}^T&=\frac{
    G^2}{r^2}\dddot{I}_{ij}\ddot{I}_{kl}Q_{ijkl}(\mathbf n)+\mathcal{O}(r^{-3}),\label{spin2he}\\
    \mathcal H_{CS}^T&=\frac{4G^2}{r^2}\ddddot{I}_{ij}\dddot{I}_{kl}Q_{ijkl}(\mathbf n)+\mathcal{O}(r^{-3}).\label{spin2he2}
\end{align}\es  The tensor structure \(Q_{ijkl}(\mathbf n)\) has been defined in \cite{Long:2024yvj}
\be 
Q^{ijkl}(\mathbf n)
=
-P^{jk}(\mathbf n)\epsilon^{ilm}n_m.
\ee The expression \eqref{spin2he} matches the helicity flux density near null infinity \cite{Long:2024yvj} up to some factors 
\be 
\frac{r^2}{8\pi G}\mathcal H_{NY}^T=\frac{dH}{dud\Omega}+\mathcal{O}(r^{-1}).\label{matching}
\ee 
The factor $r^2$ originates from the spatial volume element \(d^3x\)
and converts the volume density into a radial density per unit solid
angle
\begin{equation}
    d^3x=r^2\,dr\,d\Omega.
\end{equation} On a constant time hypersurface one has
\begin{equation}
    du=-dr.
\end{equation}
The minus sign is accounted for by reversing the radial integration
limits. The factor $1/(8\pi G)$ is the canonical gravitational
normalization appearing in the helicity flux density at $\mathscr I^+$.

The relation between the spin-2 gravitomagnetic helicity  density and the
helicity flux density should be understood as an asymptotic
matching relation in their common domain of validity, rather than as
an exact identity at arbitrary finite $t$ and $r$.
The quadrupole expression for
$\mathcal H_{\mathrm{NY}}^{T}$ is valid in the radiation zone, whereas
$dH/(du\,d\Omega)$ is defined in the asymptotic  region near
\(\mathscr I^+\). Their common matching region is therefore
\begin{equation}
\mathcal D_{\mathrm{match}}
=
\left\{
(t,r,\Omega)\ \middle|\
r\gg r_c,\quad
r\gg\lambda_{\mathrm{GW}},\quad
r\gg GM,\quad
u_i<u=t-r<u_f
\right\},
\end{equation}
where $r_c$ is the characteristic source size,
$\lambda_{\mathrm{GW}}$ is the gravitational wavelength, and
$u_i<u<u_f$ is the retarded-time interval during which the source
radiates.

Within this overlap region, both descriptions are determined by the
same leading radiative coefficient \eqref{hijTTquad}
and one obtains the asymptotic matching relation \eqref{matching}.
More precisely, the equality of the leading asymptotic coefficients is
defined by the null limit
\begin{equation}
\frac{dH}{du\,d\Omega}(u,\Omega)
=
\lim_{\substack{r\rightarrow\infty,\;t\rightarrow\infty\\
u=t-r\ \mathrm{finite}}}
\frac{r^2}{8\pi G}
\mathcal H_{\mathrm{NY}}^{T}(t,r,\Omega),
\qquad
u_i<u<u_f .\label{rehe}
\end{equation}

This limit should not be confused with the spatial infinity limit
$r\rightarrow\infty$ at fixed finite $t$. In the latter case,
$u=t-r\rightarrow-\infty$, and the observation point leaves the
radiation-shell and enters the pre-radiation region.  The complete
multipole expansion \cite{Thorne:1980ru} contains higher radiative
multipoles without altering this correspondence.

The spin-2 gravito-current density has the same quadrupolar structure
as the spin-2 gravitomagnetic density, with \(I_{ij}\) replaced by
\(\dot I_{ij}\). The latter can be compared with the gravitational
helicity flux density at \(\mathscr I^+\) considered in
\cite{Long:2024yvj}. With the same rescaling, the spin-2
gravito-current density gives
\be 
\lim_{\substack{r,t\rightarrow\infty\\u=t-r\ {\rm finite}}}
\frac{r^2}{8\pi G}\,
\mathcal H_{\mathrm{CS}}^T
=
\frac{G}{2\pi}
\ddddot{I}_{ij}\dddot{I}_{kl}
Q_{ijkl}(\mathbf n).
\ee 
This is not the same helicity flux density studied in
\cite{Long:2024yvj}.  Instead, it agrees, up to normalization,
with the density of the quantity studied
in \cite{CalderonBustillo2025Mirror}.
It is a higher time derivative,
helicity sensitive quantity and differs in physical dimension from
$dH/(dud\Omega)$ by two inverse powers of
time. The comparison is valid in the common wave zone region and
becomes an exact asymptotic matching only in the null limit
$r,t\rightarrow\infty$ with $u=t-r$ held fixed.

For a monochromatic outgoing waveform of angular frequency
\(\omega\), the leading wave-zone expressions obey
\be
\mathcal H_{\rm CS}^{T}
=4\omega^2\mathcal H_{\rm NY}^{T}.
\label{relationnYCS1}
\ee
For a circular binary, the leading gravitational wave frequency is
\(\omega_{\rm GW}=2\Omega\), and hence
\be
\mathcal H_{\rm CS}^{T}
=16\Omega^2\mathcal H_{\rm NY}^{T}.
\label{relationnYCS2}
\ee
For a general transient or multifrequency waveform there is no
universal pointwise proportionality.  The factor \(4\omega^2\)
applies frequency by frequency and, in the absence of coherent
cross-frequency terms, to the corresponding time averaged spectral
components.

For a finite duration signal, define
\bs
\begin{align}
\overline{\mathcal H_{\rm NY}^{T}}(r,\mathbf n)
&\equiv
\frac{1}{u_f-u_i}
\int_{u_i}^{u_f}du\,
\mathcal H_{\rm NY}^{T}(u,r,\mathbf n),
\\
\overline{\mathcal H_{\rm CS}^{T}}(r,\mathbf n)
&\equiv
\frac{1}{u_f-u_i}
\int_{u_i}^{u_f}du\,
\mathcal H_{\rm CS}^{T}(u,r,\mathbf n).
\end{align}
\label{spin2-time-averages}
\es

For a periodic source, \(u_f-u_i\) is replaced by one period.
Whenever the denominator is nonzero and the ratio is positive, one
may define
\be
\omega_\star^2(r,\mathbf n)
\equiv
\frac{
\overline{\mathcal H_{\rm CS}^{T}}(r,\mathbf n)}
{4\,
\overline{\mathcal H_{\rm NY}^{T}}(r,\mathbf n)},
\qquad
\lambda_\star(r,\mathbf n)
\equiv\frac{2\pi}{\omega_\star(r,\mathbf n)}.
\label{charafre}
\ee
For a monochromatic waveform,
\(\omega_\star=|\omega|\) and
\(\lambda_\star=2\pi/|\omega|\).  Both helicity densities are signed.
Opposite chiralities at different frequencies can therefore make the
denominator vanish or make the ratio negative.  In that situation
\(\omega_\star\) is not a real characteristic frequency; one must
instead define the frequency moment separately in each helicity
sector, or adopt an explicitly positive spectral weighting.

\iffalse 
which could be obtained by Helmhotz decomposition 
\be 
T_{0i}=T_{0i}^{\text{T}}+\partial_i \sigma,\quad \partial_i T_{0i}^{\text{T}}=0.
\ee It can be written as 
\be 
T_{0i}^{\text{T}}=\left(\delta_i^j-\frac{\partial_i\partial^j}{\nabla^2}\right)T_{0j}.
\ee Using Fourier transformation, we find 
\bea 
\Xi_i(t,\mathbf x)&=&16\pi G\int d^3\mathbf k e^{i\mathbf{k}\cdot\mathbf{x}}\frac{1}{\mathbf k^2}\left(T_{0i}(t,\mathbf k)-\frac{k_i k_j}{\mathbf k^2} T_{0j}(t,\mathbf k)\right)\nn\\&=&2G\int d^3\mathbf x'\frac{\delta_{ij}+\widehat R_i\widehat R_j}{R}T_{0j}(t,\mathbf x')
\eea where 
\be 
R_i=x_i-x'_i,\quad R=|\mathbf x-\mathbf x'|,\quad \widehat R_i=\frac{R_i}{R}.
\ee  

Similarly, $T_{ij}^{\text{TT}}$ is the transverse and traceless part of the stress tensor $T_{ij}$
\be 
T_{ij}^{\text{TT}}=(\delta_i^k-\frac{\partial_i\partial^k}{\nabla^2})(\delta_j^l-\frac{\partial_j\partial^l}{\nabla^2})T_{kl}.
\ee \fi
\section{Applications}
In this section, we evaluate the four gauge invariant helicity densities
introduced above for several representative gravitational configurations.
The examples are chosen to probe complementary physical regimes: sourced
and source-free fields, nonradiative and radiative sectors, and configurations
carrying orbital, intrinsic-spin, or topological structure.  Together, they
clarify which parts of the gravitational field are measured by the spin-1 and
spin-2 gravitomagnetic helicities and by their higher derivative
gravito-current counterparts.
We first consider two-body systems in the weak field and slow motion regime.
This example connects the helicity densities with the source multipole moments
and separates the constrained spin-1 field from the radiative transverse
traceless field.  In particular, it allows us to study the angular distribution
of the spin-2 densities in the wave zone and their dependence on the orbital
parameters.  We then turn to a boosted Kerr black hole, which provides a
nonradiative test of how linear momentum and intrinsic angular momentum enter
the helicity densities in a chosen asymptotic inertial frame.  The third
example is a gyratonic pp-wave, namely a null gravitational field sourced by a beam that carries both energy and intrinsic angular momentum.
It illustrates how spin-1 and spin-2 SVT sectors can coexist in a propagating
configuration.  Finally, we examine the gravitational Hopfion, a localized
source free gravitational wave with nontrivial linked tendex and vortex
structures.  In this case the asymptotically decaying vacuum constraints
eliminate the spin-1 SVT mode, so that only the two spin-2 helicity densities
remain.

Throughout this section, all helicity densities are defined on
constant time hypersurfaces and are evaluated only after the complete
SVT projection has been performed.  In particular, a
coordinate component $h_{0i}$, or a gravitoelectromagnetic vector potential
introduced in a specific coordinate system, must not be identified directly
with the gauge invariant vector mode $\Xi_i$.  For exact solutions, such as
Kerr and pp-wave geometries, the present definitions apply to their
linearization about Minkowski spacetime in the region where the weak field
approximation is valid.  
\subsection{Two-body systems}
We now evaluate the four local helicity densities for Newtonian
two-body motion\footnote{For an \(N\)-body system, closed-form trajectories
are generally unavailable. Once the trajectories are specified, the
corresponding local helicity densities follow from the formulas collected
in Appendix~\ref{moving}.}.
The calculation is performed at leading order in
the weak field and slow motion expansions.  We neglect intrinsic
spins, radiation reaction, periastron precession, and all
post-Newtonian corrections, so the relative orbit is an exact
Keplerian conic.  For a bound binary this should be understood as the
leading adiabatic approximation over one orbital period
\cite{Peters:1963ux,Peters:1964zz}.

The two stars, with masses $M_1$ and $M_2$, move in the $x$-$y$ plane and are separated by a distance $D$. In the center-of-mass frame, we use 
 $e$ to denote the orbital eccentricity: $e=0$ for circular orbits, $0<e<1$ for elliptical orbits, $e=1$ for parabolic orbits, and $e>1$ for hyperbolic orbits. The total momentum of the system is zero and the  orbital angular momentum points along $\mathbf e_z$. The
observation point is in the direction 
\be 
\mathbf n=(\sin\theta\cos\phi,\sin\theta\sin\phi,\cos\theta).
\ee To simplify notation, we introduce 
\be 
f\equiv1+e\cos\psi,
\qquad
\mathbf e_r=(\cos\psi,\sin\psi,0),\qquad
\mathbf e_\psi=(-\sin\psi,\cos\psi,0).
\ee Here $\psi$ is the true anomaly, measured in the orbital plane
from the direction of periastron. Our convention agrees with the standard Keplerian
parametrization of \cite{BrouwerClemence1961}. The unit vectors  $\mathbf e_r$ and $\mathbf e_\psi$ are the instantaneous radial and tangential unit vectors in the orbital plane.
The instantaneous separation $D$ between the two bodies is
\be 
D=\frac{\epsilon}{f}
\ee where 
$\epsilon$ is called semi-latus rectum. It equals $a$ for a circle, $a(1-e^2)$ for an ellipse,
remains finite in the parabolic limit, and equals $a(e^2-1)$ for a
hyperbola. Here $a$ is the orbital radius for a circular orbit, semi-major axis for an ellipse,
positive magnitude of the conventional semi-major axis for a
hyperbola. 

The moment conventions require some care.  Earlier we defined
\(P_i,D_{ij}\), and \(S_i\) using \(T_{0i}\).  With signature
\((-+++)\), \(T_{0i}=-T^{0i}\) at Newtonian order.  Consequently,
for those definitions,
\bs 
\begin{align}
P_i&=-\sum_A M_Av_{Ai}=0,
\label{eq:Pzero}\\
S_i&=-\sum_A M_A(\bm x_A\times\bm v_A)_i
=-\mu\sqrt{G\overline M\epsilon}\,\delta_{iz},
\label{eq:Si}\\
I_{ij}&
=\mu D^2\left(e_{ri}e_{rj}-\frac13\delta_{ij}\right),
\label{eq:Mij}\\
 D_{\langle ij\rangle}&=-\frac12\dot I_{ij}.
\label{eq:Dmoment}
\end{align}\es 
The vanishing of $P_i$ follows from the center-of-mass choice.  The quantity
$S_i$ is the current dipole and equals the conserved minus orbital angular momentum. $I_{ij}$ is the STF mass quadrupole moment. We have also defined the total mass of the binary 
\be 
\overline M=M_1+M_2
\ee and the reduced mass governing the relative Newtonian motion
\be 
\mu=\frac{M_1M_2}{M_1+M_2}.
\ee 
The direction cosines adapted to the instantaneous orbit are
\bs 
\begin{align}
n_r&\equiv\bm n\cdot\bm e_r
=\sin\theta\cos(\phi-\psi),\\
n_\psi&\equiv\bm n\cdot\bm e_\psi
=\sin\theta\sin(\phi-\psi),\\
n_z&\equiv\bm n\cdot\bm e_z=\cos\theta,
\end{align}\es
and they satisfy $n_r^2+n_\psi^2+n_z^2=1$.

Two spin-1 helicity densities are
\bs 
\begin{align}
\mathcal H_{NY}^V
&=\frac{2G^3\overline M\mu^2\epsilon}{r^5}
\frac{e\sin\psi}{f}\,n_z
+\mathcal{O}(r^{-6}),
\label{eq:HNYV-master}\\
\mathcal H_{CS}^V&
=\frac{24G^3\overline M\mu^2\epsilon}{r^7}\,n_z
\left[
\frac{e\sin\psi}{f}(2n_r^2-1)
+2n_r n_\psi
\right]
+\mathcal{O}(r^{-8}).
\label{eq:HCSV-master}
\end{align}\es 

The radiative spin-2 densities are
\bs 
\begin{align}
\mathcal H_{NY}^T
&=\frac{4G^{9/2}\overline M^{5/2}\mu^2}
{r^2\epsilon^{7/2}}
n_z f^3\mathcal F_{NY}(e;\psi,\bm n)
+\mathcal{O}(r^{-3}),
\label{eq:HNYT-master}\\
\mathcal H_{CS}^T&
=\frac{16G^{11/2}\overline M^{7/2}\mu^2}
{r^2\epsilon^{13/2}}
n_z f^6\mathcal F_{CS}(e;\psi,\bm n)
+\mathcal{O}(r^{-3})
\label{eq:HCST-master}
\end{align}\es 
where\bs  \begin{align}
\mathcal F_{NY}(e;\psi,\bm n)
={}&2(1+n_z^2)
+e\left[
(6-2n_r^2-4n_\psi^2)\cos\psi
-n_rn_\psi\sin\psi
\right]
\nonumber\\
&+e^2\left[
(3-n_r^2-2n_\psi^2)\cos^2\psi
+n_r^2-1
-n_rn_\psi\sin\psi\cos\psi
\right],
\label{eq:FNY}\\
\mathcal F_{CS}(e;\psi,\bm n)
={}&8(1+n_z^2)
+e\left[
(30-14n_r^2-16n_\psi^2)\cos\psi
-4n_rn_\psi\sin\psi
\right]
\nonumber\\
&+e^2\left[
(15-7n_r^2-8n_\psi^2)\cos^2\psi
+n_r^2-1
-4n_rn_\psi\sin\psi\cos\psi
\right].
\label{eq:FCS}
\end{align}\es 
Equations~\eqref{eq:HNYV-master}--\eqref{eq:HCST-master} are the four master
densities.  The orbit classes differ only in the allowed eccentricity, the
semi-latus rectum, and the range of the true anomaly.
One should keep in mind that the vector and tensor results are not taken in the same asymptotic limit.
This distinction is essential.
 The spin-1 expressions are large radius expansions on a fixed
$t$ hypersurface.  They require $r\gg D(t)$, where $D$ is the binary
separation, and the orbital phase is evaluated at $t$.
The spin-2 expressions are wave zone expansions with
$r,t$ large and $u=t-r$ fixed.  Their orbital phase and source
multipoles are evaluated at retarded time $u$.
To avoid cumbersome notation, the same symbol $\psi$ is used in both sectors.
It means $\psi(t)$ in every formula carrying the superscript $V$ and
$\psi(u)$ in every formula carrying the superscript $T$.

    \paragraph{Circular orbit.} For a circular orbit,
\begin{equation}
e=0,
\qquad
\epsilon=a,
\qquad
D=a,
\qquad
\psi=\Omega\tau+\psi_0,
\qquad
\Omega=\sqrt{\frac{G\overline M}{a^3}}.
\end{equation}
Here $\tau=t$ in the vector sector and $\tau=u$ in the tensor sector.  The
four densities reduce to
\bs 
\begin{align}
\mathcal H_{NY}^V&=\mathcal{O}(r^{-6}),
\label{eq:circ-NYV}\\
\mathcal H_{CS}^V&
=\frac{48G^3\overline M\mu^2a}{r^7}
n_rn_\psi n_z+\mathcal{O}(r^{-8}),
\label{eq:circ-CSV}\\
\mathcal H_{NY}^T&
=\frac{8G^{9/2}\overline M^{5/2}\mu^2}{r^2a^{7/2}}
n_z(1+n_z^2)+\mathcal{O}(r^{-3}),
\label{eq:circ-NYT}\\
\mathcal H_{CS}^T&
=\frac{128G^{11/2}\overline M^{7/2}\mu^2}{r^2a^{13/2}}
n_z(1+n_z^2)+\mathcal{O}(r^{-3}).
\label{eq:circ-CST}
\end{align}\es 
The spin-1 gravitomagnetic helicity density vanishes at the leading order. The spin-1 gravito-current helicity density may equivalently be written as
\begin{equation}
\mathcal H_{CS}^V
=\frac{24G^3\overline M\mu^2a}{r^7}
\sin^2\theta\cos\theta\,
\sin[2(\phi-\psi)]+\mathcal{O}(r^{-8}).
\end{equation}
The helicity densities for the spin-2 modes are axisymmetric.  They obey the useful monochromatic
relation \eqref{relationnYCS2}. Note that the spin-2 gravitomagnetic helicity density matches gravitational helicity flux density of \cite{Long:2024yvj} via the identity \eqref{rehe}. At the leading order, the spin-2 helicity densities are independent of time while the spin-1 gravito-current helicity density oscillates with time. In a period, the average spin-1 gravito-current helicity density vanishes 
\be 
\overline {\mathcal H_{CS}^V}=\mathcal{O}(r^{-8}).
\ee 

\paragraph{Elliptic orbit.} For an elliptic orbit,
\begin{equation}
0<e<1,
\qquad
\epsilon=a(1-e^2),
\qquad
0\leq\psi<2\pi,
\qquad
\Omega=\sqrt{\frac{G\overline M}{a^3}}.
\end{equation}
Unlike the circular orbit, 
the instantaneous expressions retain the full phase dependence. One can use the period average to smooth out the time dependence. The period
average must include the nonuniform Kepler measure
$d \tau=d\psi/\dot\psi$
\be 
\overline H=\frac{1}{T}\int_0^T d\tau \mathcal H=\frac{1}{T}\int_0^{2\pi}\frac{d\psi}{\dot\psi}\mathcal H.
\ee 
Here $T$ is the period of the orbit. Therefore, the averaged helicity densities are 
\bs \begin{align}
    \overline{ \mathcal H_{NY}^V}=&\mathcal{O}(r^{-6}),\\ 
    \overline{ \mathcal H_{CS}^V}=&\mathcal{O}(r^{-8}),\\ 
    \overline{\mathcal H_{NY}^T}=&
\frac{G^{9/2}\overline M^{5/2}\mu^2}
{2r^2a^{7/2}(1-e^2)^2}\,n_z
\times
\left[
16(1+n_z^2)
+e^2\left\{14(1+n_z^2)+5\Delta\right\}
\right]
+\mathcal O(r^{-3}),\\ 
    \overline{\mathcal H_{CS}^T}=&
\frac{G^{11/2}\overline M^{7/2}\mu^2}
{8r^2a^{13/2}(1-e^2)^5}\,n_z
\times\Big[
(1+n_z^2)
\left(1024+7328e^2+5232e^4+276e^6\right)
\nn\\&+\Delta\left(464e^2+640e^4+51e^6\right)
\Big]
+\mathcal O(r^{-3}).
\end{align}\es Here \(
\Delta\equiv n_x^2-n_y^2
=\sin^2\theta\cos2\phi\).
The averaged spin-2 gravitomagnetic helicity density agrees, up to the
normalization in Eq.~\eqref{matching}, with the averaged helicity flux
density of \cite{Long:2024yvj}. To our knowledge, the angular
distribution of the averaged spin-2 gravito-current density for a
binary has not previously been given. Both averaged densities above
contain a common factor \(n_z\). Consequently, their defining ratio
\eqref{charafre} is a \(0/0\) expression on the orbital equator
\(n_z=0\). For \(n_z\neq0\), cancellation of the common factor gives
\begin{align}
    \omega^2_*(\mathbf n)&=
\frac{\Omega^2}{16(1-e^2)^3}
\frac{
(1+n_z^2)
\left(
1024+7328e^2+5232e^4+276e^6
\right)
+\Delta
\left(
464e^2+640e^4+51e^6
\right)
}{
16(1+n_z^2)
+e^2\left[
14(1+n_z^2)+5\Delta
\right]
}.\label{omegastar2}
\end{align}
The right-hand side of Eq.~\eqref{omegastar2} is smooth at \(n_z=0\)
and will be used below as the continuous angular extension of the
off-equator ratio. 
When $e\to 0$, the characteristic frequency becomes $\omega_*(\mathbf n)=2\Omega$. For small $e$, the characteristic frequency can be expanded as 
\be 
\omega^2_*(\mathbf n)=4\Omega^2\left[1+c_2(\mathbf n)e^2+\mathcal{O}(e^4)\right].
\ee The coefficient $c_2(\mathbf n)$ is 
\be 
c_2(\mathbf n)
=
\frac{297}{32}
+\frac{9}{64}
\frac{\Delta}{1+n_z^2}.
\ee Since \be|\Delta|
=|n_x^2-n_y^2|
\leq n_x^2+n_y^2
=1-n_z^2,\quad\Rightarrow\quad -1\leq
\frac{\Delta}{1+n_z^2}
\leq1,\ee we find $c_2(\mathbf n)$ is always positive. More precisely, 
\be 
\frac{585}{64}
\leq c_2(\mathbf n)
\leq
\frac{603}{64}.
\ee Therefore, for small $e$, the characteristic frequency 
\be 
\omega_*(\mathbf n)>2\Omega.
\ee 
We have numerically checked that the right hand side of \eqref{omegastar2} is always positive for $0<e<1$.  The positivity of \eqref{omegastar2} can also be established analytically.
Set
\be
x=e^2,
\qquad
z=n_z^2,
\qquad
u=\frac{\Delta}{1+z}
=\frac{1-z}{1+z}\cos2\phi,
\qquad |u|\leq1.
\label{elliptic-xzu}
\ee
Then
\be
W(x,u)
\equiv
\frac{\omega_\star^2}{4\Omega^2}
=
\frac{1}{64(1-x)^3}
\frac{A_E(x)+B_E(x)u}{C_E(x)+5xu},
\label{elliptic-W}
\ee
where
\bs
\begin{align}
A_E(x)&=1024+7328x+5232x^2+276x^3,
\\
B_E(x)&=464x+640x^2+51x^3,
\\
C_E(x)&=16+14x.
\end{align}
\es
For \(0<x<1\),
\[
C_E+5xu\geq16+9x>0,
\qquad
A_E+B_Eu\geq A_E-B_E>0,
\]
the continuous extension satisfies \(W>0\).
The dependence of \(W\) on \(u\) is linear fractional, and
\be
\frac{\partial W}{\partial u}
=
\frac{
x(2304-19904x-16384x^2-666x^3)
}{
64(1-x)^3[C_E(x)+5xu]^2
}.
\label{W-u-derivative}
\ee
Therefore \(\partial W/\partial u=0\) does not locate a generic
interior extremum.  Instead, its sign determines which endpoint of
the allowed interval in \(u\) gives the maximum.  The nontrivial
zero occurs at
\be
x_c\simeq 0.106397,
\qquad
e_c\simeq\sqrt{x_c}=0.32619.
\label{critical-eccentricity}
\ee
Since the global values \(u=\pm1\) are attained only at \(z=0\), the
extrema of the continuous extension lie on the equator:
\[
\begin{array}{c|cc}
&W_{\max}&W_{\min}\\ \hline
0<e<e_c
&
(\theta,\phi)=(\frac{\pi}{2},0),\
(\frac{\pi}{2},\pi)
&
(\theta,\phi)=(\frac{\pi}{2},\frac{\pi}{2}),\
(\frac{\pi}{2},\frac{3\pi}{2})
\\[1mm]
e_c<e<1
&
(\theta,\phi)=(\frac{\pi}{2},\frac{\pi}{2}),\
(\frac{\pi}{2},\frac{3\pi}{2})
&
(\theta,\phi)=(\frac{\pi}{2},0),\
(\frac{\pi}{2},\pi)
\end{array}
\]
At \(e=e_c\), \(W\) is independent of \(u\), and hence of direction:
\be
W(e_c)
=
\frac{
1024+7328e_c^2+5232e_c^4+276e_c^6
}{
64(1-e_c^2)^3(16+14e_c^2)
}
\simeq2.3328,
\qquad
\omega_\star^2(e_c)
\simeq9.3312\,\Omega^2,
\qquad
\omega_\star(e_c)
\simeq3.0547\,\Omega.
\label{critical-W-value}
\ee

\begin{figure}[t]
\centering
\includegraphics[width=0.95\linewidth]
{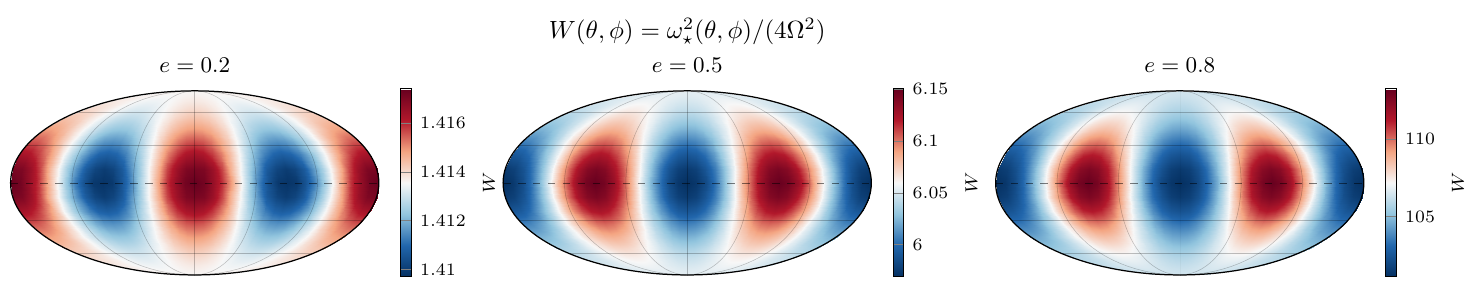}
\caption{Continuous angular extension of
\(W=\omega_\star^2/(4\Omega^2)\) for elliptic orbits.  From left to
right, \(e=0.2,0.5,0.8\).  The horizontal dashed line is the orbital
equator, and the remaining white curves are the Mollweide
graticules. Each panel uses an independent color scale to display
the angular pattern; colors should therefore not be compared as common
absolute magnitudes across panels.}
\label{We}
\end{figure}

Figure \ref{We} displays the continuous angular extension of \(W\)
for three representative eccentricities.  We use normalized
Mollweide coordinates. \be
X=\frac{\phi}{\pi}\cos v,
\qquad
Y=\sin v,
\qquad
2v+\sin2v=\pi\cos\theta,
\label{normalized-Mollweide}
\ee
with \(\phi\) increasing from left to right.  The center of the map
is \((\theta,\phi)=(\pi/2,0)\), the lateral boundaries are the common
meridian \(\phi=\pm\pi\), and \(X=\pm1/2\) on the equator corresponds
to \(\phi=\pm\pi/2\).  The reversal of the color pattern between
\(e=0.2\) and \(e=0.5,0.8\) follows from the sign change in
\eqref{W-u-derivative}.

\paragraph{ Hyperbolic orbit.} For a hyperbolic orbit, the orbital parameters satisfy
\begin{equation}
e>1,
\qquad
\epsilon=a(e^2-1),
\qquad
-A<\psi<A,
\qquad
A=\arccos\left(-\frac1e\right),
\end{equation}
where $a>0$ denotes the magnitude of the hyperbolic semi-major axis. In terms of the incoming speed $v_\infty$ and impact parameter $b$, these quantities are given by
\begin{equation}
a=\frac{G\overline M}{v_\infty^2},
\qquad
e=\sqrt{1+\frac{b^2v_\infty^4}{G^2\overline M^2}}. 
\end{equation} 
These are the standard Newtonian scattering relations and applications
of the quadrupole formalism to unbound encounters may be found, for
example, in \cite{1977ApJ...216..610T}.

\begin{figure}
    \centering 
    \includegraphics[width=6in]{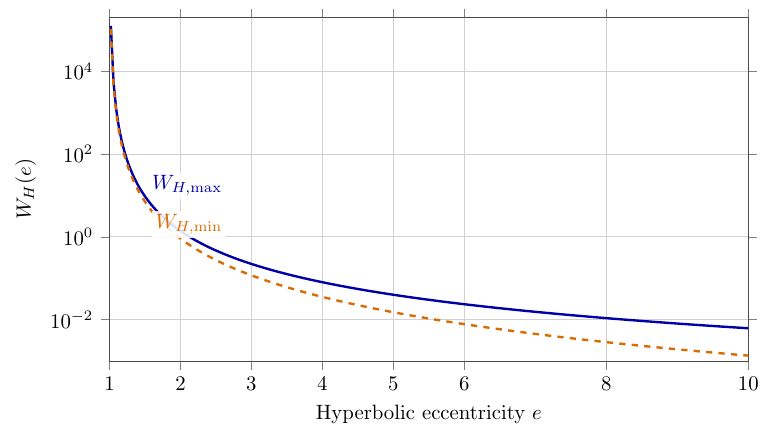}
    \caption{\centering Extrema of the characteristic frequency ratio $W_H$ for hyperbolic orbits.}\label{mon}
\end{figure}

The four helicity densities are time-dependent, and no finite orbital period exists in the hyperbolic case. 
The accumulated helicity over the entire scattering process may be defined as
\begin{equation}
\mathfrak A(r,\mathbf n)
\equiv
\int_{-\infty}^{+\infty}d\tau\,
\mathcal H(\tau,r,\mathbf n).
\end{equation}
With this definition, one obtains
\begin{align}
    \mathfrak A_{NY}^T(r,\mathbf n)&=
\frac{G^4\overline M^2\mu^2}
{2r^2\epsilon^2}
n_z\,\kappa_{\mathrm{NY}}+\mathcal{O}(r^{-3}),\\ 
    \mathfrak A_{CS}^T(r,\mathbf n)&=
\frac{G^5\overline M^3\mu^2}
{60r^2\epsilon^5e^2}
n_z
\left[
A\mathcal P_A+\chi\mathcal P_\chi
\right]+\mathcal{O}(r^{-3}),
\end{align}
where
\begin{align}
\kappa_{\mathrm{NY}}
={}&\frac{2}{3e^2}\Big[
A\big\{
15e^4\Delta
+6e^2(7e^2+8)(1+n_z^2)
\big\}
\nonumber\\
&\quad
+\chi\big\{
(12e^4+e^2+2)\Delta
+6e^2(2e^2+13)(1+n_z^2)
\big\}
\Big],
\end{align}
and the polynomials $\mathcal P_A$ and $\mathcal P_\chi$ are defined as
\begin{align} 
\mathcal P_A={}&
15e^4(51e^4+640e^2+464)\Delta
+
60e^2(69e^6+1308e^4+1832e^2+256)(1+n_z^2),
\\
\mathcal P_\chi={}&
(5891e^6+10882e^4+568e^2-16)\Delta
+
20e^2(1969e^4+6562e^2+1864)(1+n_z^2).
\end{align}
The parameter $\chi$ is defined as $\chi=\sqrt{e^2-1}$.
For the hyperbolic orbit, we may also calculate characteristic frequency 
$
\omega^2_{*,hyp}(\mathbf n)
$ and show that it is always well defined for any $e>1$. Define \[W_H(\theta,\phi)
\equiv
\frac{\omega_{\star,\mathrm{hyp}}^2(\theta,\phi)}
{4\Omega_H^2},
\qquad
\Omega_H^2=\frac{G\overline M}{a^3},\] then the maxima of $W_H$ is located at $(\theta,\phi)=(\frac{\pi}{2},\pm\frac{\pi}{2})$ for each fixed $e>1$. The minima of $W_H$ is located at $(\theta,\phi)=(\frac{\pi}{2},0)$ and $(\frac{\pi}{2},\pi)$. We denote the maxima of $W_H$ as $W_{H,max}(e)$ and the minima of $W_H$ as $W_{H,min}(e)$. They are monotonically decreasing functions of $e$, as shown in Figure \ref{mon}.
$W_{H,max}(e)$ and $W_{H,min}(e)$ obey the following power law when $e\to\infty $
\bs\begin{align}
W_{H,\max}(e)
&=
\frac{25}{64}\frac1{e^2}
+\mathcal O(e^{-3}),\\ 
W_{H,\min}(e)
&=
\frac{327\pi}{1024}\frac1{e^3}
+\mathcal O(e^{-4}).
\end{align}\es This interprets the discrepancy of the maxima and minima as $e\to\infty$. The large \(e\) limits obtained above require some explanation. In
terms of the eccentricity \(e\) and the pericenter distance \(r_{\rm p}\),
the pericenter speed is
\[
v_{\rm p}^{2}
=
\frac{G\overline M(1+e)}{r_{\rm p}}.
\]
The weak field and slow motion conditions are
\[
\frac{G\overline M}{r_{\rm p}}\ll1,
\qquad
v_p^2=\frac{G\overline M(1+e)}{r_{\rm p}}\ll1.
\]
For a hyperbolic orbit, \(e>1\), the second condition is stronger and
implies the first one.
At fixed \(r_{\rm p}\), the Newtonian result therefore applies only in
the finite interval
\be 
1<e\ll\frac{r_{\rm p}}{G\overline M}-1.
\ee
This interval is nonempty and can be parametrically large when
\(r_{\rm p}\gg G\overline M\), so that large but finite eccentricities
can consistently be considered. However, the strict limit
\(e\to\infty\) at fixed \(r_{\rm p}\) eventually gives
\(v_{\rm p}\sim1\) and leaves the domain of validity.
A controlled \(e\to\infty\) limit is possible only as a correlated
scaling limit in which \(r_{\rm p}\) also increases while
\[
\frac{G\overline M(1+e)}{r_{\rm p}}\ll1
\]
remains fixed and small. For example, one may set
\be 
r_{\rm p}
=
\frac{G\overline M}{\nu^2}(1+e),
\qquad
\nu\ll1,
\ee
which gives \(v_{\rm p}^2=\nu^2\) and
\(G\overline M/r_{\rm p}\to0\) as \(e\to\infty\). Thus, the formal
large \(e\) asymptotic expressions should be understood either within
the finite Newtonian window at fixed \(r_{\rm p}\), or in this
correlated weak field scaling limit.

We do not integrate the spin-1 large \(r\) formulas over the complete
unbound orbit.  At fixed observation radius, \(D(t)\to\infty\) as
\(|t|\to\infty\), so the condition \(r\gg D(t)\) eventually fails.
Any divergence obtained by integrating
spin helicity densities term by term is therefore an artifact of
using a nonuniform spatial multipole expansion outside its domain.
An exact treatment would require the unexpanded spin-1 field or a
matched asymptotic construction.

\paragraph{Parabolic orbit.} The parabolic orbit is obtained by taking the limits \(e\to1\), \(A\to\pi\), and \(\chi\to0\) in the hyperbolic results. In this limit, one finds
\begin{align}
\mathfrak A_{\mathrm{NY}}^T
&=
\frac{5\pi G^4\overline M^2\mu^2}
{r^2\epsilon_{\mathrm P}^2}
n_z
\left[
\Delta+6(1+n_z^2)
\right],\\
\mathfrak A_{\mathrm{CS}}^T
&=
\frac{1155\pi G^5\overline M^3\mu^2}
{4r^2\epsilon_{\mathrm P}^5}
n_z
\left[
\Delta+12(1+n_z^2)
\right].
\end{align}
Here \(\epsilon_{\mathrm P}\) denotes the semi-latus rectum of the parabolic orbit. It should be noted that, for a parabolic orbit, one has \(a\to\infty\) and \(e\to 1\); to obtain a finite parabolic limit, the semi-latus rectum \(\epsilon_{\mathrm P}=a(e^2-1)\) must be kept finite. This quantity is given by 
\begin{equation}
\epsilon_{\mathrm P}
=\frac{L^2}{G\overline M\mu^2},
\end{equation}
where \(L\) is the orbital angular momentum of the parabolic orbit.

\subsection{Boosted Kerr}
In this subsection, the word boosted Kerr means  a finite
 Lorentz transformation is applied to the Kerr geometry.  This is
not an accelerating Kerr black hole and it does not generate gravitational
radiation.  It is the same stationary spacetime described relative to a
different asymptotic inertial frame and a different family of equal-time
hypersurfaces.
In the literature, there are two different treatment of the boosted Kerr solution: 
the construction of Burinskii and
Magli \cite{Burinskii:1999ew}, and the Lorentz-boosted Cartesian Kerr-Schild solution used by Huq,
Choptuik, and Matzner \cite{Huq:2000qx}.  The former gives an analytic description for an
arbitrary magnitude and orientation of the boost, with particular emphasis
on the principal null congruence, the singular set, and ultrarelativistic
limits.  The latter gives explicit real Cartesian metric and $3+1$ data
suited to numerical apparent-horizon calculations.  For every finite
 boost the two constructions describe the same Kerr spacetime in
different variables and with different normalizations of the Kerr-Schild
null one-form.  We will use the latter to derive
the gauge invariant vector mode $\Xi_i$ in the weak field, large distance,
linear spin regime.  The result is controlled by the ADM momentum and the
intrinsic Kerr angular momentum. 

A Kerr-Schild metric is written as \cite{Debney:1969zz}
\begin{equation}
  g_{\mu\nu}=\eta_{\mu\nu}+2H \ell_\mu \ell_\nu,
  \qquad
  \eta^{\mu\nu}\ell_\mu \ell_\nu=0.
  \label{eq:KSgeneral}
\end{equation}
The decomposition is invariant under
\begin{equation}
  \ell_\mu\longrightarrow w \ell_\mu,
  \qquad
  H\longrightarrow w^{-2}H.
  \label{eq:KSrescaling}
\end{equation} For a Kerr spin along the positive $z$ axis,
\begin{align}
  H&=\frac{GMr^3}{r^4+a^2z^2},
  \label{eq:HuqH}\\
  \ell_\mu&=
  \left(
  1,
  \frac{rx+ay}{r^2+a^2},
  \frac{ry-ax}{r^2+a^2},
  \frac{z}{r}
  \right),
  \label{eq:Huql}
\end{align}
where $r$ is defined implicitly by
\begin{equation}
  \frac{x^2+y^2}{r^2+a^2}+\frac{z^2}{r^2}=1.
  \label{eq:oblate}
\end{equation}
Here $M$ is the rest mass and $J=Ma$ is the rest-frame Kerr angular
momentum.  If barred coordinates are comoving with the hole, the boosted
fields are obtained from
\begin{equation}
  H(x)=\bar H(\Lambda^{-1}x),
  \qquad
  \ell_\mu(x)=
  \frac{\partial \bar x^{\bar\alpha}}{\partial x^\mu}
  \bar\ell_{\bar\alpha}(\Lambda^{-1}x).
  \label{eq:directboost}
\end{equation}

The main result of \cite{Huq:2000qx}
is the direct $3+1$ decomposition 
\be 
ds^2=-\alpha^2dt^2+\gamma_{ij}(dx^i+\beta^i dt)(dx^j+\beta^j dt)
\ee with
\begin{align}
  \alpha&=\left(1+2H\ell_t^2\right)^{-1/2},\\
  \beta_i&=2H\ell_t\ell_i,\\
  \gamma_{ij}&=\delta_{ij}+2H\ell_i\ell_j.
  \label{eq:3plus1}
\end{align}
These fields were used to test an apparent-horizon finder for unboosted and
boosted Schwarzschild and Kerr holes.  The horizon is coordinate distorted
and Lorentz contracted, while its geometrical area remains invariant. In the rest frame, the weak field limit gives
\begin{align}
  H&=\frac{GM}{r}+\mathcal{O}(a^2/r^3),\\
  \ell_0&=1,\\
  \ell_i&=n_i-\frac{(\bm a\times\bm n)_i}{r}
  +\mathcal{O}(a^2/r^2).
  \label{eq:restweak}
\end{align}
The absence of a term linear in $a$ in $H$ follows from
Eq.~\eqref{eq:HuqH}. Choose the laboratory frame so that the hole moves with velocity $\bm v$.
To first order in $v$,
\begin{equation}
  \bar t=t-\bm v\cdot\bm x,
  \qquad
  \bar{\bm x}=\bm x-\bm v t.
\end{equation}
With
\begin{equation}
  \bm R=\bm x-\bm v t,
  \qquad
  r=|\bm R|,
  \qquad
  \bm n=\bm R/r,
\end{equation}
the boosted null one-form is
\begin{align}
  \ell_0&=1-\bm v\cdot\bm n+\mathcal{O}(v^2,av),\\
  \ell_i&=n_i-v_i-\frac{(\bm a\times\bm n)_i}{r}
  +\mathcal{O}(v^2,a^2,av).
  \label{eq:boostedlweak}
\end{align}
Substitution into $h_{\mu\nu}=2H\ell_\mu\ell_\nu$ gives
\begin{equation}
  h_{0i}
  =\frac{2GM}{r}n_i
  -\frac{2GM}{r}
   \left[v_i+n_i(\bm v\cdot\bm n)\right]
  -\frac{2G}{r^2}(\bm J\times\bm n)_i
  +\cdots .
  \label{eq:h0iboosted}
\end{equation} The first term is longitudinal because
\begin{equation}
  \frac{n_i}{r}=\partial_i\ln r.
\end{equation}
The other two displayed fields are transverse for $r\ne0$
\begin{align}
  \partial_i\left\{
  \frac{v_i+n_i(\bm v\cdot\bm n)}{r}
  \right\}=0,\qquad 
  \partial_i\left\{
  \frac{(\bm J\times\bm n)_i}{r^2}
  \right\}=0.
  \label{eq:transversechecks}
\end{align} Thus the transverse spin-1 mode is 
\begin{equation}
  \Xi_i
  =-\frac{2G}{r}
  \left[
  P_i^{\text{ADM}}+n_i(\bm P^{\text{ADM}}\cdot\bm n)
  \right]
  -\frac{2G}{r^2}\epsilon_{ijk}J_jn_k
  +\mathcal{O}(r^{-3},v^2J,a^2),
  \label{eq:Xiphysical}
\end{equation}
where $\bm P^{\text{ADM}}=M\bm v+\mathcal{O}(v^3)$ at this stage of the expansion. Comparing with \eqref{Xiexpansion}, we should identify 
\be 
 \mathbf P=-\bm P^{\text{ADM}},
  \qquad
  \mathbf  S=-\bm J
\ee and $D_{ij}=0$. The minus sign comes from the choice of convention. Thus, the spin-1 gravitomagnetic helicity density and the spin-1 gravito-current helicity density are
\bs\label{leadingp2}\begin{align}
\mathcal H_{\mathrm{NY}}^V
&=
-\frac{G^2}{r^4}
\Big[
\mathbf P\cdot\mathbf S
+
3(\mathbf P\cdot\mathbf n)(\mathbf S\cdot\mathbf n)
\Big]
+\mathcal O(r^{-5}),\\ 
\mathcal H_{\mathrm{CS}}^V
&=
\frac{4G^2}{r^6}
\Big[
\mathbf P\cdot\mathbf S
+
3(\mathbf P\cdot\mathbf n)(\mathbf S\cdot\mathbf n)
\Big]
+\mathcal O(r^{-7}).
\end{align}\es These are local, frame-dependent equal-time densities. Their nonzero value in a boosted frame
does not signal gravitational radiation. A global Lorentz transformation of Kerr has vanishing
Bondi news, so the radiative $1/r$ tensor sector vanishes 
\be 
h_{ij}^{\text{TT}}=0,\quad \mathcal H_{NY}^T=\mathcal H_{CS}^T=0
\ee  in the wave zone. Although the exact boosted Kerr geometry exists for any finite boost, the formulas above are valid under the conditions
\be 
GM/r\ll1,\qquad |a|/r\ll1,\qquad v\ll1 .
\ee 
These formulas retain the terms at first order in the weak field and the leading parts that are separately linear in \(v\) and \(a\). The displayed \(vJ\) helicity is their quadratic interference within this truncation. No global helicity integral through the black hole interior is evaluated. An integral on an exterior slice would require the corresponding inner boundary term.
\subsection{Gyratonic pp-wave}
Gyratonic pp-waves are Brinkmann geometries \cite{Brinkmann1925} generated by null matter carrying
intrinsic angular momentum \cite{Bonnor1970,
FrolovFursaev2005,FrolovIsraelZelnikov2005}.  They provide idealized
gravitational fields of spinning radiation beam pulses and ultrarelativistic
spinning sources.  In this subsection we will review the exact geometry and
then retain its first-order weak field part, which is the order required for
the quadratic helicity densities studied in this work.

Let $A,B=1,2$ label the flat transverse coordinates and define
\begin{equation}
 u=\frac{t-z}{\sqrt2},
 \qquad
 v=\frac{t+z}{\sqrt2}.
 \label{eq:gyraton-null-coordinates}
\end{equation}
The Brinkmann form of a four-dimensional gyratonic pp-wave is
\begin{equation}
 d s^2
 =-2d ud v+\delta_{AB}d x^Ad x^B
 +2m_A(u,\bm x_\perp)d ud x^A
 +H(u,\bm x_\perp)d u^2.
 \label{eq:gyratonmetric}
\end{equation}
The vector $\partial_v$ is a null Killing vector.  The metric
\eqref{eq:gyratonmetric} is invariant under the Brinkmann gauge transformation
\begin{equation}
 v\longrightarrow v+\xi(u,\bm x_\perp),\qquad
 m_A\longrightarrow m_A-\partial_A\xi,\qquad
 H\longrightarrow H-2\partial_u\xi .
 \label{eq:Brinkmanngauge}
\end{equation}
Thus $m_A$ alone is not a gauge invariant observable.  Its transverse field
strength
\begin{equation}
 f_{AB}=\partial_Am_B-\partial_Bm_A
 \label{eq:fAB}
\end{equation}
is gauge invariant.  On a non-contractible exterior region, the circulation
$\oint m_Ad x^A$ can also carry gauge invariant global information
\cite{FrolovIsraelZelnikov2005,PodolskySteinbauerSvarc2014}.

For a gyratonic source whose only nonzero components are
\begin{equation}
 T_{uu}=T_{uu}(u,\bm x_\perp),
 \qquad
 T_{uA}=T_{uA}(u,\bm x_\perp),
 \label{eq:gyratonsource}
\end{equation}
the relevant exact Ricci components are
\begin{align}
 R_{uA}
 &=\frac12\partial^Bf_{AB},
 \label{eq:RuA}
 \\
 R_{uu}
 &=-\frac12\Delta_\perp H
 +\partial_u\partial^Am_A
 +\frac14f_{AB}f^{AB},
 \label{eq:Ruu}
\end{align}
where $\Delta_\perp=\partial^A\partial_A$.  In the transverse Brinkmann
gauge $\partial^Am_A=0$, the first order Einstein equations reduce to
\begin{equation}
 \Delta_\perp H=-16\pi G T_{uu},
 \qquad
 \Delta_\perp m_A=-16\pi G T_{uA}.
 \label{eq:gyraton-linear-equations}
\end{equation}
The term $f_{AB}f^{AB}$ contributes to the exact metric at second order in the
source angular momentum.  It can be omitted consistently here because the
helicity densities are already quadratic in the first order metric. A
second order correction to $H$ would first modify them at cubic order. Gyratonic pp-waves belong to the VSI(vanishing scalar invariants) spacetime. Namely, all scalar polynomial curvature
invariants vanish \cite{PravdaPravdovaColeyMilson2002,
ColeyMilsonPravdaPravdova2004}.  A smooth matter-filled gyratonic region is
generically of Petrov type III, whereas a smooth exterior vacuum region is of
type N \cite{FrolovIsraelZelnikov2005,PodolskySteinbauerSvarc2014}.  This does
not imply that the foliation dependent, equal time quadratic helicity
densities defined below must vanish.  In particular, these densities are not
four-dimensional scalar polynomial curvature invariants.

In polar coordinates $x=\rho\cos\phi$ and
$y=\rho\sin\phi$, 
\begin{equation}
 d s^2=d\rho^2+\rho^2d\phi^2-2d ud v
 +2m_\phi(u,\rho,\phi)d ud\phi+H(u,\rho,\phi)d u^2.
 \label{eq:gyraton-polar}
\end{equation} We model the matter as a smooth gyratonic null fluid whose stress tensor
reproduces, at leading paraxial and geometric optics order, the cycle averaged
energy and spin angular momentum fluxes of a finite width, finite duration,
circularly polarized Gaussian electromagnetic beam pulse
\cite{FrolovFursaev2005}.  This is a controlled phenomenological model of the
leading Maxwell stress tensor, rather than a globally exact, non-diffracting
Maxwell solution. Let $w$ denote the effective transverse intensity radius and set
\begin{equation}
 \chi\equiv\frac{\rho^2}{w^2}.
\label{eq:gyraton-chi}
\end{equation}
We choose
\begin{align}
 T_{uu}
=\frac{\sqrt2\,\mathcal M(u)}{\pi w^2}e^{-\chi},\qquad 
 T_{uA}
 =\frac{\mathcal J(u)}{\pi w^4}
\epsilon_{AB}x_Be^{-\chi},
 \qquad \epsilon_{12}=+1.
 \label{eq:jGaussian}
\end{align} This source is conserved at the order under consideration because
$\partial^AT_{uA}=0$.

For a circularly polarized beam, let $\lambda_\gamma=\pm1$ denote the photon
helicity.  We choose the energy and spin current envelopes to satisfy
\begin{equation}
 \mathcal J(u)=\lambda_\gamma\frac{\mathcal M(u)}{\omega_0}.
 \label{eq:circularrelation}
\end{equation}
This is consistent with the integrated relation
$J=\lambda_\gamma E/\omega_0$ derived for a circularly polarized
radiation beam pulse of
\cite{FrolovFursaev2005}.  
A smooth pulse envelope is
\begin{equation}
 \mathcal M(u)=\frac{E}{\sqrt\pi\,\tau}e^{-u^2/\tau^2},
 \qquad
 \mathcal J(u)=\lambda_\gamma
 \frac{E}{\sqrt\pi\,\tau\omega_0}e^{-u^2/\tau^2},
 \label{eq:pulseprofiles}
\end{equation}
so that $\int d u\,\mathcal M=E$ is the energy of the beam.  The geometric optics and paraxial
conditions are
\begin{equation}
 \omega_0w\gg1,
 \qquad
 \omega_0\tau\gg1.
 \label{eq:paraxialconditions}
\end{equation}
The rigid transverse profile is meaningful only near the beam waist and over
propagation distances small compared with the Rayleigh scale
$z_R=\mathcal O(\omega_0w^2)$.

Define
\begin{equation}
 F(\chi)=\gamma_{\rm E}+\ln\chi-\operatorname{Ei}(-\chi),
 \qquad
 A(\chi)=1-e^{-\chi}.
 \label{eq:FAdef}
\end{equation} Here $\operatorname{Ei}(x)$ is the exponential integral
\be 
\operatorname{Ei}(x)=\text{PV}\int_{-\infty}^x \frac{e^t}{t}dt
\ee and $\gamma_E$ is the Euler constant.
The first order solution of Eq.~\eqref{eq:gyraton-linear-equations} that is
regular on the beam axis is
\begin{equation}
 H(u,\rho)=-4\sqrt2G\mathcal M(u)F(\chi),
 \qquad
 a=-4G\mathcal J(u)A(\chi) d\phi.
 \label{eq:metricpotentials}
\end{equation}
Thus
\be 
  d s^2={}-2 d u d v+ d\rho^2+\rho^2 d\phi^2
 -8G\mathcal J(u)A(\chi) d u d\phi
 -4\sqrt2G\mathcal M(u)F(\chi) d u^2 .
 \label{eq:explicitmetric}
\ee 
The solution is smooth at $\rho=0$.  At large transverse radius,
\begin{equation}
 H\sim-8\sqrt2G\mathcal M(u)\ln(\rho/w),
 \qquad
 a\longrightarrow-4G\mathcal J(u) d\phi,
 \label{eq:asymptotics}
\end{equation} and the angular momentum profile is encoded in the circulation,
\begin{equation}
 \mathcal J(u)=-\frac1{8\pi G}\oint a.
 \label{eq:gyraton-circulation}
\end{equation}

The logarithmic behavior in Eq.~\eqref{eq:asymptotics} shows that this
idealized pp-wave is not asymptotically flat. Consequently, the inverse
Laplacians entering the SVT projectors require an infrared boundary
condition or a specified global completion of the beam. In the absence
of such data, the formulas below are local, slowly varying representatives
of the projected fields rather than a unique global SVT decomposition.

On a constant time hypersurface, the nonzero first-order metric components are
\begin{equation}
 h_{00}=\frac H2,
 \quad
 h_{0A}=\frac{a_A}{\sqrt2},
 \quad
 h_{0z}=-\frac H2,
 \quad
 h_{Az}=-\frac{a_A}{\sqrt2},
 \quad
 h_{zz}=\frac H2.
 \label{eq:gyraton-metric-components}
\end{equation} The gauge invariant modes are obtained via the formula \eqref{inverse}.

Closed local coordinate expressions are especially simple in the slowly
varying  regime
\begin{equation}
\frac{w}{\tau}\ll1.
 \label{eq:gyraton-long-envelope}
\end{equation} The leading transverse contributions to the gauge invariant vector and
tensor SVT variables are
\begin{equation}
 \begin{aligned}
 \Xi_\rho&=\mathcal{O}(w/\tau),
 \\
 \Xi_\phi
&=-2\sqrt2G\mathcal J(u)A(\chi)+\mathcal{O}(w^2/\tau^2),
 \\
 \Xi_z
&=2\sqrt2G\mathcal M(u)F(\chi)+\mathcal{O}(w^2/\tau^2),
 \end{aligned}
 \label{eq:gyraton-Xi-local}
\end{equation}
and
\begin{equation}
 \begin{aligned}
 h_{\rho\rho}^{\rm TT}
 &=\frac{G\mathcal M(u)}{\sqrt2}
\left[F(\chi)-1+\frac{A(\chi)}{\chi}\right]+\mathcal{O}(w^2/\tau^2),
 \\
 h_{\phi\phi}^{\rm TT}
 &=\frac{G\mathcal M(u)\rho^2}{\sqrt2}
\left[F(\chi)+1-\frac{A(\chi)}{\chi}\right]+\mathcal{O}(w^2/\tau^2),
 \\
 h_{zz}^{\rm TT}
&=-\sqrt2G\mathcal M(u)F(\chi)+\mathcal{O}(w^2/\tau^2),
 \\
 h_{\phi z}^{\rm TT}
 &=h_{z\phi}^{\rm TT}
=2\sqrt2G\mathcal J(u)A(\chi)+\mathcal{O}(w^2/\tau^2),\\
h_{\rho z}^{\text{TT}}&=\mathcal{O}(w/\tau),\\ 
h_{\rho\phi}^{\text{TT}}&=h_{\phi\rho}^{\text{TT}}=\mathcal{O}(w/\tau).
 \end{aligned}
 \label{eq:gyraton-hTT-local}
\end{equation}
The omitted terms contain longitudinal derivatives and are
suppressed by powers of \(w/\tau\). The local expressions displayed
below therefore represent the leading slowly varying approximation to
the gauge invariant SVT variables, rather than the exact global SVT
projection of the ideal pp-wave metric.

With the spin-1 and spin-2 modes, one obtains
\bs
 \begin{align}
 \mathcal H_{\rm NY}^{V}
 &=\frac{4\lambda_\gamma G^2E^2}{\pi\omega_0\tau^2}
 e^{-2u^2/\tau^2}
 \left[
\frac{F(\chi)e^{-\chi}}{w^2}
-\frac{A(\chi)^2}{\rho^2}
 \right]+\mathcal{O}(w^2/\tau^2),
 \\
 \mathcal H_{\rm CS}^{V}
 &=\frac{32\lambda_\gamma G^2E^2}
 {\pi\omega_0\tau^2w^4}
 e^{-2u^2/\tau^2-\chi}+\mathcal{O}(w^2/\tau^2),
 \\
 \mathcal H_{\rm NY}^{T}
 &=\frac{\lambda_\gamma G^2E^2}
 {2\pi\omega_0\tau^2w^2}
e^{-2u^2/\tau^2}\mathcal G_T(\chi)+\mathcal{O}(w^2/\tau^2),
 \\
 \mathcal H_{\rm CS}^{T}
 &=\frac{4\lambda_\gamma G^2E^2u^2}
 {\pi\omega_0\tau^6w^2}
e^{-2u^2/\tau^2}\mathcal G_T(\chi)+\mathcal{O}(w^4/\tau^4),
 \end{align}
 \label{eq:gyraton-four-local-densities}
\es
where
\begin{equation}
\mathcal G_T(\chi)
=[-3F(\chi)-1+q(\chi)][2e^{-\chi}-q(\chi)]
+q(\chi)[8A(\chi)-3F(\chi)+1-q(\chi)]
 \label{eq:gyraton-FT}
\end{equation} with \begin{equation}
q(\chi)=\frac{A(\chi)}{\chi}.
\end{equation}
Thus, in the slow envelope expansion
\(w/\tau\ll1\), the leading contributions scale as
\begin{equation}
\left(
\mathcal H_{\rm NY}^{V},
\mathcal H_{\rm CS}^{V},
\mathcal H_{\rm NY}^{T},
\mathcal H_{\rm CS}^{T}
\right)
=
\left(
O(1),O(1),O(1),O(w^2/\tau^2)
\right).
\end{equation} All odd powers of \(w/\tau\) vanish for the
axisymmetric rigid-profile configuration considered here. Therefore,
the first corrections to the first three densities are of order
\(O(w^2/\tau^2)\), whereas the first correction to
\(\mathcal H_{\rm CS}^{T}\) is of order
\(O(w^4/\tau^4)\).
All four densities reverse sign when the photon helicity
$\lambda_\gamma$ is reversed.  Since a Gaussian has exponentially small but
nonzero support at every finite $\rho$, the spin-1 gravito-current density is
not exactly zero outside a sharp beam radius. Instead, it is exponentially
localized in the beam core. 

For the  spin-2 density, a monochromatic TT mode obeys
\begin{equation}
 \mathcal H_{\rm CS}^{T}=4\omega^2\mathcal H_{\rm NY}^{T}.
\end{equation}
Thus the pointwise characteristic frequency depends on time
\begin{equation}
 \omega_{\star,p}^2
 \equiv\frac{\mathcal H_{\rm CS}^{T}}
 {4\mathcal H_{\rm NY}^{T}}=\frac{2u^2}{\tau^4}.
 \label{eq:gyraton-omegaT-def}
\end{equation}
Its retarded time
average is
\begin{align}
{\omega}_{\star}^2
 &=
 \frac{\displaystyle\int_{-\infty}^{+\infty} d u\,
 \mathcal H_{\rm CS}^{T}}
 {\displaystyle4\int_{-\infty}^{+\infty} d u\,
 \mathcal H_{\rm NY}^{T}}
 =\frac1{2\tau^2}.
 \label{eq:gyraton-omegaT-averaged}
\end{align}
Hence the accumulated spin-2 ratio measures the inverse duration of the
gravitational waveform, \be \omega_{\star}=\frac{1}{\sqrt2\tau},\ee  rather than
the optical carrier frequency $\omega_0$.  The carrier frequency controls the
amplitude through $\mathcal J=\lambda_\gamma\mathcal M/\omega_0$ and cancels
from both characteristic ratios.

The derivative hierarchy between the two helicities in each spin sector can
be used to define a characteristic scale.   For a transverse vector helicity eigenmode satisfying
$\bm\nabla\times\bm\Xi_\sigma=\sigma k\bm\Xi_\sigma$, the definition of spin-1 helicity densities gives
\begin{equation}
 \mathcal H_{\rm CS}^{V}=-2k^2\mathcal H_{\rm NY}^{V}.
\end{equation}
The corresponding local characteristic reduced wave length is therefore
\begin{equation}
 \lambda_*^2=
-\frac{2\mathcal H_{\rm NY}^{V}}
 {\mathcal H_{\rm CS}^{V}}.
 \label{eq:gyraton-kappaV-def}
\end{equation}
For the Gaussian gyraton profile this becomes
\begin{equation}
\lambda^2_*
 =
\frac
 {w^2\left[A(\chi)^2/\chi-F(\chi)e^{-\chi}\right]}{4e^{-\chi}}.
 \label{eq:gyraton-kappaV-local}
\end{equation}
This characteristic wave length  probes the transverse structure and is of order $w$ away
from zeros of $\mathcal H_{\rm NY}^{V}$.

Note that the two-dimensional transverse logarithmic
Green function implies
\begin{equation}
 \mathcal H_{\rm NY}^{V}
 \sim-\frac{4G^2\mathcal M\mathcal J}{\rho^2},
 \qquad
 \mathcal H_{\rm NY}^{T}
 \sim\frac{5G^2\mathcal M\mathcal J}{\rho^2},
 \qquad \rho\gg w.
\end{equation}
Consequently, the corresponding transverse integrals are infrared sensitive
and require an outer matching radius since the ideal four-dimensional pp-wave is
not asymptotically flat in the transverse direction.

%\subsection{Kerr C-metric}
%\cite{Hong:2004dm,Griffiths:2005se}

\subsection{Gravitational Hopfion}
The term gravitational Hopfion does not identify a unique metric
field.  Several related, but inequivalent, constructions occur in the
literature. 
Penrose transform constructions generate massless spin-2 fields whose
electric and magnetic Weyl tensors possess linked tendex and vortex
structures.  The type N solution and its relation to linked gravitational
radiation were studied in \cite{Swearngin:2013sks,Thompson_2014}. Moreover,
type N, D, and III Hopfions were subsequently classified in
\cite{Thompson:2014owa}. In \cite{Smolka:2018rup}, it is shown that 
a reduced complex scalar can be used to reconstruct the gauge invariant
spin-2 data and generalized gravitational Hopfions. 
The 
complex anti-self-dual Kerr-Schild solutions obtained from
Plebanski's second-heavenly  construction \cite{Plebanski:1975wn} include the Sparling-Tod metric \cite{Sparling:1981nk,Tod:1982mmp}.
Its single copy is the electromagnetic Hopfion, and its curvature realizes a
Weyl double copy construction \cite{
Sabharwal:2019ngs}. These gravitational Hopfions are direct generalizations of the Hopfions in electromagnetism  \cite{Arrayas:2017sfq} and various other branches of physics
\cite{Gladikowski:1996mb,2003JGP....46..125U}.

We set the characteristic length of the Hopfion to unity.  Introduce
\begin{equation}
D_+=r^2-(t+i)^2=r^2+d^2,
\qquad
d=1-i t,
\qquad 
\ell_i=\bigl(x+i y,-i(x+i y),z-i d\bigr)
\label{eq:ell}
\end{equation}
and the complex
spatial seed \footnote{More explicitly, Eq.~(1) and the \(p=q=1\) specialization stated
immediately before Eq.~(5) in \cite{Sabharwal:2019ngs} provide the
Robinson principal spinor and the \(D_+^{-3}\) profile, respectively.
Equations~(10)--(11) of that reference
embed the corresponding null Maxwell spinor in a complex anti-self-dual
Kerr-Schild metric. After the imaginary-time translation \(t\to t+i\)
described below its Eq.~(15), the spatial part linear in the Kerr-Schild
parameter gives Eq.~\eqref{eq:q-seed}, up to the overall normalization
absorbed into \(\kappa\).}
\begin{equation}
q_{ij}(t,\mathbf x)=\frac{4\kappa\,\ell_i\ell_j}{D_+^3},
\qquad \kappa\in\mathbb R.
\label{eq:q-seed}
\end{equation}
It carries the $p=q=1$ principal spinor and denominator structure of the
complex anti-self-dual Kerr-Schild Hopfion.  The linearized regime further requires
\(|h_{ij}^{\rm TT}|\ll1\), which, under the unit length convention, implies taking the amplitude \(\kappa\) to be sufficiently small. The metric field is kept at \(O(\kappa)\), and the helicity densities at \(O(\kappa^2)\). We do not assert that \(\eta_{\mu\nu}+h_{\mu\nu}\) is the corresponding exact nonlinear Einstein metric.
The physical tensor field  is generated by the seed \(q_{ij}\)
\begin{equation}
h_{ij}^{\text{TT}}=\text{Re}\left[\hat{\Pi}_{ij,kl} q_{kl}\right].
\label{eq:real-TT-field}
\end{equation}
The complex seed \(q_{ij}=O(\kappa)\) is the spatial part of the
perturbation obtained by linearizing the complex type N Kerr-Schild
Hopfion solution about its flat background. It is not by itself a complete
four-dimensional metric perturbation and is not yet transverse and traceless. The field used in
the calculations is produced only after the spatial transverse and traceless projection and the
real-part operation in Eq.~\eqref{eq:real-TT-field}. 
Use the Fourier convention
\begin{equation}
\widetilde f(\mathbf k)=\int d^3\mathbf x\,
e^{-i\mathbf k\cdot\mathbf x}f(\mathbf{x}),
\qquad
f(\mathbf x)=\int\frac{d^3{\mathbf k}}{(2\pi)^3}
e^{i{\mathbf k}\cdot{\mathbf x}}\widetilde f({\mathbf k}).
\label{eq:Fourier-convention}
\end{equation}
For $\text{Re}\ d>0$,
\begin{equation}
\mathcal F\left[(r^2+d^2)^{-3}\right]
=\frac{\pi^2}{4d^3}(1+dk)e^{-dk},
\qquad k=|\mathbf k|.
\label{eq:denominator-transform}
\end{equation}
Multiplication by $x_i$ in Eq.~\eqref{eq:q-seed} becomes differentiation
with respect to $k_i$.  After those derivatives and the algebraic Fourier
projector are applied, all longitudinal and trace pieces cancel, giving
\begin{equation}
\widetilde{h}_{ij}^{\text{TT}}(t,\mathbf k)
=-\pi^{2}\kappa k e^{-k}\cos(kt)\,
w_i(\mathbf{n})w_j(\mathbf{n}),
\qquad
\mathbf n=\frac{\mathbf k}{k},
\label{eq:TT-Fourier-result}
\end{equation}
where
\begin{equation}
\bm w(\mathbf n)=
\left(
i n_y-n_xn_z,
-i n_x-n_yn_z,
1-n_z^2
\right).
\label{eq:w-vector}
\end{equation}
The polarization vector satisfies
\begin{equation}
n_iw_i=0,
\qquad
w_iw_i=0,
\qquad
i(\mathbf n\times\bm w)_i=w_i,
\qquad
w_i^*w_i=2(1-n_z^2).
\label{eq:w-properties}
\end{equation}
Consequently, Eq.~\eqref{eq:TT-Fourier-result} is transverse and traceless.
It also obeys the vacuum wave equation.
A useful exact inverse transform is
\begin{equation}
h_{ij}^{\text{TT}}
=\text{Re}\left[-\frac{\kappa}{2}\mathcal L_i\mathcal L_j u(d,r)\right],
\label{eq:coordinate-TT-field}
\end{equation}
where
\begin{equation}
u(d,r)
=-\frac{d}{r}\arctan\!\left(\frac{r}{d}\right)
-\frac12\log(d^2+r^2),
\label{eq:u-potential}
\end{equation}
and
\begin{equation}
\mathcal L_x=\partial_x\partial_z-\partial_d\partial_y,
\qquad
\mathcal L_y=\partial_y\partial_z+\partial_d\partial_x,
\qquad
\mathcal L_z=-(\partial_x^2+\partial_y^2).
\label{eq:L-operators}
\end{equation}
The branch in Eq.~\eqref{eq:u-potential} is the analytic continuation from
$\text{Re}\ d>0$.  Formally,
\begin{equation}
\widetilde u(d,k)=\frac{2\pi^2e^{-dk}}{k^3}.
\end{equation}
The potential $u$ by itself has an infrared ambiguity, but
$\mathcal L_i\mathcal L_j u$ is infrared finite.  Equation
\eqref{eq:u-potential} is the representative selected by the Fourier
boundary condition in Eq.~\eqref{eq:Fourier-convention}.  In particular,
replacing its first term by \(d\arctan(d/r)/r\) without the compensating
\(-\pi d/(2r)\) term changes the TT field.

At fixed $t$, Eq.~\eqref{eq:coordinate-TT-field} is smooth at the origin and
falls as $h_{ij}^{\text{TT}}=\mathcal{O}(r^{-4})$.  The surface terms generated by the
spatial integrations by parts used below therefore vanish at spatial
infinity.

For real $\kappa$, we have
\begin{equation}
\dot h_{ij}^{\text{TT}}(0,\mathbf x)=0,
\qquad
\mathcal H_{CS}^{T}(0,\mathbf x)=0.
\label{eq:CS-zero-t0}
\end{equation}
The spin-2 gravitomagnetic helicity density remains nonzero.  Let
\begin{equation}
c=\frac{z}{r}=\cos\theta,
\qquad
\alpha(r)=\arctan r.
\end{equation}
The exact result is
\begin{equation}
\mathcal H_{NY}^{T}(0,\mathbf x)
=\frac{\kappa^2}{16r^{10}(1+r^2)^4}\mathcal N(r,c),
\label{eq:NY-t0}
\end{equation}
where
\begin{align}
\mathcal N(r,c)={}&
-18(1+r^2)^4(45c^4-10c^2+13)\alpha^2
+12r(1+r^2)\alpha
\Bigl[
18c^4r^6+171c^4r^4+360c^4r^2+135c^4
\nonumber\\
&\hspace{27mm}
+12c^2r^6+42c^2r^4-80c^2r^2-30c^2
+18r^6+75r^4+104r^2+39
\Bigr]
\nonumber\\
&+2r^2
\Bigl[
108c^4r^6-801c^4r^4-1350c^4r^2-405c^4
-440c^2r^6-302c^2r^4+300c^2r^2+90c^2
\nonumber\\
&\hspace{20mm}
-148r^6-385r^4-390r^2-117
\Bigr],
\label{eq:N-polynomial}
\end{align}
with $\alpha=\alpha(r)$.  The apparent singularity in
Eq.~\eqref{eq:NY-t0} is removable.  Indeed, 
\bs
\begin{align}
\mathcal H_{NY}^{T}(0,\mathbf x)
&=\frac{96}{25}\kappa^2+\mathcal{O}(r^2),
\qquad r\longrightarrow0,
\label{eq:NY-origin}
\\
\mathcal H_{NY}^{T}(0,\mathbf x)
&=\frac{9\pi\kappa^2}{4r^9}
(3c^4+2c^2+3)+\mathcal{O}(r^{-10}),
\qquad r\longrightarrow\infty.
\label{eq:NY-infinity}
\end{align}
\es
Direct integration gives
\begin{equation}
H_{NY}^T=\int d^3\mathbf x\,\mathcal H_{NY}^{T}(0,\mathbf x)
=\frac{\pi^2\kappa^2}{2}.
\label{eq:NY-t0-integral-check}
\end{equation}

For general $t$, we find the non-vanishing spin-2 gravitomagnetic and gravito-current helicities
\begin{subequations}
\label{eq:integrated-results}
\begin{align}
H_{NY}^{T}(t)
=\frac{\pi^2\kappa^2}{4}
\left[1+\text{Re}(1-i t)^{-6}\right],
\label{eq:HNY-integrated}
\\
H_{CS}^{T}(t)
=\frac{21\pi^2\kappa^2}{2}
\left[1-\text{Re}(1-i t)^{-8}\right].
\label{eq:HCSkin-integrated}
\end{align}
\end{subequations}

We now verify the energy-helicity inequalities directly for the Hopfion defined by
Eqs.~\eqref{eq:TT-Fourier-result}--\eqref{eq:w-properties}. The
positive frequency amplitude belongs to a single circular polarization
sector. Introducing the normalized tensor
\[
e_{ij}^{(+)}(\mathbf n)
=
\frac{w_i(\mathbf n)w_j(\mathbf n)}
{\sqrt{2}(1-n_z^2)},
\qquad
e_{ij}^{(+)*}e_{ij}^{(+)}=2,
\]
the mode amplitudes satisfy
\[
|h_+(\mathbf k)|^2
=
\frac{\pi^4\kappa^2}{2}
k^2e^{-2k}(1-n_z^2)^2,
\qquad
h_-(\mathbf k)=0.
\]
The negative frequency complex conjugate required to make the metric
real does not constitute an independent opposite-helicity mode.

Using
\[
\int d\Omega_{\mathbf k}\,(1-n_z^2)^2
=
\frac{32\pi}{15},
\]
the angle-integrated energy spectrum becomes
\[
\mathscr E_{\rm Hopf}(k)
=
\frac{k^4}{8\pi G(2\pi)^3}
\int d\Omega_{\mathbf k}\,|h_+(\mathbf k)|^2
=
\frac{\pi\kappa^2}{60G}k^6e^{-2k}.
\]
Consequently,
\[
E_{\rm Hopf}
=
\int_0^\infty dk\,\mathscr E_{\rm Hopf}(k)
=
\frac{\pi\kappa^2}{60G}\frac{6!}{2^7}
=
\frac{3\pi\kappa^2}{32G}.
\]
Since
\[
\overline{\operatorname{Re}(1-it)^{-n}}=0,
\qquad n>1,
\]
their long-time averages of the helicities are
\[
\overline{H_{NY}^{T}}
=
\frac{\pi^2\kappa^2}{4},
\qquad
\overline{H_{CS}^{T}}
=
\frac{21\pi^2\kappa^2}{2}.
\]

The right-hand side of the Nieh-Yan inequality \eqref{NY-energy-inequality} evaluates to
\be 
8\pi G\int_0^\infty\frac{dk}{k}\,
\mathscr E_{\rm Hopf}(k)
=
8\pi G\frac{\pi\kappa^2}{60G}\frac{5!}{2^6}
=
\frac{\pi^2\kappa^2}{4}=\left|\overline{H_{NY}^{T}}\right|.
\ee 
Similarly, one can varity \eqref{CS-energy-inequality}.
The time average is essential. For example,
\[
H_{NY}^{T}(0)=\frac{\pi^2\kappa^2}{2},
\]
which is twice the corresponding spectral bound
\(\pi^2\kappa^2/4\). The additional contribution is the
positive-negative-frequency interference term and is removed by the
average. The inequalities therefore constrain the averaged or dephased
helicities, not the instantaneous quantities in
Eqs.~\eqref{eq:HNY-integrated} and \eqref{eq:HCSkin-integrated}.

The vector modes of the above gravitational Hopfion vanish. However, the absence of the spin-1 densities does not imply that the gravitational
Hopfion lacks nontrivial linked structure. In the Hopfion literature, the
linking refers to the tendex and vortex structures determined by the
electric and magnetic parts of the Weyl tensor, rather than to an
independent SVT vector mode. The nonvanishing parity odd spin-2 pairings
found here probe the chirality of the same radiative TT sector from which
those curvature tensors are constructed. Nevertheless,
\(H_{\rm NY}^{T}\) and \(H_{\rm CS}^{T}\) should not, without an additional
topological argument, be identified directly with a Hopf invariant or a
linking number.

\section{Conclusion and discussion}
In this work, we have investigated the helicity structures induced on a
spacelike hypersurface by two four-dimensional topological densities: the
Pontryagin and the Nieh-Yan density.  Our construction combines the
descent of these topological terms with the SVT decomposition
of linearized gravity.  It therefore provides a systematic way of separating
the helicity structures carried by the gauge invariant spin-1 and spin-2
sectors of the gravitational field. 

At quadratic order in the metric perturbations, the descendants of
the Pontryagin and Nieh-Yan terms can be decomposed into gauge invariant bulk
contributions and codimension-two surface terms.  After the surface
improvements have been separated, four helicity structures emerge
\bs 
\begin{align}
\mathcal H_{\rm NY}^{V}
&=
-\frac14\Xi_i\epsilon_{ijk}\partial_j\Xi_k,
&
\mathcal H_{\rm NY}^{T}
&=
\frac14 h_{ij}^{\rm TT}(Ch^{\rm TT})_{ij},
\\
\mathcal H_{\rm CS}^{V}
&=
-\frac12\epsilon_{ijk}
(\partial_j\Xi_k)\nabla^2\Xi_i,
&
\mathcal H_{\rm CS}^{T}
&=
\dot h_{ij}^{\rm TT}(C\dot h^{\rm TT})_{ij}.
\end{align}\es 
 We have referred to the one derivative
Nieh-Yan expressions as the spin-1 and spin-2 gravitomagnetic helicities,
and to the higher derivative Pontryagin expressions as the corresponding
gravito-current helicities.

The spin-2 gravito-current helicity density must be distinguished from the complete spin-2
Pontryagin descendant,
\begin{equation}
\mathcal H_{\rm CS,0}^{T}
=
\frac12
\left[
\dot h_{ij}^{\rm TT}(C\dot h^{\rm TT})_{ij}
+
h_{ij}^{\rm TT}\nabla^2(Ch^{\rm TT})_{ij}
\right].
\end{equation}
The replacement
\(h_{ij}^{\rm TT}\rightarrow\dot h_{ij}^{\rm TT}\) relates the
Nieh-Yan induced spin-2 gravitomagnetic  helicity density to the spin-2 gravito-current helicity density, but does not transform it directly into the complete
Pontryagin descendant.  Relating these two expressions requires the field
equations, integrations by parts, and suitable boundary conditions.

Although these quantities descend from four-dimensional topological
densities, their local representatives on a spatial hypersurface should not
themselves be regarded as topological invariants.  They are gauge invariant
pseudoscalar densities at the linearized level, but their integrals can depend
on the hypersurface, the boundary conditions, and the treatment of
codimension-two surface terms.  In particular, they need not be quantized and
need not be conserved for a general sourced or time-dependent configuration.

We have analyzed these helicities from two complementary perspectives.  The
first is the plane-wave or mode expansion. This analysis makes particularly transparent the close
correspondence between magnetic helicity and the
spin-2 gravitomagnetic helicity.
The second perspective is the source multipole expansion.  In linearized
Einstein gravity, the gauge invariant vector mode is determined by the
transverse momentum density,
whereas the transverse traceless tensor mode is sourced by
\(T_{ij}^{\rm TT}\) and propagates according to a retarded wave equation.
The multipole expansion of \(\Xi_i\) relates the spin-1 densities to the total
momentum, current dipole, quadrupole, and the remaining irreducible
moments of the momentum distribution. The wave zone expansion of
\(h_{ij}^{\rm TT}\), on the other hand, relates the spin-2 densities to time
derivatives of the source mass quadrupole and to higher radiative multipoles.
Care is required because the transverse projections \(T_{0i}^{\rm T}\) and
\(T_{ij}^{\rm TT}\) are spatially nonlocal and need not have compact support,
even when the original stress tensor does.

These general results were illustrated by several physically distinct
configurations.  For two-body systems, the spin-1 densities probe the
constrained field generated by orbital momentum and angular momentum, while
the spin-2 densities describe the helicity distribution of quadrupolar
gravitational radiation.  Bound and unbound trajectories provide different
notions of averaging: periodic averaging for elliptic orbits and accumulated
helicity along the complete encounter for parabolic and hyperbolic orbits.
The corresponding angular distributions also provide a natural setting for
the helicity weighted characteristic frequency.
 A particularly noteworthy feature of the elliptic orbit result is the
existence of a critical eccentricity
\begin{equation}
e_{\rm c}\simeq 0.32619.
\end{equation}
At \(e=e_{\rm c}\), the angular dependences of the period-averaged
spin-2 gravitomagnetic and gravito-current  helicity densities become
proportional.  Their ratio is therefore independent of the observation
direction, and the normalized characteristic frequency satisfies
\begin{equation}
W(\theta,\phi)
\equiv
\frac{\omega_\star^2(\theta,\phi)}{4\Omega^2}
=
2.33279.
\end{equation}
The all-sky distribution of \(W\) is consequently uniform at the critical
eccentricity, so that there is no distinguished direction of maximum or
minimum characteristic frequency.  This does not imply that either
helicity density, or the gravitational radiation itself, is isotropic. Instead,
their angular anisotropies cancel in the ratio defining
\(\omega_\star\).  As \(e\) crosses \(e_{\rm c}\), the sign of the
remaining angular-anisotropy coefficient reverses, interchanging the
directions in which \(W\) attains its maxima and minima.

A boosted Kerr black hole provides a complementary nonradiative example.  It
shows how the total momentum and intrinsic spin of a compact object combine
in the spin-1 sector when the field is described relative to a chosen
asymptotic inertial frame.  The gyratonic pp-wave illustrates a null
configuration carrying both energy and intrinsic angular momentum, for which
spin-1 and spin-2 SVT components can coexist. Interestingly, the two helicity sectors consequently probe complementary scales of the same
source.  The spin-1 characteristic  length measures the transverse beam width \(w\), whereas
the spin-2 characteristic frequency measures the inverse pulse duration \(\tau^{-1}\).  The
energy \(E\), Newton's constant \(G\), the polarization  \(\lambda_\gamma\), and the
carrier frequency \(\omega_0\) determine the amplitudes of the helicity
densities but cancel from these characteristic scale ratios.  The helicity
ratios therefore separate the geometry of the source from its overall
strength: the vector sector resolves its transverse vortical structure,
while the tensor sector resolves its temporal variation.  Finally, the type N gravitational
Hopfion provides a localized source-free radiative configuration.  Under the
vacuum constraints and asymptotically decaying boundary conditions, its
spin-1 mode vanishes and only the spin-2 helicity densities remain.  These
metric helicities are related to, but should not be identified directly with,
the linking of the tendex and vortex fields constructed from the Weyl
curvature.

Several directions deserve further investigation.

\begin{enumerate}
\item
The relation between equal-time helicity densities and radiative helicity
fluxes at \(\mathscr I^+\) should be developed more systematically.
A spatial density on a \(t=\mathrm{const}\) slice and a flux density on
\(\mathscr I^+\) are different observables.  Their relation requires a
controlled matching limit in the region where the wave zone and
null infinity expansions are simultaneously valid.

\item
The codimension-two surface terms should be studied for finite regions and
open boundaries.  The electromagnetic analogy suggests that a gravitational
counterpart of relative helicity may be necessary when the relevant fields
have nonvanishing flux through the boundary.

\item
The four densities can be extracted from analytical waveforms and
numerical binary black hole simulations.  Their angular 
distributions and time dependence may provide interesting observations on circular polarization, precession,
parity violation, and helicity memory that are complementary to the usual
energy and angular momentum fluxes.
\item The relation between the Pauli-Lubanski
operator and the spin-2 helicity density suggests a
representation-theoretic extension
to free massless fields of arbitrary integer spin.  Let
\(\varphi_{i_1\cdots i_s}^{\rm TT}\) be a symmetric, transverse, and
traceless rank-\(s\) field on a spatial hypersurface.  Its generalized
symmetric curl may be defined by
\begin{equation}
(C\varphi)_{i_1\cdots i_s}
=
\epsilon_{kl(i_1}\partial_k
\varphi_{i_2\cdots i_s)l}.
\end{equation}
With the normalized symmetrization used here, the temporal
Pauli-Lubanski operator satisfies
\begin{equation}
W^0_{(s)}=sC.
\end{equation}
On circularly polarized plane waves,
\begin{equation}
C\varphi^{(\pm s)}
=
\pm k\,\varphi^{(\pm s)},
\qquad
W^0_{(s)}\varphi^{(\pm s)}
=
\pm sk\,\varphi^{(\pm s)},
\end{equation}
which reproduces the physical helicities \(\pm s\).  This observation
naturally leads to a hierarchy of parity odd quadratic functionals,
\begin{equation}
H_s^{(n)}
=
\int d^3\bm x\,
\varphi_{i_1\cdots i_s}^{\rm TT}
(-\nabla^2)^n
(C\varphi^{\rm TT})_{i_1\cdots i_s},
\qquad
n=0,1,\ldots .
\end{equation}
The \(n=0\) member generalizes the magnetic or gravitomagnetic helicity,
whereas the higher-\(n\) members measure progressively higher signed
moments of the helicity spectrum.  The cases \(s=1\) and \(s=2\) studied
in this work are the first two members of this possible higher spin
hierarchy. The Pauli-Lubanski construction alone, however, fixes only the helicity
operator and its spectrum.  It does not determine the normalization,
conservation law, boundary completion, or topological origin of
\(H_s^{(n)}\).  These questions should first be investigated for free
Fronsdal fields and then, where possible, in frame-like higher spin
theories on Minkowski or anti-de Sitter backgrounds. 

A related open problem is whether the higher spin quadratic helicities can
be obtained from the descent of genuine higher spin characteristic forms.
For an ordinary gauge connection, Chern-Weil theory distinguishes clearly
between a bulk characteristic class, its boundary Chern-Simons or
transgression form, and the functional induced on a chosen hypersurface.
For metric-like higher spin fields, by contrast, the generalized curvature
carries several antisymmetric index pairs and does not automatically define
an ordinary differential form characteristic class.

It is important to distinguish four notions in such an extension:
a bulk topological characteristic class, a boundary descendant, a
gauge invariant helicity functional, and an integer-valued linking or Hopf
invariant.  The results of the present work establish a relation between the
first three notions for spin 1 and spin 2 under specified boundary
conditions, but they do not imply that the resulting helicity integrals are
integer-valued topological invariants.  Establishing a genuine linking
interpretation, especially for gravitational Hopfions and their
higher spin analogues, remains an interesting open problem.
\end{enumerate}

 \vspace{3pt}
{\bf Acknowledgments.} 
The work of J.L. is supported by NSFC Grant No. 12575074.
\appendix
\section{Boundary terms at spatial infinity}\label{falloff0}
 Here we consider a weak field and slowly moving binary system in its
center-of-mass frame,
\begin{equation}
P_i=0,
\qquad
\int d^3\mathbf x\,x_iT_{00}=0.
\end{equation}
The limit considered in this subsection is
\begin{equation}
t=\text{finite},
\qquad
r\longrightarrow\infty.
\end{equation}
It approaches spatial infinity \(i^0\), rather than  $\mathscr I^+$,
because
\begin{equation}
u=t-r\longrightarrow-\infty.
\end{equation} At leading Newtonian order, the two gauge invariant scalar modes behave as
\begin{align}
\Phi(t,\mathbf x)
&=
-\frac{G\overline M}{r}
+\mathcal O(r^{-3}),
\\
\Theta(t,\mathbf x)
&=
\frac{2G\overline M}{r}
+\mathcal O(r^{-3}),
\end{align}
where \(\overline M=M_1+M_2\) is the total mass of the binary system. The mass-dipole terms are absent in the
center-of-mass frame. Consequently,
\begin{equation}
\Phi=\mathcal O(r^{-1}),
\qquad
\Theta=\mathcal O(r^{-1}),
\end{equation}
while their leading time-dependent contributions arise from the mass
quadrupole:
\begin{equation}
\dot\Phi,\dot\Theta=\mathcal O(r^{-3}),
\qquad
\partial_i\Phi,\partial_i\Theta=\mathcal O(r^{-2}).
\end{equation}

In a general inertial frame, the spin-1 mode has the leading behavior\footnote{One can find this expression in the next section.}
\begin{equation}
\Xi_i
=
\frac{2G}{r}
\left[
P_i+n_i(\mathbf P\cdot\mathbf n)
\right]
+\mathcal O(r^{-2}).
\end{equation}
Thus, \(\Xi_i=\mathcal O(r^{-1})\) when the total momentum is nonzero.
In the center-of-mass frame, the \(r^{-1}\) term vanishes and one obtains
\begin{equation}
\Xi_i
=\mathcal O(r^{-2}),
\end{equation}
together with
\begin{equation}
\partial_j\Xi_i=\mathcal O(r^{-3}),
\qquad
\partial_j\partial_k\Xi_i
=
\nabla^2\Xi_i
=
\mathcal O(r^{-4}).
\end{equation}

The leading wave zone expression for the spin-2 mode is
\begin{equation}
h_{ij}^{\rm TT}(t,r\mathbf n)
=
\frac{2G}{r}
\Pi_{ij,kl}(\mathbf n)
\ddot I_{kl}(u)
+\mathcal O(r^{-2}),
\qquad
u=t-r.
\end{equation}
Its behavior at spatial infinity depends on the boundary data at
\(u\to-\infty\). For an idealized Keplerian binary that has been periodic
for all past times,
\begin{equation}
\ddot I_{ij}(u)=\mathcal O(1),
\end{equation}
and hence
\begin{equation}
h_{ij}^{\text{TT}}=\mathcal O(r^{-1}).
\end{equation}
However, \(r h_{ij}^{\rm TT}\) then oscillates as \(r\to\infty\), and in
general does not possess a smooth limit at \(i^0\).
For an isolated system that becomes stationary in the remote past, one may
instead impose
\begin{equation}
I_{ij}^{(2)}(u)\longrightarrow0,
\qquad
u\longrightarrow-\infty.
\end{equation}
It follows that
\begin{equation}
h_{ij}^{\text{TT}}=o(r^{-1}).
\end{equation}
With a regular asymptotic expansion, this behavior is usually strengthened
to \(\mathcal O(r^{-2})\) or faster.
For the idealized, eternally periodic binary in the center-of-mass frame,
the six modes can therefore be summarized as
\begin{equation}
\left(
\Phi,\Theta,\Xi_{+1},\Xi_{-1},h_{+2},h_{-2}
\right)
=
\left(
\mathcal O(r^{-1}),
\mathcal O(r^{-1}),
\mathcal O(r^{-2}),
\mathcal O(r^{-2}),
\mathcal O(r^{-1}),
\mathcal O(r^{-1})
\right).
\end{equation}
For an isolated system with stationary remote-past data, the last two
entries should instead be replaced by \(o(r^{-1})\).

We first impose regular spatial-infinity boundary conditions. In particular,
the coefficient of the leading \(r^{-1}\) tensor field is assumed to be
stationary as \(u\to-\infty\). The relevant derivative falloffs are
\bs\begin{align}
h_{ij}^{\rm TT}
&=\mathcal O(r^{-1}),
&
\partial_kh_{ij}^{\rm TT}
&=\mathcal O(r^{-2}),
&
\partial_k\partial_lh_{ij}^{\rm TT}
&=\mathcal O(r^{-3}),
\\
\dot h_{ij}^{\rm TT}
&=\mathcal O(r^{-2}),
&
\partial_k\dot h_{ij}^{\rm TT}
&=\mathcal O(r^{-3}),
\\
\Xi_i
&=\mathcal O(r^{-2}),
&
\partial_j\Xi_i
&=\mathcal O(r^{-3}),
&
\partial_j\partial_k\Xi_i
&=\mathcal O(r^{-4}),
\\
\dot\Theta
&=\mathcal O(r^{-3}),
&
\partial_i\dot\Theta
&=\mathcal O(r^{-4}).
\end{align}\es
Substitution into the five boundary densities gives\bs 
\begin{align}
\mathcal B^r_{\Xi\Xi}
&=\mathcal O(r^{-6}),
&
\mathcal B^r_{hh}
&=\mathcal O(r^{-4}),
\\
\mathcal B^r_{\dot\Theta\Xi}
&=\mathcal O(r^{-6}),
&
\mathcal B^r_{\dot h\Xi}
&=\mathcal O(r^{-5}),
&
\mathcal B^r_{\Theta h}
&=\mathcal O(r^{-4}).
\end{align}\es 
Since the area element on a large sphere is
\begin{equation}
dS=r^2d\Omega,
\end{equation}
the corresponding surface contributions behave as
\begin{equation}
\begin{array}{c|ccccc}
&
\mathcal B^r_{\Xi\Xi}
&
\mathcal B^r_{hh}
&
\mathcal B^r_{\dot\Theta\Xi}
&
\mathcal B^r_{\dot h\Xi}
&
\mathcal B^r_{\Theta h}
\\ \hline
r^2\mathcal B^r
&
\mathcal O(r^{-4})
&
\mathcal O(r^{-2})
&
\mathcal O(r^{-4})
&
\mathcal O(r^{-3})
&
\mathcal O(r^{-2})
\end{array}.
\end{equation}
They consequently vanish at spatial infinity:
\begin{equation}
\lim_{r\to\infty}
\int_{S_r}dS\,\mathcal B^r_{\alpha}=0,
\qquad
\alpha\in
\left\{
\Xi\Xi,hh,\dot\Theta\Xi,\dot h\Xi,\Theta h
\right\}.
\end{equation}

This conclusion does not follow from
\(h_{ij}^{\rm TT}=\mathcal O(r^{-1})\) alone. For an eternally periodic
binary,
\begin{equation}
h_{ij}^{\rm TT}
=
\frac{F_{ij}(u,\mathbf n)}{r}
+\mathcal O(r^{-2}),
\qquad
u=t-r,
\end{equation}
and radial derivatives acting on \(u\) do not improve the radial falloff:
\begin{equation}
\partial h^{\rm TT},
\quad
\partial^2h^{\rm TT},
\quad
\dot h^{\rm TT},
\quad
\partial\dot h^{\rm TT}
=
\mathcal O(r^{-1}).
\end{equation}
A conservative power counting then gives
\begin{align}
\mathcal B^r_{\Xi\Xi}
&=\mathcal O(r^{-6}),
&
\mathcal B^r_{hh}
&=\mathcal O(r^{-2}),
\\
\mathcal B^r_{\dot\Theta\Xi}
&=\mathcal O(r^{-6}),
&
\mathcal B^r_{\dot h\Xi}
&=\mathcal O(r^{-3}),
&
\mathcal B^r_{\Theta h}
&=\mathcal O(r^{-3}).
\end{align}
All surface terms still vanish except for the tensor-tensor term, which may
have a finite limit. Its leading local contribution is
\begin{equation}
\mathcal B^r_{hh}
=
-\frac{1}{2r^2}
\epsilon^{ijk}n_j
F_{i\ell}(u,\mathbf n)
\partial_u^2F_{k\ell}(u,\mathbf n)
+\mathcal O(r^{-3}).
\end{equation}
Therefore, the vanishing of the complete codimension-two boundary term
requires a stationary remote past condition, a no incoming radiation
condition with suitable regularity, or an explicit condition setting the
tensor flux through spatial infinity to zero.

\section{Pauli-Lubanski operator and the helicity of spin-1 field}\label{pauli1}
This section discuss the relation between Pauli-Lubanski operator and the magnetic helicity in  electromagnetism.  Applying the temporal
Pauli-Lubanski component to a four-vector  and subsequently performing the
Helmholtz decomposition\footnote{On a fixed spatial hypersurface, decompose the electromagnetic field
\begin{equation}
 a_0=\varphi,
 \qquad
 a_i=a_i^{\mathrm T}+\partial_i\psi,
 \qquad
 \partial_ia_i^{\mathrm T}=0.
 \label{eq:Helmholtz}
\end{equation}
Under the gauge transformation
\begin{equation}
 a_\mu\longrightarrow a_\mu+\partial_\mu\alpha,
\end{equation}
the variables transform according to
\begin{equation}
 \delta\varphi=\dot\alpha,
 \qquad
 \delta\psi=\alpha,
 \qquad
 \delta a_i^{\mathrm T}=0.
\end{equation}
The reduced gauge invariant variables may thus be chosen as
\begin{equation}
 \Psi=\varphi-\dot\psi,
 \qquad
 a_i^{\mathrm T}.
\end{equation}} selects the transverse vector potential
\be (W^0\mathbf a^{\mathrm T})_i=(\bm\nabla\times\mathbf a^{\mathrm T})_i=b_i.\ee 
This gives a gauge invariant, but spatially nonlocal, representative
\be \widetilde h_{m}=\mathbf a^{\mathrm T}\cdot\mathbf b\label{tildehm}\ee  
of the usual magnetic helicity density.
Its integral agrees with \(\int\mathbf a\cdot\mathbf b\, d^3x\) for closed,
periodic, or sufficiently decaying fields. The transverse field is obtained using the nonlocal projector
\begin{equation}
 a_i^{\mathrm T}
 =
 \hat P_{ij}a_j.
\end{equation}  With the gauge invariant density \eqref{tildehm}, we may define the integrated quantity,
\begin{equation}
 \widetilde{H}_{\mathrm m}
 =
 \int_V d^3x\,
 \mathbf a^{\mathrm T}\cdot\mathbf b
 =
 \int_V d^3x\,
 a_i^{\mathrm T}(W^0_{(1)}a^{\mathrm T})_i
 \label{eq:Hm-transverse}
\end{equation}
The standard magnetic helicity 
\be 
H_m=\int_V d^3x\ \mathbf a\cdot\mathbf b \label{eq:standard-Hm}
\ee 
 plays a central role in force-free magnetic fields and in the
topological description of linked flux tubes.  Substituting
\(\mathbf a=\mathbf a^{\mathrm T}+\bm\nabla\psi\) gives
\begin{align}
 H_{\mathrm m}
 =
 \int_V d^3x\,
 \mathbf a^{\mathrm T}\cdot\mathbf b
 +
 \int_V d^3x\,
 \bm\nabla\psi\cdot\mathbf b
=
 \widetilde H_m
 +
 \int_{\partial V} d^2y \sqrt{q} r_i\,
 \psi\,b^i,
 \label{eq:Hm-boundary}
\end{align}
where \(\bm\nabla\cdot\mathbf b=0\) has been used.  
Whenever one of the following standard conditions holds:
\begin{equation}
 \begin{aligned}
 &r_i b_i\big|_{\partial V}=0
 \quad\text{or periodic boundary conditions},
 \\
 &\text{or sufficiently rapid falloff at spatial infinity}.
 \end{aligned}
 \label{eq:magnetic-boundary-conditions}
\end{equation}
Equation~\eqref{eq:Hm-transverse} is therefore not a new global magnetic
helicity.  It is the transverse-potential representative of the standard
functional \eqref{eq:standard-Hm}. 

At the density level,
\begin{equation}
 \mathbf a \cdot\mathbf b
 -
 \mathbf a^{\mathrm T}\cdot\mathbf b 
 =
 \bm\nabla\cdot(\psi\mathbf b).
 \label{eq:density-improvement}
\end{equation}
Thus \(\mathbf a\cdot\mathbf b\) is local but gauge dependent, whereas
\(\mathbf a^{\mathrm T}\cdot\mathbf  b\) is gauge invariant after fixing the
Helmholtz boundary problem but nonlocal as a functional of the original
potential.  This use of transverse potentials is standard in reduced
phase space and duality based descriptions of electromagnetism
\cite{Calkin1965,DeserTeitelboim1976,Stewart2003,
CameronBarnettYao2012,BliokhBekshaevNori2013}.
Since 
\begin{equation}
 \bm\nabla\times\mathbf b
 =
 \bm\nabla\times
 (\bm\nabla\times\mathbf a^{\mathrm T})
 =
 -\nabla^2\mathbf a^{\mathrm T},
\end{equation}
it follows that
\begin{equation}
 \mathbf  a^{\mathrm T}
 =
 -\nabla^{-2}(\bm\nabla\times\mathbf b).
 \label{eq:AT-from-B}
\end{equation}
On \(\mathbb R^3\), with fields decaying at infinity, this becomes
\begin{equation}
 \mathbf a^{\mathrm T}(\bm x)
 =
 \frac{1}{4\pi}
 \int d^3\bm y\,
 \frac{
 \mathbf b(\bm y)\times(\bm x-\bm y)
 }{
 |\bm x-\bm y|^3
 }.
 \label{eq:Biot-Savart-AT}
\end{equation}
The transverse representation of magnetic helicity can hence be written
entirely in terms of \(\mathbf b\) as the nonlocal Gauss-type integral
\begin{equation}
\widetilde H_{\mathrm m}
 =
 \frac{1}{4\pi}
 \int d^3\bm x\, d^3\bm y\,
 \frac{
 \mathbf b(\bm x)\cdot
 [\mathbf b(\bm y)\times(\bm x-\bm y)]
 }{
 |\bm x-\bm y|^3
 }= \frac{1}{4\pi}
 \int d^3\bm x\, d^3\bm y\,
 \frac{
 \left(\mathbf b(\bm x)\times
 \mathbf b(\bm y)\right)\cdot (\bm x-\bm y)
 }{
 |\bm x-\bm y|^3
 }.
 \label{eq:Gauss-helicity}
\end{equation}
This expression makes the connection with flux tube linking explicit
\cite{Moffatt1969}.
\section{Helicity densities for moving particles}\label{moving}
For point particles with worldlines $z_A^\mu(t)=(t,\bm y_A(t))$, the
stress tensor on a Minkowski background is
\begin{align}
 T^{\mu\nu}(t,\bm x)
 &=\sum_A m_A\int d\tau_A\,
 u_A^\mu u_A^\nu\,
 \delta^{(4)}\bigl(x-z_A(\tau_A)\bigr) \notag\\
 &=\sum_A \frac{m_Au_A^\mu u_A^\nu}{u_A^0}
 \delta^{(3)}\bigl(\bm x-\bm y_A(t)\bigr),
 \label{eq:Tpp}
\end{align}
where the four-velocity is $u_A^\mu=\gamma_A(1,\bm v_A)$ and
$\gamma_A=(1-\bm v_A^2)^{-1/2}$.  Thus
\begin{align}
 T^{00}&=\sum_A m_A\gamma_A\delta_A,
 &T^{0i}&=\sum_A m_A\gamma_Av_A^i\delta_A,
 &T^{ij}&=\sum_A m_A\gamma_Av_A^iv_A^j\delta_A,
 \label{eq:Tcomponents}
\end{align}
with $\delta_A=\delta^{(3)}(\bm x-\bm y_A)$.
According to the formula \eqref{sourcetoXi}, 
\be 
\Xi_i=2G\int d^3\mathbf x' \frac{\delta_{ij}+N_i N_j}{R}T_{0j}(t,\mathbf x')=2G\sum_A \frac{p_{Ai}+N_{Ai}p_{Aj}N_{Aj}}{R_A}
\ee 
where 
\be 
p_{Ai}=-m_A \gamma_A v_{Ai},\quad R_A=|\mathbf x-y_A(t)|, \qquad
\mathbf N_A=\frac{\mathbf x-\mathbf y_A(t)}{R_A}.
\ee
Here $\mathbf N_A$ is the unit vector pointing from particle $A$ to the
field point, and $R_A$ is the corresponding distance. 
This expression should be understood as the leading post-Newtonian
term of the spin-1 field, rather than as a complete solution sourced
by the matter stress tensor alone. In the conserved effective source
of the post-Minkowskian formulation,
\begin{equation}
 \tau^{\mu\nu}
 =
 \sqrt{-g}T_{\rm pp}^{\mu\nu}
 +t_{\rm grav}^{\mu\nu},
 \qquad
 \partial_\mu\tau^{\mu\nu}=0,\label{taupM}
\end{equation}
the leading momentum density is supplied by the point particle matter
current, whereas interaction and gravitational field corrections enter
at higher post-Newtonian orders. Consequently, the above expression is
sufficient for the leading spin-1 helicity densities of a slowly moving
binary away from the particle worldlines.

 Therefore, the spin-1 gravitomagnetic and gravito-current helicity densities are 
\bs\label{largerH}\begin{align}
    \mathcal H_{NY}^V&
 =
-2G^2\sum_{A\not=B}
\frac{
\left[
\mathbf p_A+
\mathbf N_A(\mathbf p_A\cdot\mathbf N_A)
\right]\cdot
(\mathbf p_B\times\mathbf N_B)}
{R_AR_B^2}
, \label{eq:HNYexact}\\ 
    \mathcal H_{CS}^V&=
-8G^2\sum_{A\not=B}
\frac{
(\mathbf p_A\times\mathbf N_A)\cdot
\left[
\mathbf p_B
-3\mathbf N_B(\mathbf p_B\cdot\mathbf N_B)
\right]}
{R_A^2R_B^3}
. \label{eq:HCSexact}
\end{align}\es 

The summation is over $A\not=B$.
The $A=B$ self-contribution vanishes algebraically at every regular field
point. In other words, nonzero terms are interference terms between distinct particle fields.
At the worldlines the point-particle fields are singular, so an integrated
helicity requires excision, a finite-size model, or a regularization
prescription. One can check that the point-particle expression \eqref{largerH} for the spin-1 constraint field is
consistent with the binary calculation at leading post-Newtonian order.

For \(r\gg |\bm y_A(t)|\), the large-\(r\) expansion of
Eqs.~\eqref{eq:HNYexact}--\eqref{eq:HCSexact} reproduces the general
asymptotic form \eqref{leadingp} with the identifications
\be 
D_{\langle ij\rangle}
=
\sum_A y_{A\langle i}p_{Aj\rangle},
\qquad
P_i=\sum_Ap_{Ai},
\qquad
S_i=\sum_A\epsilon_{ijk}y_{Aj}p_{Ak}.
\ee 
Next we turn to the radiative modes.  In the wave zone, where \(r\)
and \(t\) are large while the retarded time \(u=t-r\) is held fixed, the
transverse traceless metric perturbation takes the form
\be 
h_{ij}^{\mathrm{TT}}
=
\frac{4G}{r}
\Pi_{ij,kl}(\mathbf n)
\sum_A
\left.
\frac{
m_A\gamma_Av_{Ak}v_{Al}
}{
1-\mathbf n\cdot\mathbf v_A
}
\right|_{t_A}
+\mathcal O(r^{-2}).\label{wavezoneh}
\ee where $t_A=u+\mathbf n\cdot\mathbf y_A(t_A)$. Above equation is the leading
matter contribution to the far-zone TT field. However, its interpretation
requires some care. The TT wave equation is not independent of the remaining linearized
Einstein equations. By the linearized Bianchi identity, a consistent
source must satisfy
\begin{equation}
 \partial_\mu T^{\mu\nu}=0.
\end{equation}
For the point particle matter stress tensor one instead finds
\begin{equation}
 \partial_\mu T_{\rm pp}^{\mu\nu}
 =
 \sum_A m_A\int d\tau_A\,
 a_A^\nu\,
 \delta^{(4)}
 \bigl(x-z_A(\tau_A)\bigr)
 \label{eq:div-Tpp}
\end{equation} where $a_A$ is the acceleration of the particle $A$. Consequently, Eq.~\eqref{wavezoneh} can be used
directly for freely moving particles in Minkowski spacetime. It can
also describe the asymptotic incoming and outgoing inertial states of
a scattering process. In such regions the velocities are constant,
and the resulting field is an asymptotic boosted Coulomb or memory
field rather than radiation produced during the interaction. For a self-gravitating binary, the role of the conserved source is
played by the post-Minkowskian effective stress tensor
\eqref{taupM}
where $t^{\mu\nu}_{\text{grav}}$ contains the gravitational interaction
terms. At leading post-Newtonian order, source conservation converts
the far-zone stress integral into the second derivative of the mass
quadrupole,
\begin{equation}
 h_{ij}^{\rm TT}
 =
 \frac{2G}{r}
 \Pi_{ij,kl}(\mathbf n)
 \ddot I_{kl}(u)
 +\mathcal O(r^{-2},\text{higher PN}).
\end{equation}
Accordingly, the spin-2 helicity densities of a bound binary must be
computed from the quadrupole waveform rather than from
Eq.~\eqref{wavezoneh}. More precisely, according to Eq.\eqref{defIij}, 
\be 
h_{ij}^{\rm TT}
=
\frac{4G}{r}\Pi_{ij,kl}(\mathbf n)
\sum_A m_A
\left(
v_{A\langle k}v_{Al\rangle}
+y_{A\langle k}a_{Al\rangle}
\right)
+\mathcal O(r^{-2},\text{higher PN}).
\ee In the slow motion limit, the above expression distinguishes from \eqref{wavezoneh} by the term associated with the acceleration. With this expression, we find 
\bs\begin{align}
\mathcal H_{\rm NY}^{T}
={}&
\frac{4G^2}{r^2}
\sum_{A,B}m_Am_B
\left(
3v_{A\langle i}a_{Aj\rangle}
+y_{A\langle i}j_{Aj\rangle}
\right)
\left(
v_{B\langle k}v_{Bl\rangle}
+y_{B\langle k}a_{Bl\rangle}
\right)
Q_{ijkl}(\mathbf n)
+\mathcal O(r^{-3}),\\ 
\mathcal H_{\rm CS}^{T}
={}&
\frac{16G^2}{r^2}
\sum_{A,B}m_Am_B
\left(
3a_{A\langle i}a_{Aj\rangle}
+4v_{A\langle i}j_{Aj\rangle}
+y_{A\langle i}s_{Aj\rangle}
\right)
\left(
3v_{B\langle k}a_{Bl\rangle}
+y_{B\langle k}j_{Bl\rangle}
\right)
Q_{ijkl}(\mathbf n)
+\mathcal O(r^{-3}).
\end{align}\es

\iffalse 
In the slow motion limit, we find 
\bs\begin{align}
    \mathcal H_{NY}^T
={}&
\frac{4G^2}{r^2}
\sum_{A,B}m_Am_B
\Big\{
\left[
\mathbf v_A\cdot\mathbf v_B
-(\mathbf n\cdot\mathbf v_A)
(\mathbf n\cdot\mathbf v_B)
\right]
\mathbf n\cdot
(\mathbf v_B\times\mathbf a_A)
\nn\\
&\quad+
\left[
\mathbf a_A\cdot\mathbf v_B
-(\mathbf n\cdot\mathbf a_A)
(\mathbf n\cdot\mathbf v_B)
\right]
\mathbf n\cdot
(\mathbf v_B\times\mathbf v_A)
\Big\}
+\mathcal O(r^{-3},\text{higher PN}),\\ 
 \mathcal H_{CS}^T={}&
\frac{16G^2}{r^2}
\sum_{A,B}m_Am_B
\Bigg\{
2\left[
\mathbf a_A\cdot\mathbf a_B
-(\mathbf n\cdot\mathbf a_A)
(\mathbf n\cdot\mathbf a_B)
\right]
\mathbf n\cdot
(\mathbf v_B\times\mathbf a_A)
\nn\\
&\quad+
2\left[
\mathbf a_A\cdot\mathbf v_B
-(\mathbf n\cdot\mathbf a_A)
(\mathbf n\cdot\mathbf v_B)
\right]
\mathbf n\cdot
(\mathbf a_B\times\mathbf a_A)
\nn\\
&\quad+
\left[
\mathbf v_A\cdot\mathbf a_B
-(\mathbf n\cdot\mathbf v_A)
(\mathbf n\cdot\mathbf a_B)
\right]
\mathbf n\cdot
(\mathbf v_B\times\mathbf j_A)
\nn\\
&\quad+
\left[
\mathbf j_A\cdot\mathbf a_B
-(\mathbf n\cdot\mathbf j_A)
(\mathbf n\cdot\mathbf a_B)
\right]
\mathbf n\cdot
(\mathbf v_B\times\mathbf v_A)
\nn\\
&\quad+
\left[
\mathbf v_A\cdot\mathbf v_B
-(\mathbf n\cdot\mathbf v_A)
(\mathbf n\cdot\mathbf v_B)
\right]
\mathbf n\cdot
(\mathbf a_B\times\mathbf j_A)
\nn\\
&\quad+
\left[
\mathbf j_A\cdot\mathbf v_B
-(\mathbf n\cdot\mathbf j_A)
(\mathbf n\cdot\mathbf v_B)
\right]
\mathbf n\cdot
(\mathbf a_B\times\mathbf v_A)
\Bigg\}
+\mathcal O(r^{-3},\text{higher PN}).
\end{align}\es \fi 
In the quadrupole formulas above, which are derived from \(I_{ij}(u)\) in Eq.~\eqref{defIij}, all worldline quantities \(\mathbf y_A,\mathbf v_A,\mathbf a_A,\mathbf j_A,\mathbf s_A\), and likewise those with label \(B\), are evaluated at the common retarded time \(u\). Here \(j_{Ai}=\dot a_{Ai}\) is the jerk, and \(s_{Ai}=\dot j_{Ai}\) is the snap of particle \(A\), where the dot denotes \(d/du\) in these formulas. The particle dependent retarded times \(t_A\) defined below Eq.~\eqref{wavezoneh} belong to the unexpanded matter field expression and are not used in the quadrupole representation.
%Here \(a_{Ai}=\dot{v}_{Ai}\) and \( j_{Ai}=\dot{a}_{Ai}\)
%are the acceleration and jerk of particle \(A\), evaluated at the
%retarded time \(t_A\). The corresponding quantities for particle \(B\)
%are evaluated at \(t_B\). 
\iffalse  In the slow motion limit used here, the
differences between the retarded times of different particles produce
only higher-order corrections, so at leading order all dynamical
variables may be regarded as evaluated at the common retarded time \(u\). For a single particle of mass \(m\), the above expressions reduce to 
\bs\begin{align}
\mathcal H_{\mathrm{NY}}^T
={}&
\frac{4G^2m^2}{r^2}
\left[
\mathbf v^2-(\mathbf n\cdot\mathbf v)^2
\right]
\mathbf n\cdot(\mathbf v\times\mathbf a)
+\mathcal O(r^{-3}),\\ 
\mathcal H_{\mathrm{CS}}^T
={}&
\frac{16G^2m^2}{r^2}
\Bigg\{
\left[
2\mathbf a^2
-2(\mathbf n\cdot\mathbf a)^2
-\mathbf v\cdot\mathbf j
+(\mathbf n\cdot\mathbf v)
(\mathbf n\cdot\mathbf j)
\right]
\mathbf n\cdot(\mathbf v\times\mathbf a)
\nn\\
&\quad+
\left[
\mathbf v\cdot\mathbf a
-(\mathbf n\cdot\mathbf v)
(\mathbf n\cdot\mathbf a)
\right]
\mathbf n\cdot(\mathbf v\times\mathbf j)
\nn\\
&\quad+
\left[
\mathbf v^2-(\mathbf n\cdot\mathbf v)^2
\right]
\mathbf n\cdot(\mathbf a\times\mathbf j)
\Bigg\}
+\mathcal O(r^{-3}).
\end{align}\es We note that the same combination
\(\mathbf n\cdot(\mathbf v\times\mathbf a)\) has appeared in the
analogous electromagnetic helicity fluxes studied in
\cite{Heng:2025kmr}.\fi
\bibliography{refs}

\end{document}